\documentclass[journal]{IEEEtran}
\usepackage{cite}
\usepackage{cite}
\usepackage{amssymb,amsfonts}
\usepackage{algorithmic}
\usepackage{graphicx}
\usepackage{textcomp}
\usepackage{subfigure}
\usepackage{setspace}
\usepackage{enumitem}
\usepackage{algorithm}
\usepackage{multirow}
\usepackage{textcomp}
\usepackage[flushleft]{threeparttable}
\usepackage{booktabs,tabularx, tabu}
\usepackage{gensymb}
\usepackage{fixltx2e}
\usepackage{dblfloatfix}
\usepackage{amsmath}
\DeclareMathOperator*{\argmin}{argmin}

\makeatletter
\newcommand*{\rom}[1]{\expandafter\@slowromancap\romannumeral #1@}
\makeatother

\makeatletter
\DeclareRobustCommand{\textsupsub}[2]{{%
  \m@th\ensuremath{%
    ^{\mbox{\fontsize\sf@size\z@#1}}%
    _{\mbox{\fontsize\sf@size\z@#2}}%
  }%
}}
\makeatother

\ifCLASSINFOpdf
\else
\fi
\begin{document}
%
\title{Deep Convolutional Neural Network for \\ Identifying Seam-Carving Forgery}


\author{Seung-Hun Nam, Wonhyuk Ahn, In-Jae Yu, Myung-Joon Kwon, Minseok Son, and Heung-Kyu Lee

\thanks{ S.-H. Nam, W. Ahn, I.-J. Yu, M.-J. Kwon, M. Son, and H.-K. Lee are with the School of Computing, Korea Advanced Institute of Science and Technology (KAIST), Daejeon 34141, South Korea (e-mail: nam1202@kaist.ac.kr). \emph{Corresponding author: Heung-Kyu Lee.}}%
\thanks{This work has been submitted to the IEEE for possible publication. Copyright may be transferred without notice, after which this version may no longer be accessible.}}
%
%

\markboth{PREPARED FOR IEEE TRANSACTIONS ON}
{Nam \MakeLowercase{\textit{et al.}}: Deep Convolutional Neural Network for Identifying Seam-Carving Forgery}
%



\maketitle

\begin{abstract}
Seam carving is a representative content-aware image retargeting approach to adjust the size of an image while preserving its visually prominent content.
To maintain visually important content, seam-carving algorithms first calculate the connected path of pixels, referred to as the seam, according to a defined cost function and then adjust the size of an image by removing and duplicating repeatedly calculated seams.
Seam carving is actively exploited to overcome diversity in the resolution of images between applications and devices; hence, detecting the distortion caused by seam carving has become important in image forensics.
In this paper, we propose a convolutional neural network (CNN)-based approach to classifying seam-carving-based image retargeting for reduction and expansion.
To attain the ability to learn low-level features, we designed a CNN architecture comprising five types of network blocks specialized for capturing subtle signals.
An ensemble module is further adopted to both enhance performance and comprehensively analyze the features in the local areas of the given image.
To validate the effectiveness of our work, extensive experiments based on various CNN-based baselines were conducted.
Compared to the baselines, our work exhibits state-of-the-art performance in terms of three-class classification (original, seam inserted, and seam removed).
In addition, our model with the ensemble module is robust for various unseen cases.
The experimental results also demonstrate that our method can be applied to localize both seam-removed and seam-inserted areas.
\end{abstract}


%
\IEEEpeerreviewmaketitle

\section{Introduction}
\label{seam_sec_intro}
\IEEEPARstart{W}{ith} the recent spread of mobile devices, including smartphones and tablet computers, and the use of social networking services, sharing images has become a familiar phenomenon.
The majority of users go through the process of resizing a given image to their preferred size and aspect ratio before sharing the image \cite{resizing}.
In addition, resizing is generally used in the process of overcoming incompatibility between modules because the size and aspect ratio provided by each device and application is different \cite{resizing2,seam1,seam2,seam3,seam4}.
To adjust the size of the image to fit the target size, traditional resizing techniques (e.g., linear scaling and center cropping) have been actively employed in various tasks \cite{icip}.
However, these approaches, which only consider geometric constraints, have the disadvantage that the visually prominent areas of images can be distorted or discarded during the resizing process \cite{cho, dong}.

\begin{figure*}[t]%
\centering%
\subfigure[]{%
  \includegraphics[height=2.3in]{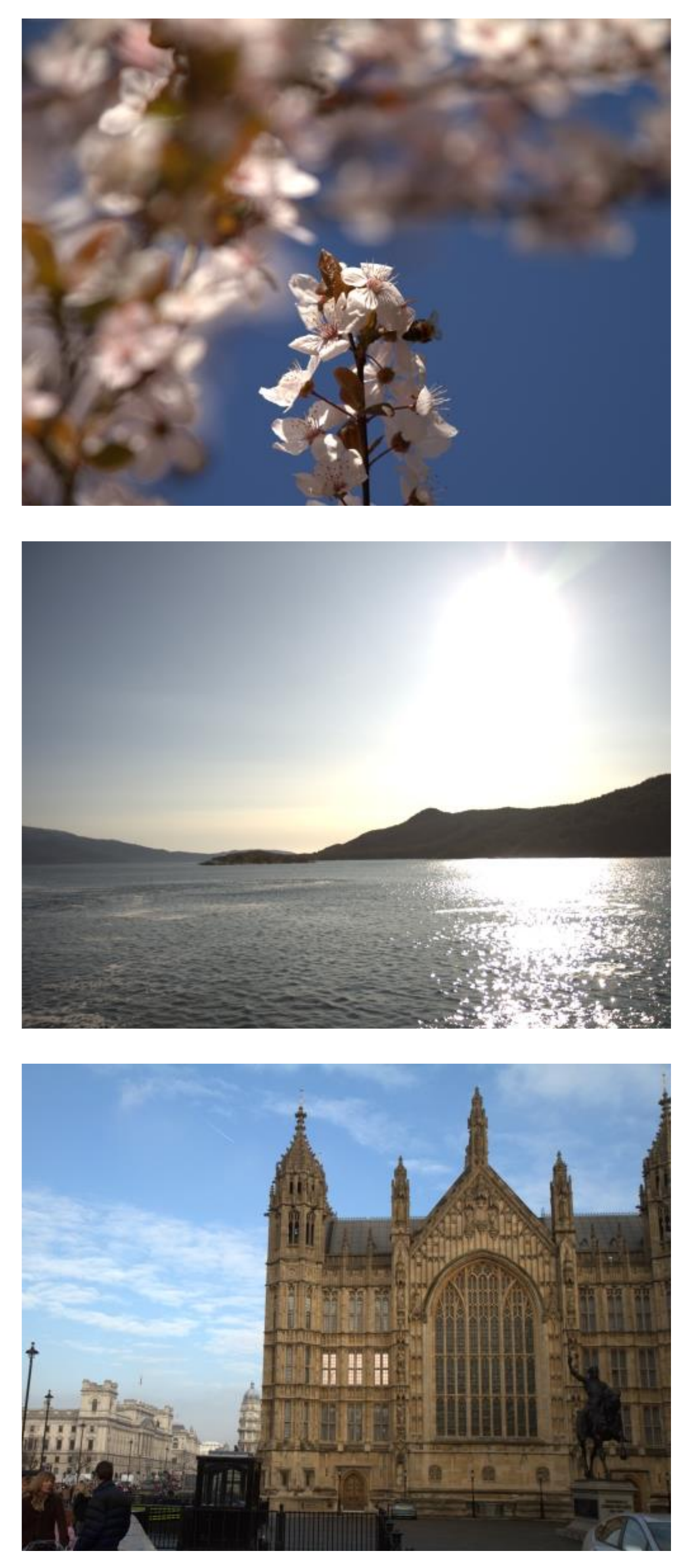}%
  \label{fig1:ex:a}
}\hspace{-3mm}
\subfigure[]{%
  \includegraphics[height=2.3in]{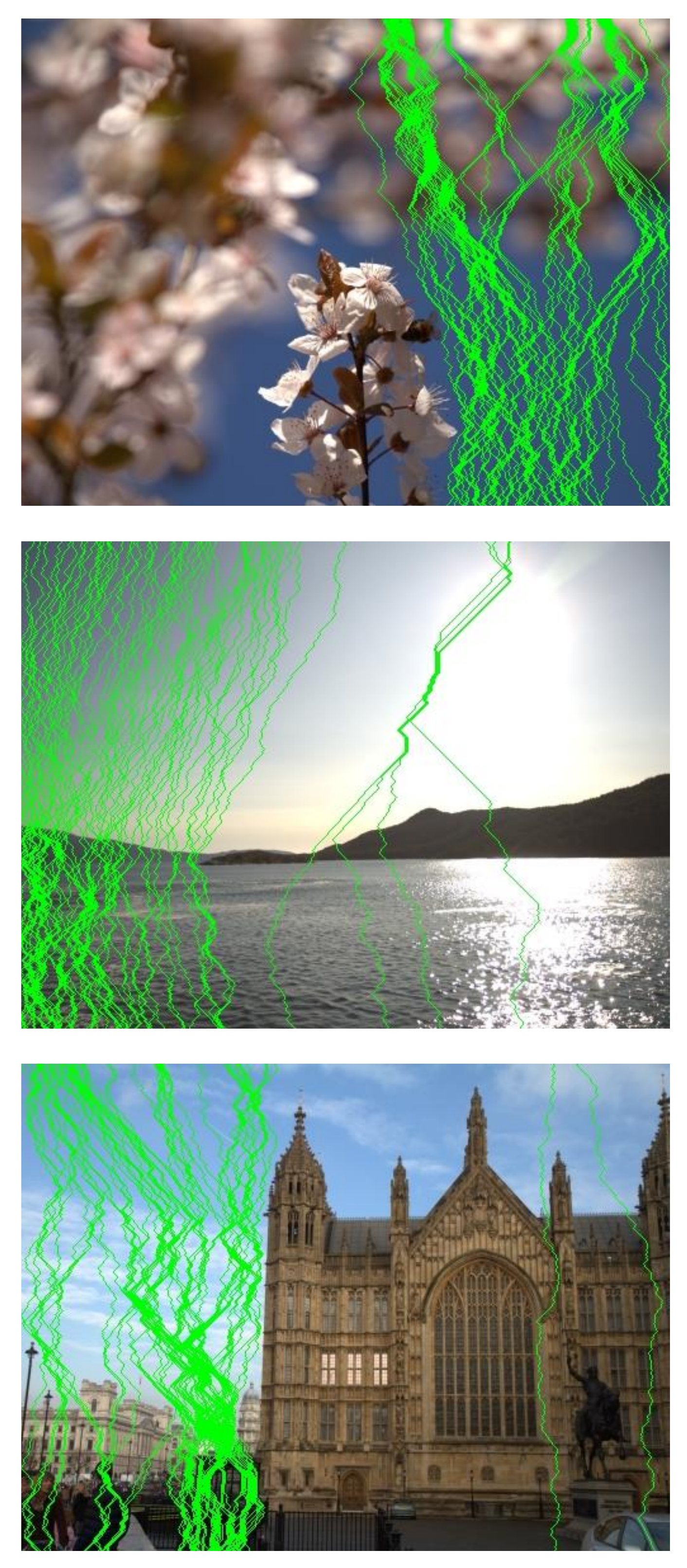}%
  \label{fig1:ex:b}
}\hspace{-3mm}
\subfigure[]{%
  \includegraphics[height=2.3in]{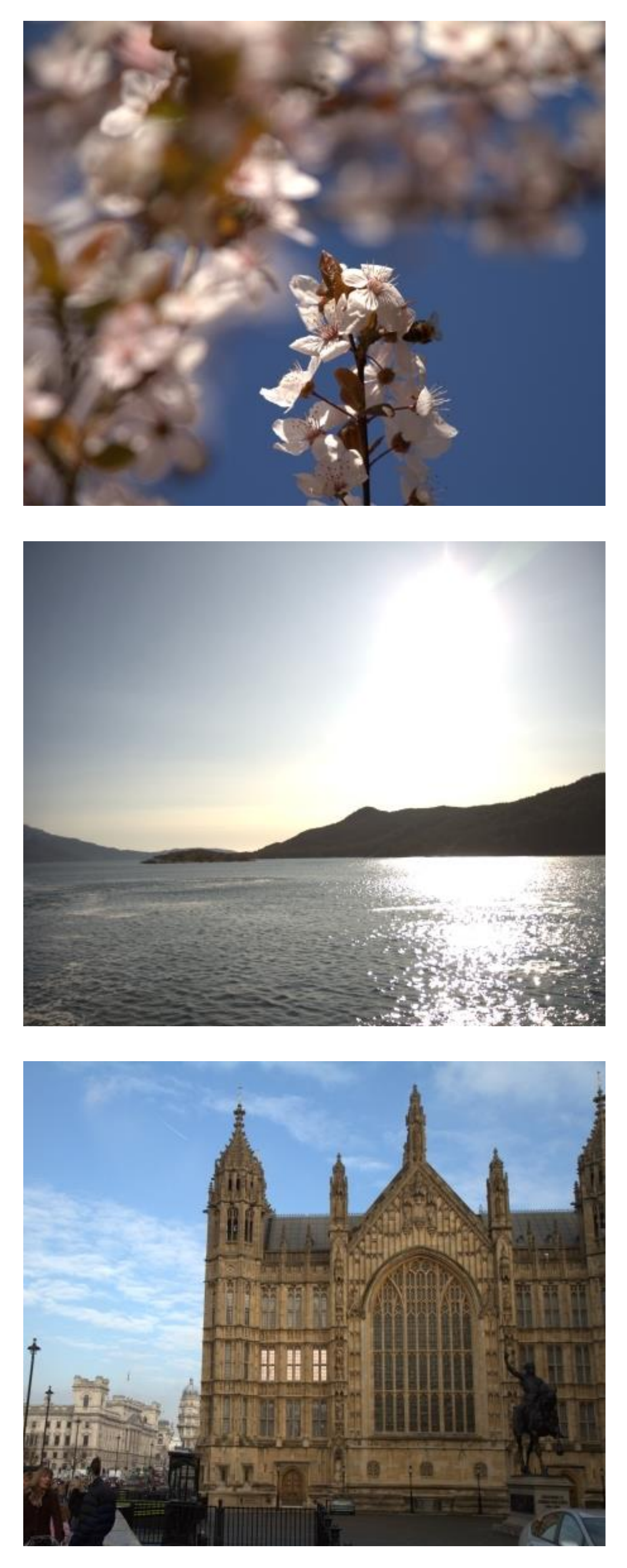}%
  \label{fig1:ex:c}
}\hspace{-3mm}
\subfigure[]{%
  \includegraphics[height=2.3in]{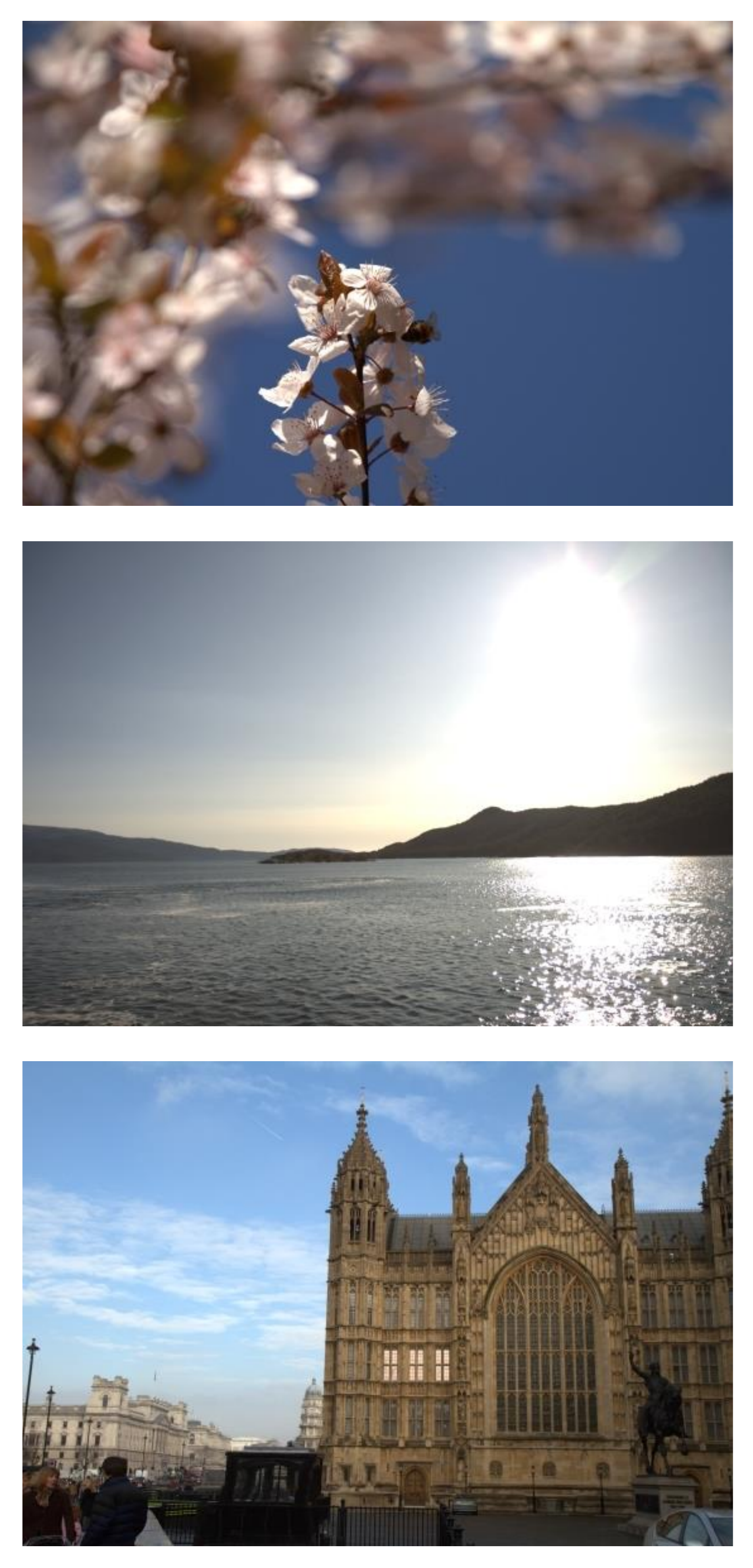}%
  \label{fig1:ex:d}
}\hspace{-3mm}
\subfigure[]{%
  \includegraphics[height=2.3in]{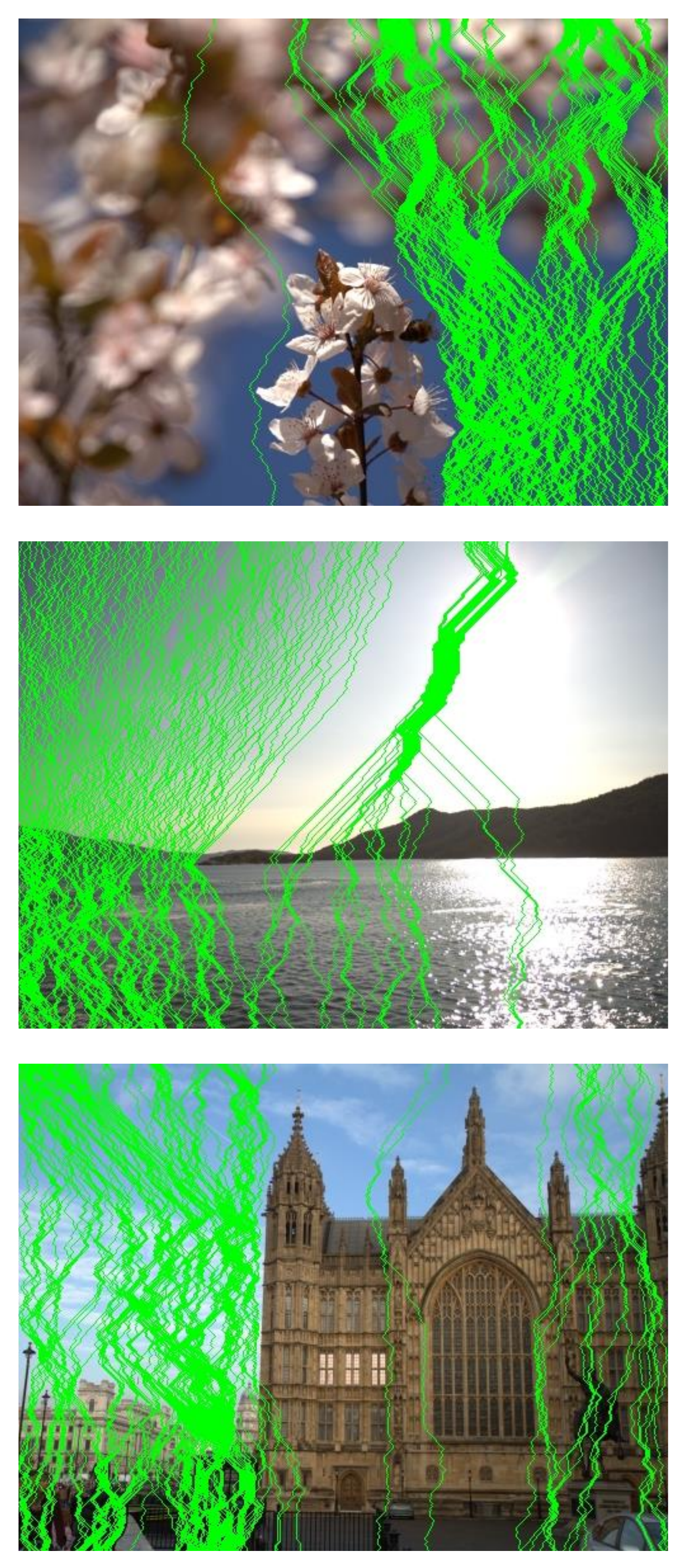}%
  \label{fig1:ex:e}
}\hspace{-3mm}
\subfigure[]{%
  \includegraphics[height=2.3in]{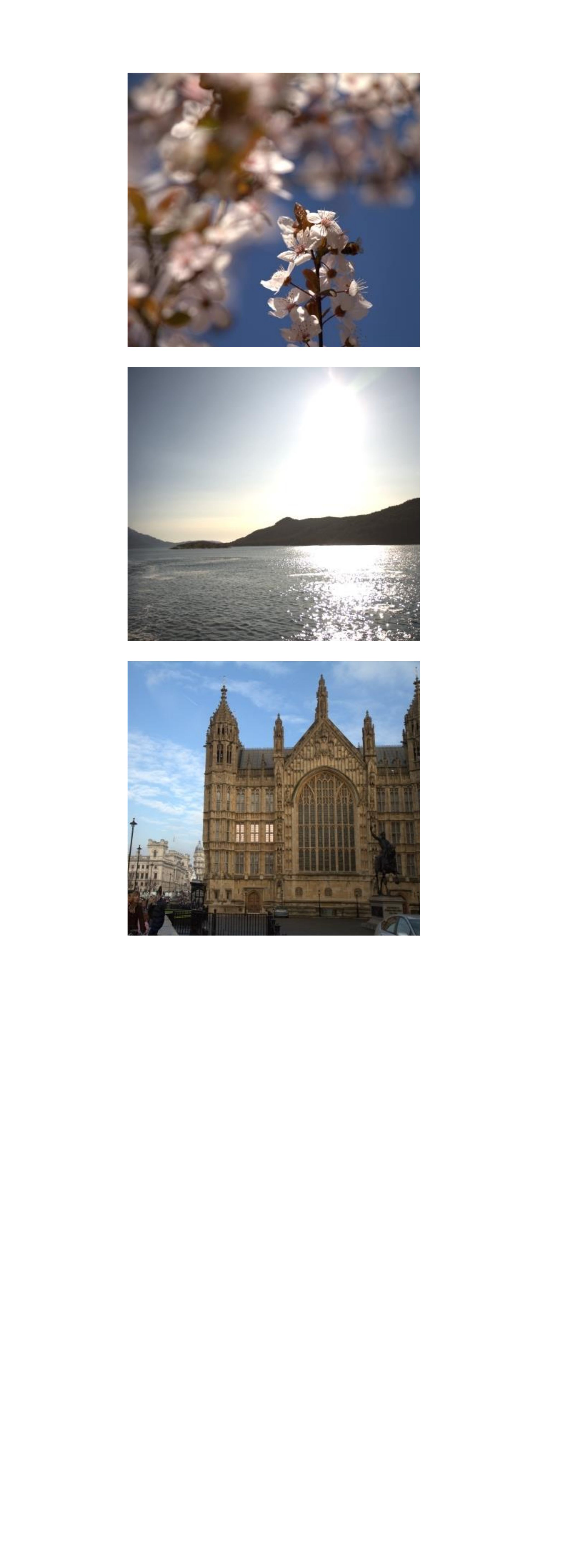}%
  \label{fig1:ex:f}
}\hspace{-3mm}
\subfigure[]{%
  \includegraphics[height=2.3in]{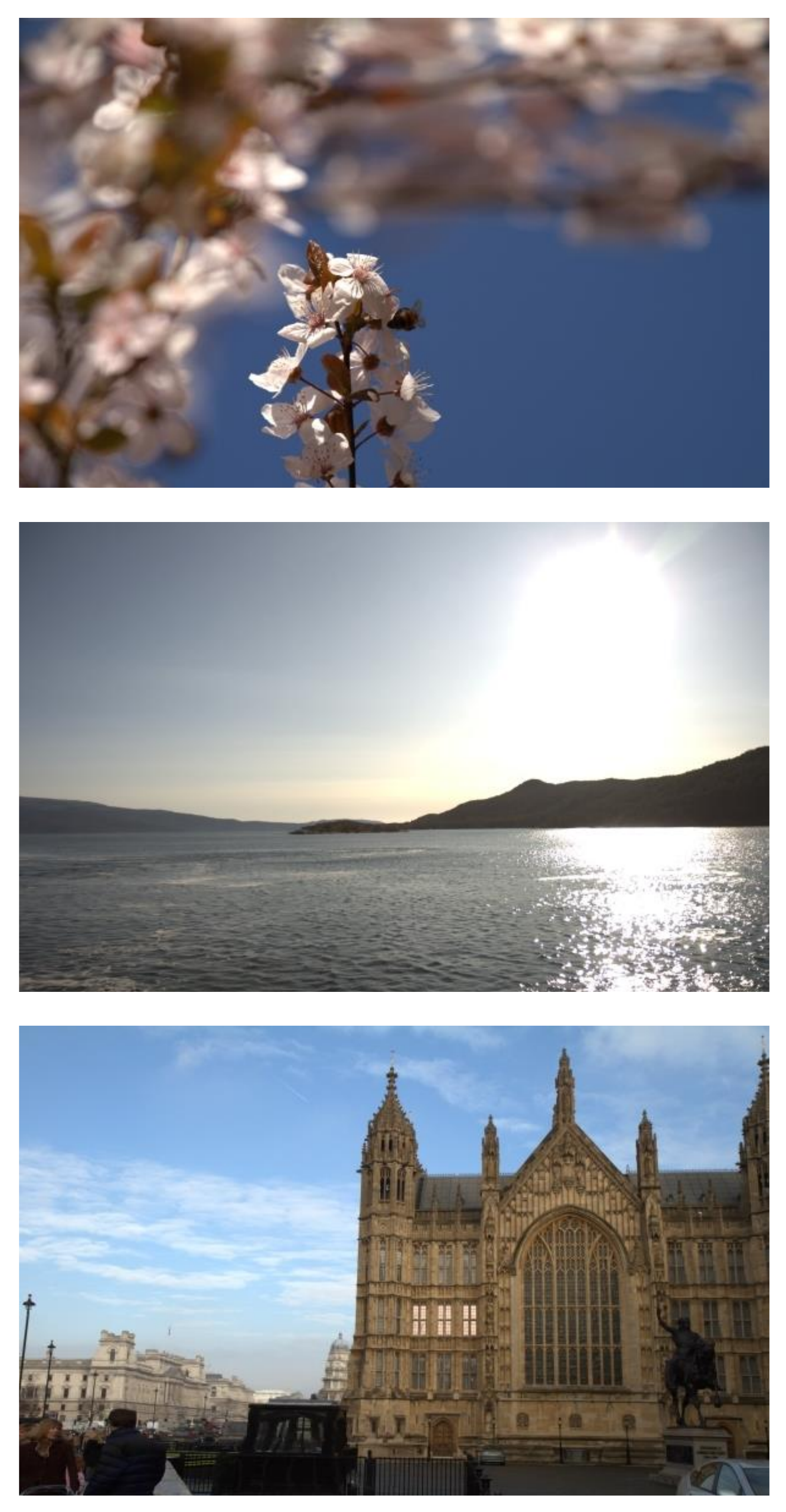}%
  \label{fig1:ex:g}
}
\caption{Examples of content-aware image retargeting using the seam-carving method \cite{seam1}: (a) original images $(512\times384)$, (b) visualization of computed 10\% seams marked in green, (c) 10\% seam-removed images $(461\times384)$, (d) 10\% seam-inserted images $(563\times384)$, (e) visualization of computed 20\% seams marked in green, (f) 20\% seam-removed images $(410\times384)$, and (g) 20\% seam-inserted images $(614\times384)$.}
\label{fig1}%
\end{figure*}

\begin{figure}[b]%
\centering%
\subfigure[]{%
  \includegraphics[scale=0.128]{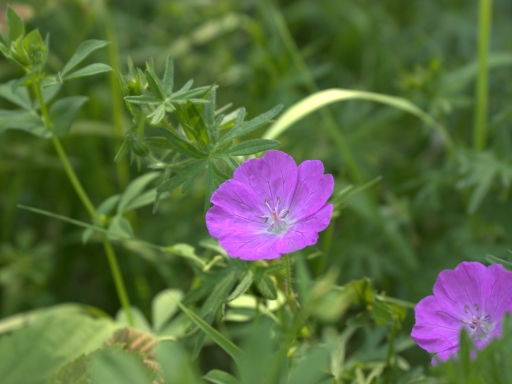}%
  \label{fig2:ex:a}
}
\subfigure[]{%
  \includegraphics[scale=0.128]{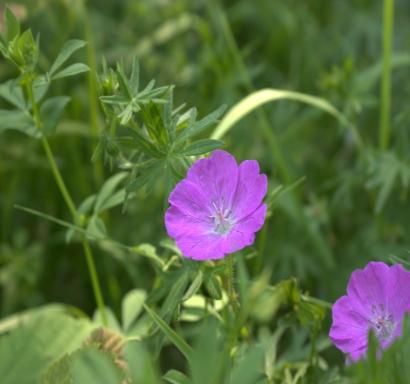}%
  \label{fig2:ex:b}
}
\subfigure[]{%
  \includegraphics[scale=0.128]{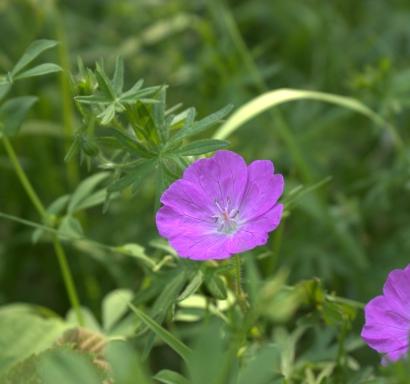}%
  \label{fig2:ex:c}
}
\subfigure[]{%
  \includegraphics[scale=0.128]{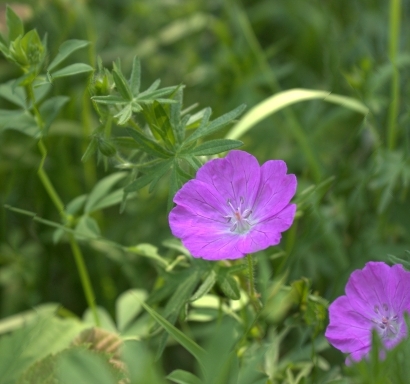}%
  \label{fig2:ex:d}
}
\caption{Examples of image resizing techniques: (a) an original image $(512\times384)$, (b) linear scaling results $(410\times384)$, (c) center cropping results $(410\times384)$, and (d) 20\% seam-removed image $(410\times384)$ based on the seam-carving method \cite{seam1}. Seam-carving-based image retargeting tends to preserve the prominent content of the given image, and our goal is to design a forensic method to classify seam-carved images that are resized naturally.}
\label{fig2}%
\end{figure}

To address this issue, content-aware image retargeting, also known as content-based image resizing, is introduced \cite{seam1,seam2,seam3,seam4,icip,cho,dong}.
Unlike the conventional approaches, this promising technique allows us to adjust the size of an image while preserving important content, as illustrated in Fig.~\ref{fig1}.
As a specific example, as shown in Fig.~\ref{fig2:ex:b}, when applying linear scaling that reduces or expands at an equal rate along one axis, the aspect ratio of the area in the original image is transformed to distort the identity of the main object \cite{resizing,resizing2}.
When applying center cropping to crop part of the original image, as depicted in Fig.~\ref{fig2:ex:c}, some of the prominent content (i.e., the small flower on the right side) can be lost \cite{cho,dong}.
Content-aware image retargeting is a technique that reduces the visual quality deterioration caused by image resizing by maintaining the original content of areas where prominent objects exist and allowing a distortion of relatively less important areas (Fig.~\ref{fig2:ex:d}).

In other words, content-aware image retargeting aims to preserve as much important content as possible, and a representative approach to this is the seam-carving technique \cite{cho,dong,retargeting_survey}.
Seam-carving-based image retargeting computes energy or saliency maps for a given image and preferentially selects seams with small amounts of energy \cite{seam1,seam2,seam3,seam4}.
The computed seams, represented in green, form a connected path of pixels.
The computed seams are located in the relatively less important areas, as illustrated in Figs.~\ref{fig1:ex:b} and \ref{fig1:ex:e}.
By removing seams or inserting duplicated existing seams according to the priority determined based on the energy value, the image size can be adjusted while preserving important objects.
Figs.~\ref{fig1:ex:c} and \ref{fig1:ex:f} and Figs.~\ref{fig1:ex:d} and \ref{fig1:ex:g} are examples of seam-removed and seam-inserted images, respectively.

Seam-carving-based image retargeting to generate natural resized scenes can be deliberately exploited to distort or remove original content; therefore, detecting artifacts of seam carving has become an important topic in image forensics \cite{seam_wei,seam_yin}.
Fig.~\ref{fig1} reveals that seam carving may not leave visual clues for the human visual system and subtly alters the underlying statistics of an image.
In addition, it is challenging to model and analyze artifacts caused by seam carving through only resized images because the parts where the seam insertion and seam removal occur are different according to the content and area in the image \cite{icip,seam_wei,seam_yin}.
In other words, compared to linearly scaled images with periodic characteristics, it is more difficult to classify retargeted images generated from the seam-carving method because the computed seams are scattered globally according to the inherent characteristics of the given image.

This paper proposes a convolutional neural network (CNN)-based forensic approach to classifying seam-carved images with three-class classification: original, seam insertion, and seam removal.
This work is an extended version of our previous work \cite{icip}, which was presented at the \textit{IEEE International Conference on Image Processing (ICIP) 2019} and was referred to as LFNet.
In \textit{ICIP 2019}, we focused on ideas and concepts for learning subtle local artifacts caused by seam-carving-based image retargeting.
In this paper, we propose a network architecture with improved low-level feature learning (ILFNet), which is more sophisticated than the architecture that we initially proposed in \cite{icip}.
We expect the components for local residual learning and local feature fusion employed in the residual dense block (RDB) to help our model learn forensic features caused by seam carving.
The effectiveness of our work is demonstrated through extensive experiments based on the BOSSbase \cite{boss} and UCID \cite{ucid} datasets.
In addition, an ensemble module to improve the classification performance of CNN-based classifiers is introduced.
Our main contributions are summarized as follows: 

\begin{figure*}[t]%
\centering%
\subfigure[]{%
  \includegraphics[height=3.5in]{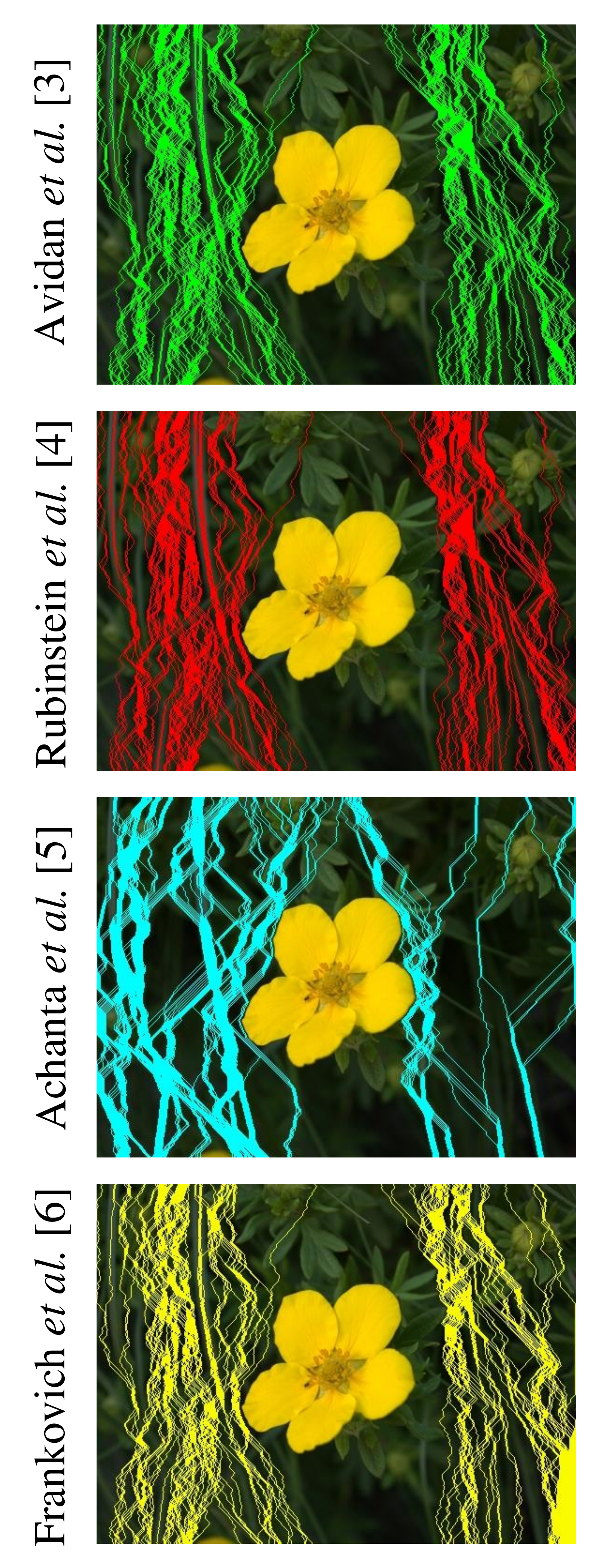}
  \label{fig3:ex:a}
}\hspace{-3.0mm}
\subfigure[]{%
  \includegraphics[height=3.5in]{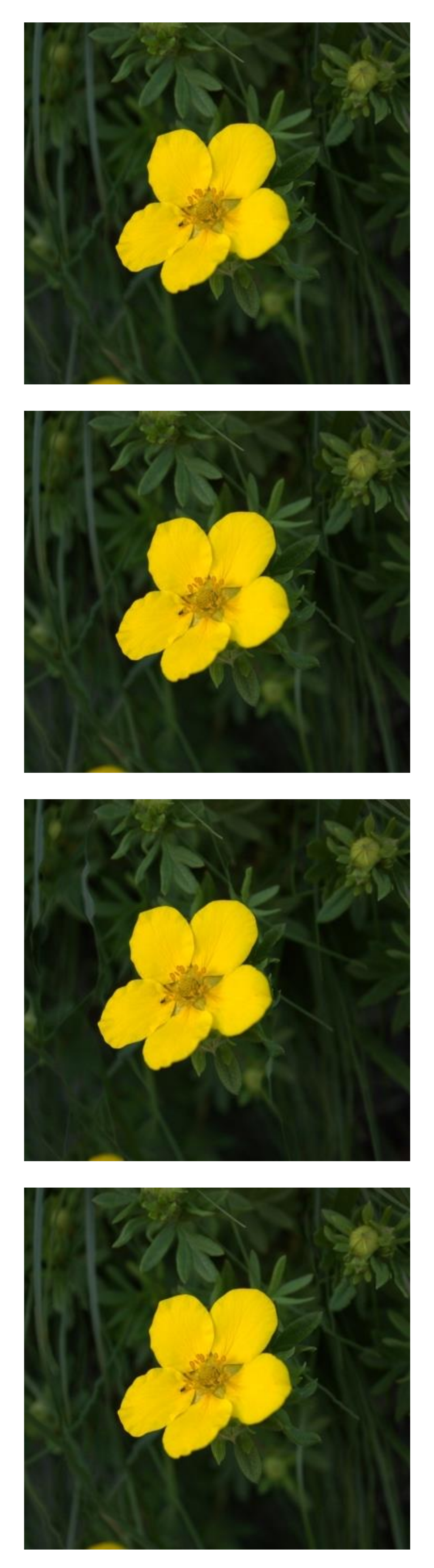}
  \label{fig3:ex:b}
}\hspace{-3.0mm}
\subfigure[]{%
  \includegraphics[height=3.5in]{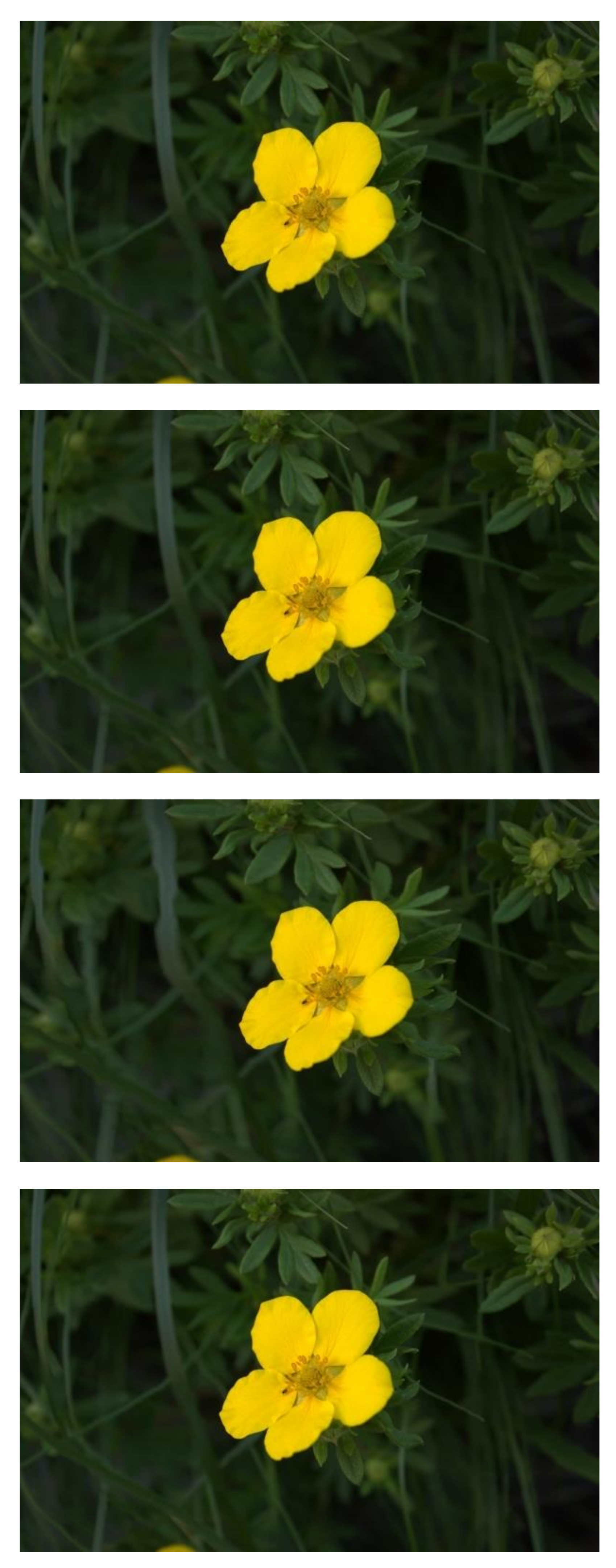}%
  \label{fig3:ex:c}
}\hspace{-3.0mm}
\subfigure[]{%
  \includegraphics[height=3.5in]{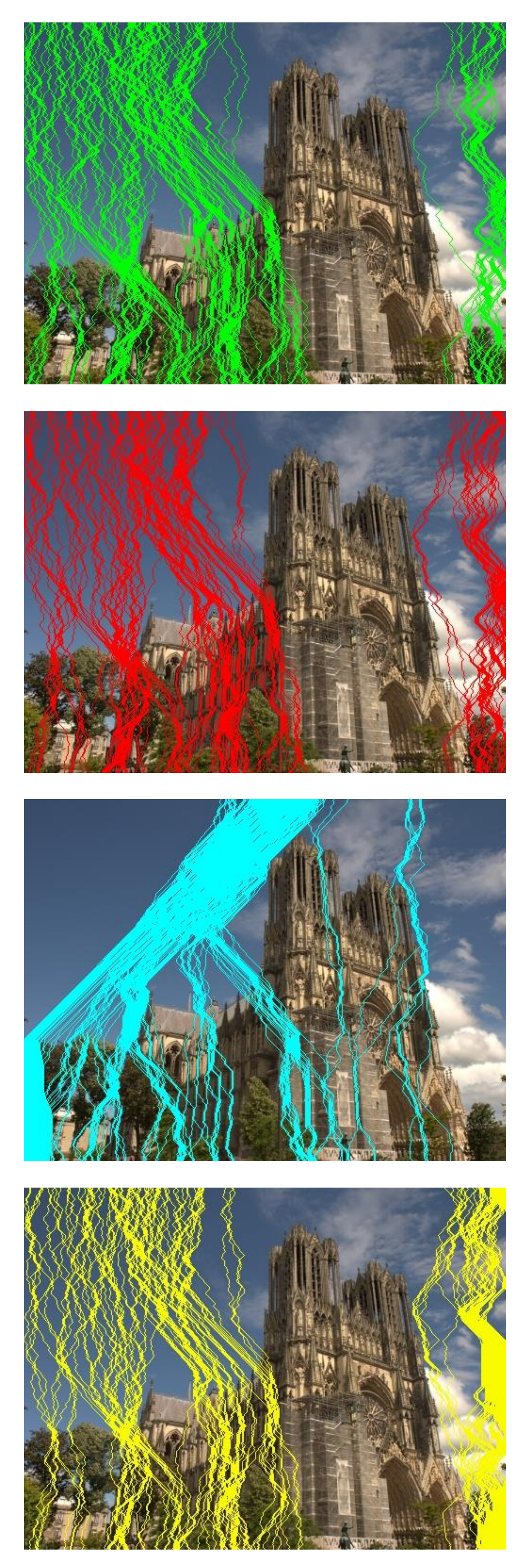}%
  \label{fig3:ex:d}
}\hspace{-3.0mm}
\subfigure[]{%
  \includegraphics[height=3.5in]{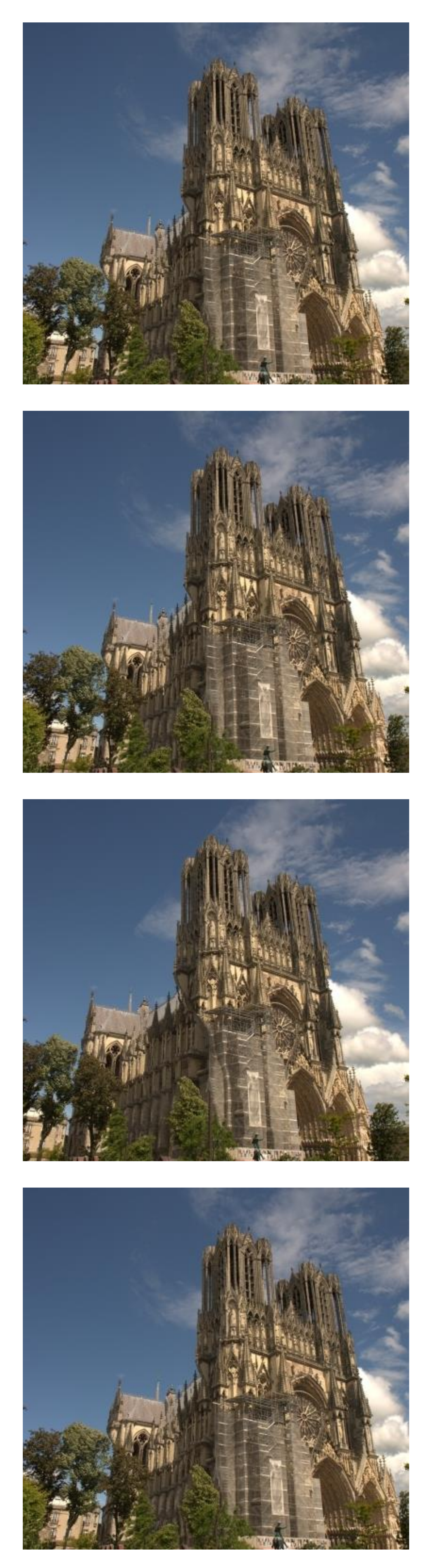}%
  \label{fig3:ex:e}
}\hspace{-3.0mm}
\subfigure[]{%
  \includegraphics[height=3.5in]{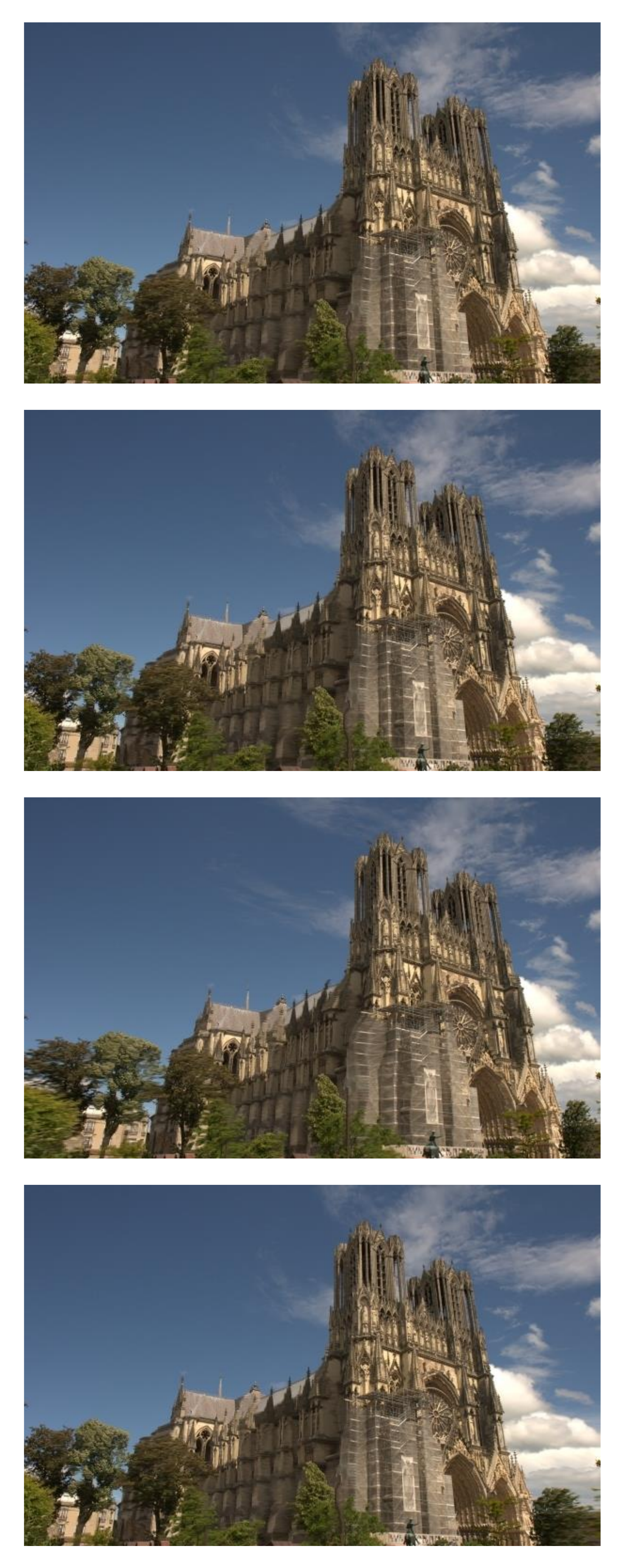}%
  \label{fig3:ex:f}
}
\caption{Examples of content-aware image retargeting generated by four types of seam-carving algorithms: (a) and (d) are original images $(512\times384)$ with visualization of the computed 20\% seams using \cite{seam1,seam2,seam3,seam4}, (b) and (e) are 20\% seam-removed images $(410\times384)$, and (c) and (f) are 20\% seam-inserted images $(614\times384)$. Green, red, cyan, and yellow connected paths of pixels represent seams computed through seam-carving algorithms \cite{seam1,seam2,seam3,seam4}.}
\label{fig3}%
\end{figure*}

\begin{itemize}
 \item Compared to CNN-based approaches \cite{xception,resnet,bayar,henet,h-vgg,yenet,icip,srnet} and the handcrafted feature-based approach \cite{seam_ryu}, the proposed ILFNet exhibits state-of-the-art performance.
 \vspace{2mm}
 \item This work is the first attempt to classify artifacts caused by four types of seam-carving algorithms \cite{seam1,seam2,seam3,seam4}.
 \vspace{2mm}
 \item The ensemble module of this study improves the detection performance of CNN-based classifiers without further training.
 \vspace{2mm}
 \item The superiority of the proposed ILFNet is validated based on extensive experiments, including seam-carving forgery classification, robustness testing against unseen cases (e.g., saving format, seam-carving algorithm, noise addition, and retargeting ratios), and localization.
\end{itemize}
\vspace{2mm}

The remainder of this paper is organized as follows.
Section \rom{2} presents the seam-carving methods and reviews relevant existing work classifying low-level features.
The proposed methodology is presented in Section \rom{3}, and the performance of the proposed method is demonstrated in Section \rom{4}.
Finally, Section \rom{5} concludes this paper.

\section{Related Work}
In this section, we review seam-carving methods and previous forensic approaches related to our work.

\subsection{Seam-carving-based Image Retargeting}
\label{seam_sec_seamcarving}

Seam carving is a representative content-aware image retargeting approach for adjusting the size of an image while preserving its visually important objects \cite{seam_survey}.
Various seam-carving algorithms have been introduced, and this section provides a review of four types of approaches \cite{seam1,seam2,seam3,seam4}.
Avidan and Shamir first proposed a concept of seam carving that is an image operator for identifying the pixels with the lowest energy in ascending order \cite{seam1}.
The goal of seam carving is to compute a monotonic and connected path of low energy pixels (i.e., a seam) in an image.
As depicted in the first row of Figs.~\ref{fig3:ex:a} and \ref{fig3:ex:d}, the vertical seam represented by the green monotonic line crosses the image from top to bottom and contains only one pixel in each row \cite{seam1}.
The ordering of the seams can be determined by the energy function defining the importance of the pixels.

The measure of energy used in \cite{seam1} is defined by the $L_{1}$-norm of the gradient:

\begin{equation}
\begin{aligned}
\label{Eq0_1}
e_{g}(I)=\left|\frac{\partial}{\partial x}I\right|+\left|\frac{\partial }{\partial y}I\right|,
\end{aligned}
\end{equation}

\noindent where $I$ denotes the grayscale intensity of the image with a size of $W\times H$.
Given an $e_{g}$, the optimal vertical seam $\mathbf{\hat{s}}$ that minimizes the total energy $E$ of a seam $\mathbf{s}$ can be obtained: $\mathbf{\hat{s}} = \argmin_\mathbf{s}{E(\mathbf{s})} = \argmin_\mathbf{s}{\sum_{k=1}^H e_{g}(I(\mathbf{s}_k))}$, where $k$ is an index of $\mathbf{s}$, which is the path of $H$ connected pixels.
With the dynamic programming approach, an optimal $\mathbf{\hat{s}}$ can be found by updating the cumulative energy matrix $\mathbf{m}$ for all possible connected seams: 

\begin{equation}
\begin{aligned}
\label{Eq0_2}
&\mathbf{m}(i,j)=e_{g}(i,j)+\\
&min(\mathbf{m}(i-1,j-1),\mathbf{m}(i,j-1),\mathbf{m}(i+1,j-1)),
\end{aligned}
\end{equation}

\noindent where $(i,j)$ indicates the location of a particular pixel.
At the end of this process, $\mathbf{\hat{s}}$ is obtained by backtracking from the minimum element in the last row of $\mathbf{m}$.
By repeatedly removing and inserting the minimum-cost seam, the size of the image can be reduced or enlarged while maintaining visually prominent content (see the seam-carved and seam-inserted examples in the first row of Fig.~\ref{fig3}).

In \cite{seam2}, Rubinstein \textit{et al.} noted that the original operator in \cite{seam1} only focuses on finding seams with the minimum energy cost, ignoring energy that is re-introduced by joining previously nonadjacent pixels.
To address this issue, the authors presented a forward energy criterion for finding the optimal seam by measuring the effect of seam carving on the retargeted image:
\begin{equation}
\label{Eq1}
\mathbf{m}(i,j)=e_{g}(i,j)+min
\begin{cases}
C_{L}(i,j)+\mathbf{m}(i-1,j-1) \\
C_{U}(i,j)+\mathbf{m}(i,j-1) \\
C_{R}(i,j)+\mathbf{m}(i+1,j-1).\\
\end{cases}
\end{equation}

\noindent Here, $C_L$, $C_U$, and $C_R$ are the three possible vertical seam-step costs for pixel $(i,j)$, and these costs are computed as follows:

\begin{equation}
\label{Eq2}
\begin{cases}
C_L(i,j) = C_U(i,j) + |I(i,j-1)-I(i-1,j)| \\
C_U(i,j) = |I(i+1,j)-I(i-1,j)| \\
C_R(i,j) = C_U(i,j) + |I(i,j-1)-I(i+1,j)|.\\
\end{cases}
\end{equation}

\noindent With newly added cost terms, seam removal that introduces the least amount of energy into the retargeted images is possible.

Unlike previous work \cite{seam1,seam2} relying on a gradient map of intensity, Achanta \textit{et al.} introduced a saliency map-based seam-carving algorithm \cite{seam3}. 
The saliency value is computed by evaluating the Euclidean distance of the average of all Lab pixel vectors of the original image $\mathbf{I}$ with each pixel value of the Gaussian blurred image $\mathbf{I}_G$ with a $5\times 5$ kernel: $e_{s}(i,j) = ||\mathbf{I}_\mu-\mathbf{I}_G(i,j)||$, where $\mathbf{I}_\mu$ represents the average of all pixel vectors in the Lab color space.
Inspired by Equation (\ref{Eq2}), the authors in \cite{seam3} further presented color information-based cost terms by replacing the scalar differences of the grayscale intensity $I$ into vector distances of $\mathbf{I}$ in the Lab color space.
After applying $e_{s}(i,j)$ and the newly defined cost function to Equation (\ref{Eq1}), the optimal seam is found using the dynamic programming approach \cite{seam1,seam2}.

In \cite{seam4}, Frankovich and Wong extended the backward and forward energy cost functions in \cite{seam1,seam2} by incorporating an absolute energy cost function in the optimization process.
As in the absolute energy cost case, the optimal seam can be calculated using the dynamic programming process by updating the following cumulative energy matrix:
\begin{equation}
\label{Eq3}
\mathbf{m}(i,j)=e_{a}(i,j)+min\begin{cases}
C_{L}(i,j)+\mathbf{m}(i-1,j-1) \\
C_{U}(i,j)+\mathbf{m}(i,j-1) \\
C_{R}(i,j)+\mathbf{m}(i+1,j-1),\\
\end{cases}
\end{equation}
where $e_{a}(i,j)$ is equal to $e_{g}(i,j)+|e_{g}(i+1,j)-e_{g}(i,j)|+|e_{g}(i,j+1)-e_{g}(i,j)|$.
The newly designed cost function penalizes seam candidates that cross areas of local extrema that characterize regions containing a high concentration of key features.

Fig.~\ref{fig3} illustrates the retargeted results generated by the seam-carving approaches \cite{seam1,seam2,seam3,seam4} observed in this section.
The seams computed via \cite{seam1,seam2,seam3,seam4} are represented as the connected paths of green, red, cyan, and yellow, respectively.
Because a difference exists in the function of determining the energy and saliency value, the form of seams corresponding to 10\% of the image width computed using each technique is different (see the first and fourth columns in Fig.~\ref{fig3}).
Approaches in \cite{seam2,seam4}, which are extended from \cite{seam1} using $e_{g}$, generally calculate seams similar to the results of \cite{seam1}, whereas the seam-carving method in \cite{seam3} using the newly defined $ e_{s}$ calculates relatively different forms of seams compared to those in \cite{seam1,seam2,seam4}.

In this paper, the seams have various characteristics due to the inherent properties of the content (e.g., the shape of the object and background) and the predefined rule of each seam-carving algorithm.
To deal with this issue, we designed a network specialized for learning and capturing low-level features so that manipulation identification can be performed even in areas where few traces of artifacts are caused by seam-carving forgery.
Considering the case in which seam-carving traces are scattered throughout the image, we further aim to improve the classification performance by comprehensively analyzing the results of multiple local patches through the ensemble module.
More details are provided in Section \rom{3}.

\begin{figure*}[t]
\centering{\includegraphics[width=1.0\linewidth]{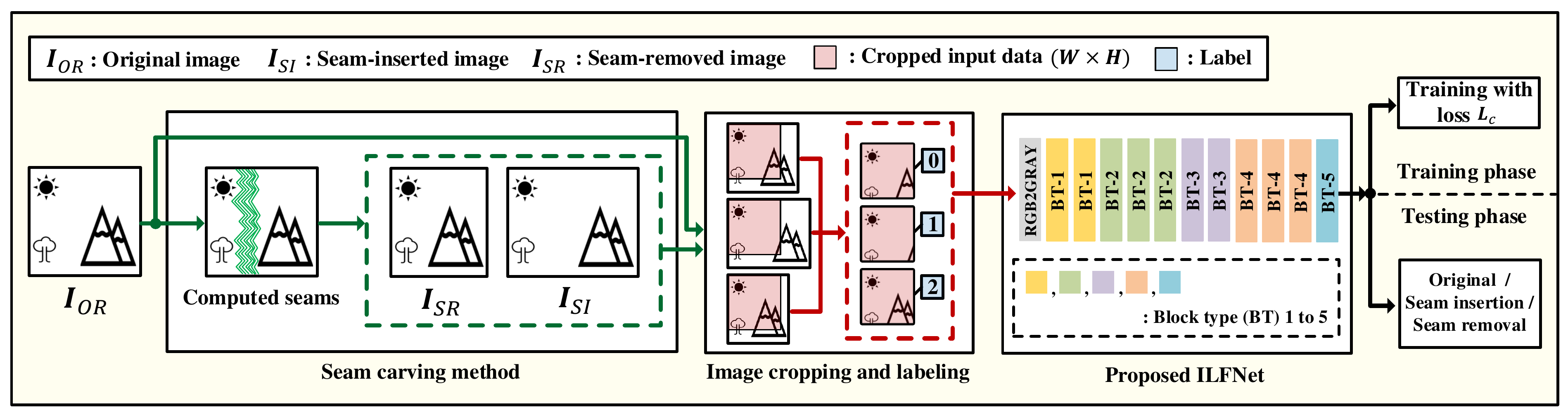}}
\caption{Overview of the forensic approach classifying seam-carving-based image retargeting. In the process of training the proposed ILFNet, the mini-batch consists of randomly selected original, seam-inserted, and seam-removed images. In the process of testing, the trained model with the classification loss $\mathcal{L}_{c}$ enables three-class classification for a given suspicious image.}
\label{fig4}
\end{figure*}

\subsection{Seam-carving Artifact Detection}
In the sections below, conventional handcrafted feature-based (i.e., non-CNN-based) approaches for classifying seam-carved images and CNN-based classifiers for capturing local artifacts caused by various manipulations are covered.

\subsubsection{Conventional Handcrafted Feature-based Approach}

To capture the inherent statistical changes caused by seam removal or seam insertion in retargeted images, handcrafted feature-based approaches \cite{seam_sarkar,seam_fillion,seam_ryu,seam_liu,seam_srmq1,seam_ccjrm,seam_liu2,seam_wei,seam_yin} have been presented.
Sarkar \textit{et al.} proposed a forensic approach \cite{seam_sarkar} exploiting 324-dimensional Markov features (i.e., Shi-324), consisting of a 2D difference histogram in the discrete cosine transform domain and a support vector machine (SVM) framework.
In \cite{seam_fillion}, Fillion and Sharma demonstrated that an SVM-based trained model employing the hand-designed features based on energy bias, seam behavior, and wavelet absolute moments is suitable for detecting seam-carving forgery.

In \cite{seam_wei}, Wei \textit{et al.} introduced an SVM-based approach using $2 \times2$ blocks (called a mini square) and three types of patch transition probability matrices.
To highlight the local texture artifacts, Yin \textit{et al.} produced a set of features by combining half-seam features, energy features, and noise-based features from the local binary pattern domain \cite{seam_yin}.
The author of \cite{seam_liu2} revealed that a set of directional derivative-based and Gabor residual-based features generally performed well in a given forensic task.
In \cite{seam_liu}, Liu and Chen presented an approach using calibrated neighboring joint density and demonstrated that an ensemble classifier \cite{seam_ensemble} with rich models (e.g., CC-JRM \cite{seam_ccjrm} and SRMQ1 \cite{seam_srmq1} features) for image steganalysis is effective for seam-carved forgery detection.
In \cite{seam_ryu}, Ryu \textit{et al.} presented a feature vector that combines energy features, seam features, and noise features for exploring artifacts of seam removal and analyzed the relationship among neighboring pixels to estimate the seam insertion.

The described conventional approaches have shown acceptable performance, but they do not fully meet the needs of forensics for seam-carving detection because forensic traces can be lost during the generation of handcrafted features \cite{srnet,bayar}.
In addition, in some cases, two independent algorithms are required to detect seam insertion and seam removal \cite{icip,seam_survey}.
In other words, two tests should be undergone to authenticate an image, which leads to a high false alarm rate.
To address the inherent problem of these hand-designed feature-based approaches, forensic techniques using deep learning frameworks to let the network automatically learn forensic features have been proposed.
This is described in the next section.

\subsubsection{Convolutional Neural Network-based Approach}
Inspired by high-level vision tasks (e.g., ImageNet \cite{imagenet} classification and object detection) that have achieved significant advances using deep learning, various approaches to CNN-based multimedia forensics have been proposed \cite{deep_forensic1,deep_forensic2,deep_forensic3}.
While CNNs for computer vision are capable of learning features from the data, in their general form, they tend to learn high-level features of the content of a given image \cite{bayar,srnet}.
To address this issue, CNN-based forensic approaches \cite{bayar,icip,resnet,xception,henet,h-vgg,yenet,srnet} have been designed to learn forensic features while suppressing the content of the image by exploiting preprocessing layers or network components specialized for learning low-level features.

In \cite{bayar}, Bayar and Stamm introduced a constrained convolution layer that forces the CNN model, called BayarNet in this paper, to learn prediction error filters that produce low-level forensic features.
In \cite{resnet}, He \textit{et al.} suggested a residual network (ResNet) based on a skip-connection for residual learning, which had a positive effect on improving the performance of the CNN for forensics \cite{icip,deep_forensic1} and steganalysis \cite{srnet}.
In particular, the revised ResNet-34 (rResNet) without the initial pooling layer to prevent the loss of noise-like features exhibited stable and outstanding performance in capturing seam-carved forgery, as introduced in \cite{icip}.
In \cite{h-vgg}, Nam \textit{et al.} proposed H-VGG in combination with VGGNet \cite{vgg} with high-pass filtering (HPF), and the model successfully detected double compression artifacts in the decoded intra-coded frame (I‑frame) of H.264 video.

Regarding detecting the relocated I-frame in the H.264 video, He \textit{et al.} presented HeNet, consisting of a component for extracting high-frequency features to eliminate the influence of diverse video content \cite{henet}.
To derive results specialized for deepfake detection, R{\"o}ssler \textit{et al.} \cite{faceforensics} constructed a FaceForensics dataset consisting of fake videos and experimentally demonstrated that Xception \cite{xception} effectively detects artifacts that occur during the generation of a fake face.
Boroumand \textit{et al.} presented SRNet \cite{srnet}, which is the first end-to-end framework for steganalysis in both the spatial and JPEG domains. 
In addition, SRNet, where the pooling layer is excluded from the network blocks in the early and middle stages, is effective for exploring low-level artifacts.
Ye \textit{et al.} proposed a CNN-based method, referred to as YeNet, in which a preprocessing layer based on HPF is placed at the front of the network for seam-carved image detection \cite{yenet}.

Inspired by CNN-based approaches covered in this section, we aim to design a CNN architecture specialized in micro-signal detection.
In particular, without the aid of heuristic components (e.g., a preprocessing layer and hand-designed features), we focused on the framework of learning forensic features in an end-to-end fashion.
To do this, five types of network blocks were introduced in our work, and the proposed ILFNet based on the advantages obtainable in each block can effectively detect traces caused by seam carving. 
To reveal the effectiveness of our work, we conducted extensive experiments comparing the conventional approach \cite{seam_ryu} and CNN-based approaches \cite{xception,resnet,bayar,henet,h-vgg,yenet,icip,srnet}, and the detailed description of ILFNet is provided in the next section.

\begin{figure*}[t]
\centering{\includegraphics[width=1.0\linewidth]{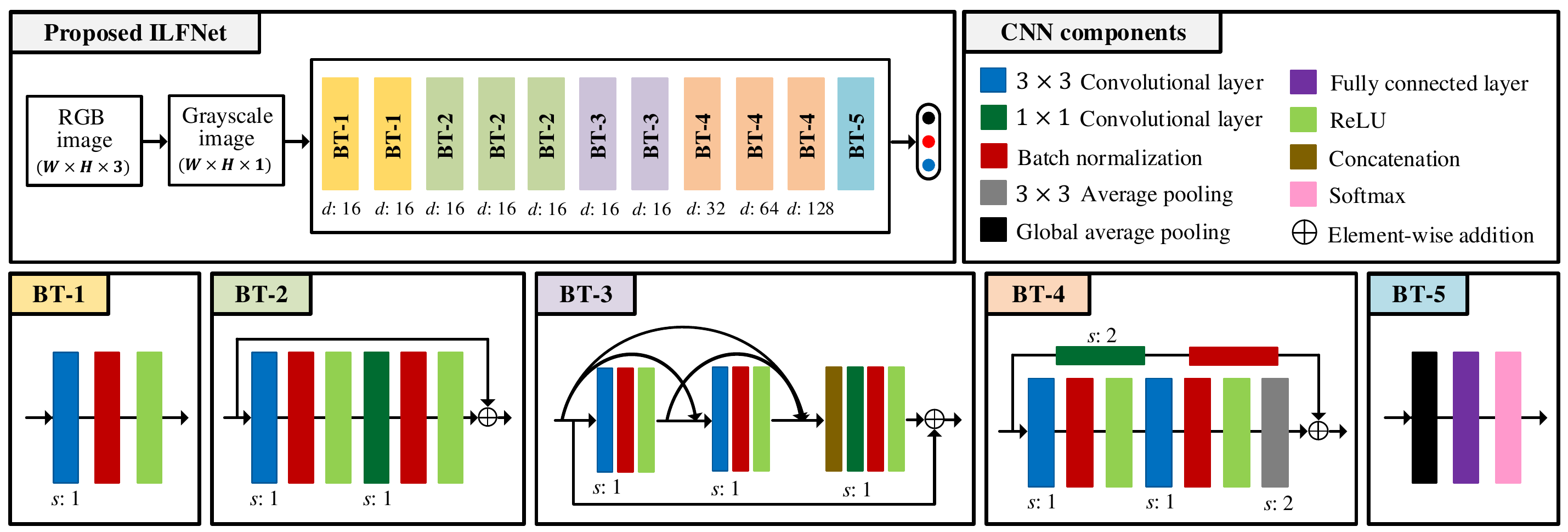}}
\caption{Architecture of the proposed ILFNet for classifying low-level artifacts caused by seam-carving-based image retargeting: $\emph{d}$ and $\emph{s}$ indicate the number of output feature maps and stride of each layer, respectively.}
\label{fig5}
\end{figure*}

\section{Proposed Framework}

This paper focuses on designing a CNN-based framework to capture local artifacts caused by seam carving.
Learning fine-grained forensic features requires a different approach from CNNs that are specialized for learning content-dependent features.
To overcome this obstacle, we explored forensic features through the proposed ILFNet formed by considering the role of each network block. 
As illustrated in Fig.~\ref{fig4}, our architecture consists of five block types (BTs) from BT-1 to BT-5.
Unlike previous work using heuristic components, the proposed work automatically learns forensic features in an end-to-end fashion.
Next, we present the motivation for our approach and the detailed descriptions related to the proposed ILFNet.

\subsection{Motivation and Strategy}
\label{seam_sec_motivation}
Distinguishing between original, seam-inserted, and seam-removed images can be regarded as a three-class classification problem.
Given the set of training data ${(x^{1},y^{1}),...,(x^{N},y^{N})}$ of $N$ samples, $x$ represents the image, and $y$ denotes its corresponding class (0: original image $I_{OR}$, 1: seam-inserted image $I_{SI}$, and 2: seam-removed image $I_{SR}$).
Fig.~\ref{fig4} depicts the overview of our framework for classifying seam-carving forgery.
Fig.~\ref{fig4} illustrates that $I_{SR}$ and $I_{SI}$ are generated by removing or duplicating the less visually important areas of $I_{OR}$ through the seam-carving algorithm. 
The tendency of the calculated seam expressed in green varies according to the given image content and the energy calculation approach of the seam-carving algorithm (Section~\ref{seam_sec_seamcarving}).
In addition, the CNN-based classification of seam-carving artifacts has an obstacle in handling data of different sizes (e.g., the size of $I_{OR}$, $I_{SR}$, and $I_{SI}$) at the same time.

We first considered an approach to resize data to the same size using scaling, but it has the disadvantage that local texture artifacts caused by seam carving can be lost.
In addition, employing interpolation-based scaling causes unintended traces into the given image \cite{resizing}.
Thus, data samples with a size of $W\times H$ were generated using cropping rather than scaling \cite{icip}.
Because the intrinsic content of the natural image is very diverse, we judged that cropping a sufficiently large area at a fixed location would include various cases of seam-carving artifacts in the cropped sample.
That is, because the form of the calculated seams is affected by the content of the image, such as the object and background, various cases of local artifacts can be observed in the obtained samples. 
As illustrated in the middle part of Fig.~\ref{fig4}, input data of the proposed network were generated by cropping an area of $W\times H$ from the upper left of the images including $I_{OR}$, $I_{SR}$, and $I_{SI}$.

Unlike the paired mini-batch training methodology \cite{paired} focusing on the difference between the paired original and its corresponding manipulated data, we constructed the mini-batch through random sampling from the training set.
Through this, the proposed model was induced to consider various cases (e.g., content information and user-preferred ratio parameters for retargeting) for each iteration in the training phase.
Before inputting it into the network, the input in the RGB color space is converted to grayscale ($W\times H\times3\rightarrow W\times H\times1$) in our work.
The proposed ILFNet consists of five types of network blocks from BT-1 to BT-5.
Through the combination of network blocks, the following fundamental abilities for forensic feature learning are included in ILFNet: (i) local texture artifact learning, (ii) refined feature learning via local feature fusion, and (iii) hierarchical feature learning and classification.

Unlike existing approaches exploiting preprocessing \cite{h-vgg,henet,yenet} and hand-designed features \cite{deep_forensic2,deep_forensic3}, the proposed network learns and extracts subtle traces of seam carving from input in an end-to-end fashion.
To do this, we first placed network blocks for extracting low-level features, also known as residual noise \cite{srnet} and prediction residuals \cite{bayar}, in shallow layers of the network.
Because fine-grained artifacts caused by manipulation are vulnerable to destruction by the pooling layer \cite{mislnet,sr1,sr2}, the pooling layer of the suppressing noise-like signal is excluded from the first to fifth network blocks comprising BT-1 and BT-2.
In particular, BT-2 is added with a skip-connection to help propagate gradients to the upper layers \cite{srnet}, which has proven effective at residual learning \cite{resnet}.
We expected that the front segment for extracting low-level features would play a role similar to the high-pass filter, which is verified by visualizing the feature map in Section \rom{4}.

Next, we further improve the classification ability by applying RDB \cite{sr1} for super-resolution into BT-3, constituting the middle segment.
The super-resolution is a task for reconstructing high-resolution images from low-resolution images \cite{sr2} by improving textured details.
It differs from the given classification task in that it is intended for image restoration, but a common point exists from the perspective of dealing with low-level signals.
Because RDB is specialized in extracting abundant features, we expected sub-components for local residual learning and local feature fusion in BT-3 to help our model learn meaningful features from feature maps generated from BT-2.
In addition, inspired by \cite{icip,srnet}, we kept the number of feature maps $d$ of the first to seventh network blocks constant.

Finally, the last segment comprising BT-4 and BT-5 is used for dimensionality reduction of the feature maps generated from the middle segment and performs three-class classification.
The higher-level features are learned from the lower-level features obtained from the front and middle segments through consecutively placed blocks of BT-4. 
For BT-5, global average pooling (AvgPool) \cite{gap} is exploited to replace the numerous neurons of the fully connected (FC) layers to mitigate the chance of overfitting.
Based on the experiments, we empirically determined the number and arrangement of network blocks that constitute the proposed ILFNet.
The network automatically explores forensic features that are difficult to learn with training from randomly initialized parameters in an end-to-end fashion.

\subsection{Network Architecture}

Fig.~\ref{fig5} depicts a detailed configuration for each network block constituting the proposed ILFNet.
The network consists of 11 network blocks using five BTs, as displayed in the figure.
In this section, each block type is described, and then details of the differences from our previous work \cite{icip} are provided.
Finally, we introduce a loss function for training our model and an ensemble module for further performance improvement in the testing phase.
Moreover, BT-1 is composed of a $3\times 3$ convolutional (Conv) layer with stride 1, which is followed by batch normalization \cite{batch} to alleviate the potential of overfitting and uses the rectified linear unit (ReLU) \cite{relu} as an activation function.
On content classification tasks, the AvgPool layer, which is a representative pooling layer, is employed to reinforce the content \cite{srnet} and reduce the dimensionality of the feature maps, but it suppresses subtle signals by averaging the adjacent information \cite{icip,mislnet}.
Therefore, if the AvgPool layer is placed in the initial layers, it prevents the network from learning subtle pixel-value dependent low-level features.
Inspired by the insight and approaches in \cite{icip,sr1,srnet}, we prevented the noise-like signal from disappearing by excluding the pooling layer from the BT-1 configuration.
As depicted in Fig.~\ref{fig5}, we induced the extraction of shallow features from the input data by placing two BT-1s in the early layers of ILFNet.

In addition, BT-2 is designed to improve the ability of ILFNet to extract forensic features using a skip-connection \cite{resnet,resnet2}.
Like BT-1, the pooling layer is excluded in BT-2.
The main stream of BT-2 consists of two Conv layers, each of which is followed by batch normalization, and ReLU as a nonlinear activation function.
Unlike the approaches in \cite{bayar,henet} in which a $1\times1$ Conv layer is placed on a deeper layer, we sequentially placed a $3\times3$ Conv layer that learns the relationship between neighboring elements and a $1\times1$ Conv layer that learns the association between the feature maps in the shallow layer.
As displayed in Fig.~\ref{fig5}, the feature map input into BT-2 is reused using the skip-connection for residual learning.
The skip-connection is proposed to alleviate the vanishing gradient problem that adversely affects the convergence of deep-structured CNNs \cite{resnet,resnet2}, and this component helps propagate the gradient to the upper layers.
Thus, local residual learning through the skip-connection can help in learning local texture artifacts caused by seam carving.
We expect that the proposed ILFNet can learn low-level features through the front segment comprising consecutive BT-1s and BT-2s.

Inspired by RDB in \cite{sr1}, which addressed low-level signals, the structure of BT-3 constituting the middle segment was determined.
Because RDB is specialized in extracting abundant features \cite{sr2}, we expected sub-components for local residual learning and local feature fusion in BT-3 to help ILFNet learn refined features from feature maps generated from the previous block. 
In this work, a lightweight version of RDB, consisting of two connected BT-1s followed by the component for local feature fusion, is adopted.
Fig.~\ref{fig5} indicates that the architecture of BT-3 not only enables the feature maps of the previous block to connect with each Conv layer of current the BT-3 but also learns comprehensively abundant local features through local feature fusion \cite{sr1}.
Based on concatenation and $1\times 1$ Conv layers, the extracted local features are fused.
Then, the local residual learning through the skip-connection is performed.
Inspired by \cite{icip,srnet}, we kept the number of feature maps $d$ of the first to seventh network blocks constant $(16)$ and the growth rate of the $3\times3$ Conv layers in BT-3 was set to 16.
In summary, through the abilities of BT-3 in terms of abundant local feature extraction and comprehensive feature learning, the proposed ILFNet can learn and explore refined forensic features.

For hierarchical feature learning, three consecutive BT-4s are placed in front of the last segment.
To learn and extract a higher-level representation of the previously learned features, BT-4 consists of Conv layers for higher feature learning and pooling layer reduction for feature extraction.
The main path of BT-4 employed two $3\times 3$ Conv layers, each of which is followed by batch normalization and ReLU, and the AvgPool layer ($3\times3$ kernel and stride 2) was applied to the last layer for the dimensionality reduction of the feature maps.
Inspired by \cite{icip,xception}, the skip-connection has a $1\times1$ Conv layer with a stride of 2 to perform element-wise addition, and this component enables the fusion of feature representations of multiple resolutions.
We increase the number of filters $d$ by a factor of 2 whenever BT-4 is inserted into ILFNet.

Finally, the condensed feature maps are directly passed to BT-5 designed for three-class classification through consecutive dimensionality reduction.
For BT-5, constituting the last segment, global AvgPool is exploited to replace the numerous neurons of the FC layers to alleviate overfitting and improve the generalization ability \cite{gap}.
Next, BT-5 consists of an FC layer that has three output neurons and a softmax layer.
In our work, $\hat{y}$ denotes the output through the FC layer.
The proposed ILFNet consisting of a combination of BT-1 to BT-5 with unique characteristics and purposes automatically explores the forensic features for seam-carving forgery in an end-to-end fashion.

\subsection{Differences From the Original Work}
\label{seam_sec_diff}

In \cite{icip}, we proposed a network for low-level feature learning, referred to as LFNet.
Compared to our original work, the new architecture was designed based on a considerable design refinement, and this design choice has been motivated by extensive experiments.
Unlike the original network block for learning subtle signals, in which the $3\times 3$ Conv layer and $1\times 1$ Conv layer are sequentially placed on BT-1, we believe that it is important to learn the relationship between the neighboring pixel elements in the shallow layers.
Therefore, to focus on learning low-level features based on the relationship between adjacent pixels, only the $3\times 3$ Conv layer was employed for BT-1 of ILFNet.
In addition, the proportion of BT-2 adopting residual learning was increased, and the proportion of BT-1 without a skip-connection was reduced.

In particular, the new architecture contains a middle segment for local feature fusion and refined feature learning.
Inspired by the RDB addressing a low-level signal for super-resolution in \cite{sr1}, we newly adopted BT-3 to learn comprehensively refined features from lower-level features.
With the sub-components of BT-3 (e.g., the contiguous memory mechanism for passing local features, $1\times 1$ Conv layer-based feature fusion, and skip-connection for local residual learning), the proposed ILFNet can learn refined and higher features from shallow features obtained through the front segment.
For BT-4, for learning hierarchical features, the $3\times 3$ AvgPool layer with a stride of 2 was employed instead of the max-pooling (MaxPool) layer employed in the original architecture.
This was determined based on the insight obtained from \cite{sr1,yenet,henet} and the performance analysis experiments for the pooling layer type (i.e., AvgPool and MaxPool layers).
From BT-1 to BT-4, batch normalization \cite{batch} was used to alleviate the overfitting problem.
Lastly, from BT-2 to BT-4, shortcut connections were employed to help propagate the gradient to the higher layer.

\subsection{Loss Function}

In this section, the loss function for three-class classification is defined, where $\hat{y}_{j}$ refers to the output for which the input is class $j$ among three classes, where $j = \{0, 1, 2\}$.
In our work, the original, seam-inserted, and seam-removed images were set to Class 0, Class 1, and Class 2, respectively.
The probability $P(\hat{y}=j)$ can be computed from $\hat{y}_{j}$ using the following softmax function: $P(\hat{y}=j)=\frac{e^{\hat{y}_{j}}}{\sum_{j=0}^{2}e^{\hat{y}_{j}}}$, where ${y}$ is a one-hot vector and ${y}_{j}$ denotes the specific class.
If a given image corresponds to Class 0, ${y}$ is defined as ${y}=[{y}_{0}; {y}_{1}; {y}_{2}]=[1; 0; 0]$.
The classification loss $\mathcal{L}_{c}$ was computed from the cross-entropy as follows: $\mathcal{L}_{c} = -\sum_{j=0}^{2}{y}_{j}\log{P(\hat{y}=j)}$.
The proposed ILFNet trained with the defined $\mathcal{L}_{c}$ classifies suspicious test images into three classes (i.e., original, seam insertion, and seam removal) with high accuracy.

\begin{figure}[t]
\centering{\includegraphics[width=0.98\linewidth]{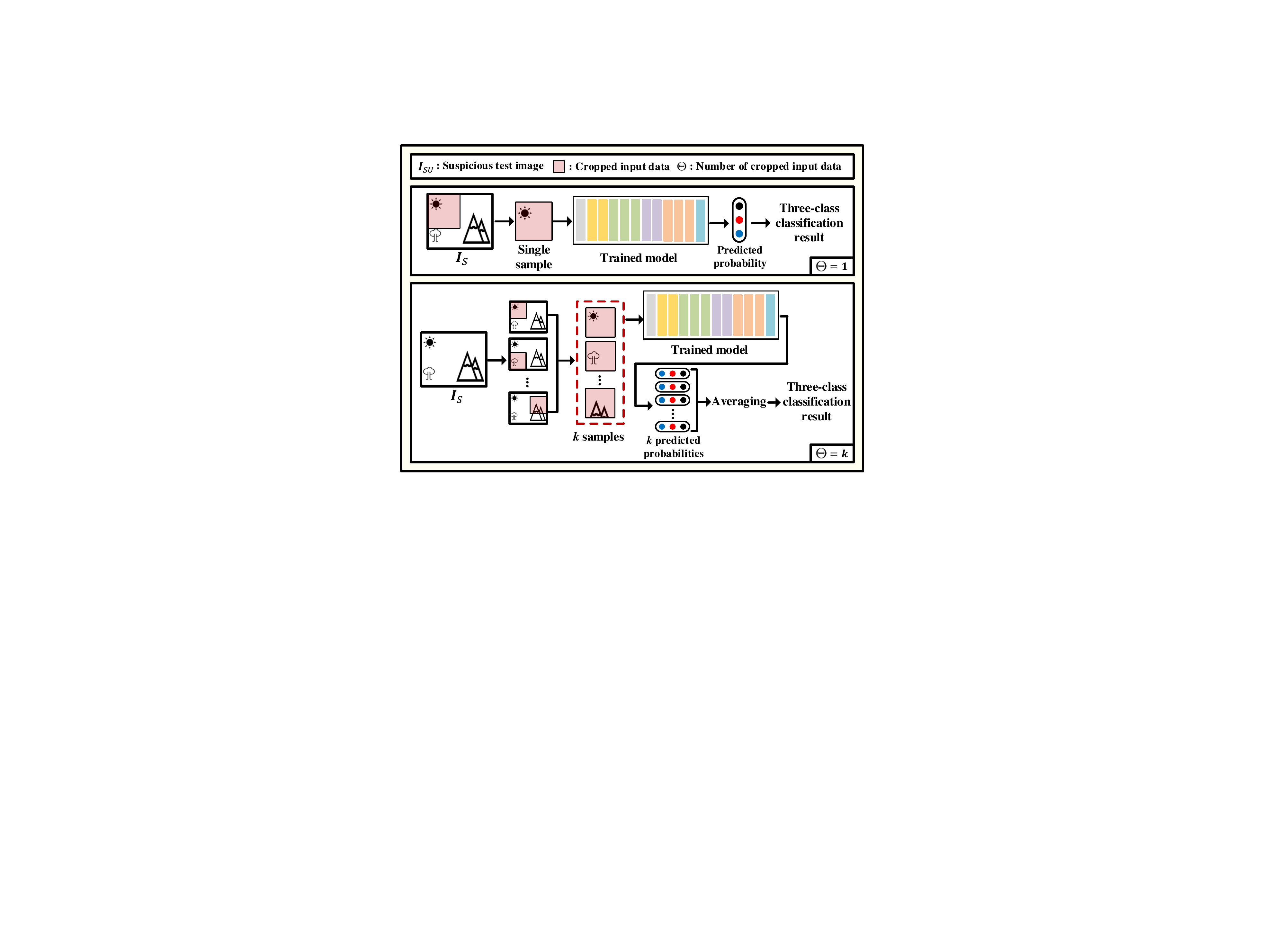}}
\caption{Overview of the ensemble module to improve the performance of the trained model for classifying seam-carving forgery in the testing phase.}
\label{fig6}
\end{figure}

\subsection{Ensemble Module}
\label{seam_sec_ensemble}
To enhance the performance of our trained model in the testing phase, we propose an ensemble module.
As described in Section~\ref{seam_sec_motivation}, our approach crops an area of $W\times H$ from the upper left of the input data before inputting it into the network in the training process.
This approach also applies to the testing phase, which is illustrated at the top of Fig.~\ref{fig6}.
Here, $\Theta$ indicates the number of patches sampled by applying cropping to the suspicious test image $I_{S}$.
Inspired by the data augmentation-based self-ensemble \cite{self_ensemble}, we aimed to improve the classification performance by generating multiple samples from a given $I_{S}$, providing them to the trained model to obtain multiple outputs, considering them comprehensively (see bottom of Fig.~\ref{fig6}).

\begin{figure*}[t]%
\centering%
\subfigure[]{%
  \includegraphics[width=0.4\linewidth]{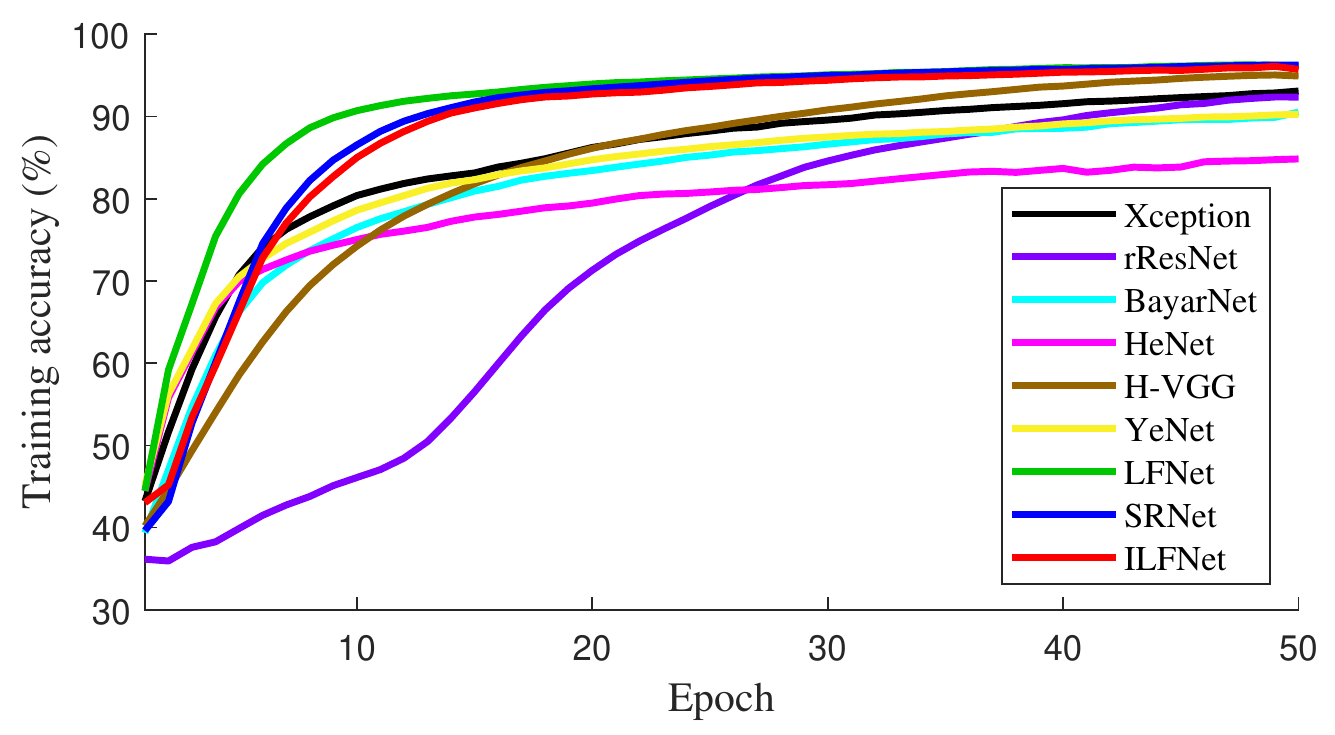}
  \label{fig7:ex:a}
}\hfil
\subfigure[]{%
  \includegraphics[width=0.4\linewidth]{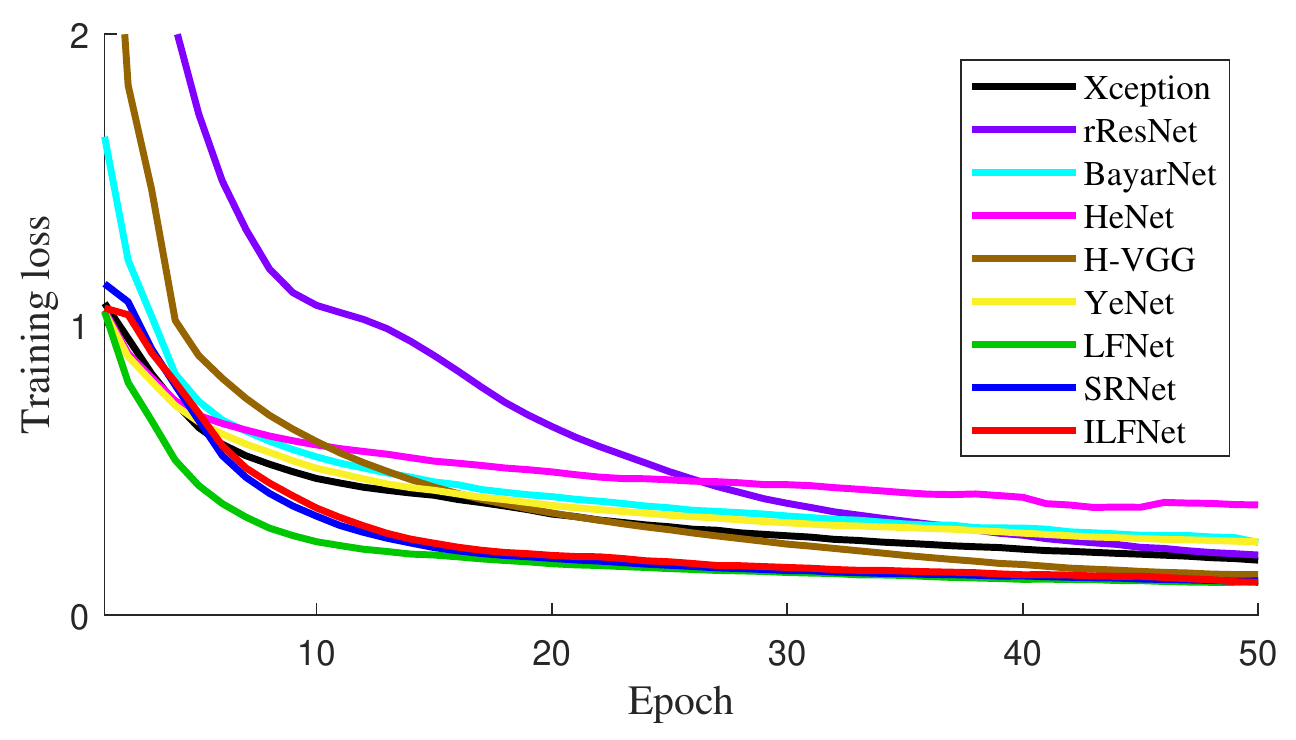}
  \label{fig7:ex:b}
}
\caption{Training accuracy (a) and training loss (b) tendencies for each network for 50 epochs.}
\label{fig7}%
\end{figure*}

If $\Theta$ is equal to $k$, the proposed ensemble module in the testing phase acquires $k$ samples from $I_{S}$ with a size of $W_{S}\times H_{S}$.
The upper left coordinate for generating the $i$-th sample $I_{S_{i}}$, represented by $(r_{x_{i}}, r_{y_{i}})$, is uniformly sampled according to:
\begin{equation}
    \begin{aligned}
        r_{x_{i}} \sim U(0, W_{S}-W),\quad r_{y_{i}} \sim U(0, H_{S}-H),
    \end{aligned}
    \label{eq_samplebox}
\end{equation}
where $U$ stands for uniform distribution, and the area of size $W\times H$ is cropped based on $(r_{x_{i}}, r_{y_{i}})$, where $i=\{0,...,k-1\}$. 
We exceptionally fix a case where the $(r_{x_{0}}, r_{y_{0}})$ is equal to $(0,0)$, which is the same as the case with $\Theta=1$.
If the value of $i$ is between 1 and $k-1$, then $(r_{x_{i}}, r_{y_{i}})$ and its corresponding $I_{S_{i}}$ is obtained based on Equation (\ref{eq_samplebox}).
In the case of $\Theta=k$, the trained model takes $k$ patches sampled from $I_{S}$ as input, and $k$ predicted probabilities are generated.
We averaged the $k$ predicted probabilities to determine the final output score.

This simple ensemble module does not require additional training of the separate model.
Through the ensemble module, we expected the proposed forensic framework with the ensemble module and multiple samples to comprehensively explore local texture artifacts due to seam-carving scattered throughout a given $I_{S}$.
In addition, we found that the ensemble module provides an additional performance gain for both our work and comparative CNN-based approaches in terms of classifying seam-carving artifacts.
The effectiveness of the ensemble module is covered in detail in the following section.

\section{Experiments}

To assess the performance of our ILFNet for classifying seam-carving forgery, we conducted a set of experiments and analysis.
This section provides detailed descriptions of the experimental setup and the results of the extensive experiments.

\subsection{Dataset}
\label{seam_sec_dataset}
In the experiments, the BOSSbase \cite{boss} and UCID \cite{ucid} datasets were used to generate 10,000 original images with a size of $512\times384$.
Inspired by the various uses of JPEG compression reported in \cite{deep_forensic1,deep_forensic2,deep_forensic3}, the obtained original images were saved using JPEG compression with a quality factor of 100.
Based on the algorithm \cite{seam1}, the original images were retargeted using vertical seam removal from 10\% to 50\% in 10\% steps, resulting in 50,000 seam-carved images in total.
Similarly, the original images were enlarged using vertical seam insertion \cite{seam1} from 10\% to 50\% in 10\% steps, resulting in 50,000 seam-inserted images.
Like the original images, the generated seam-removed and seam-inserted images were also saved using JPEG compression with a quality factor of 100.
In total, 110,000 images with resolutions in the range from $256\times384$ to $768\times384$ were obtained.
We divided the images into three sets for training, validation, and testing (with an $8:1:1$ ratio).
In the training and testing process, the ratio of image data corresponding to each class was maintained at $1:1:1$ for the three-class classification.
Before being input into the network, the sample size for the $W\times H$ cropped from the images in the generated dataset was set to $256\times256$.

To demonstrate the effectiveness of our approach, we generated additional testing sets for experiments assessing the robustness in the unseen cases (i.e., seam-carving algorithms \cite{seam2,seam3,seam4}, retargeting ratios for 4\% to 8\% in 2\% steps, noise signal addition for interfering classification, and uncompressed image format, such as BMP).
In addition, we designed experiments regarding the horizontal seam-carving classification and spatial localization and created testing sets for them.
In these cases, the training based on a new methodology was applied to the model.
The description of additionally generated testing sets is introduced in detail in each experimental section.

\subsection{Training Settings}
\label{seam_sec_training_setting}
We built our network using the TensorFlow framework and ran the experiments using NVIDIA GeForce RTX 2080 Ti.
In the experiments, we use the Adam optimizer \cite{adam} with a learning rate of $10^{-3}$, momentum coefficients $\beta_{1}=0.9$ and $\beta_{2}=0.999$, and the numerical stability constant $\epsilon=10^{-8}$.
The size of the mini-batch was set to 24.
In the training process, the mini-batch was constructed by randomly sampling the data corresponding to the training set.
The proposed ILFNet is trained for 50 epochs, and the best model is selected as the one that maximizes the classification accuracy on the validation set.

\begin{table*}[t]
\caption{Performance evaluation of ILFNet and comparative convolutional neural networks for three-class classification on various retargeting ratios $(\%)$.}
    \centering
    \footnotesize
    \begin{tabu} to \linewidth{X[0.8,c] X[1.0,c] X[1.0,c] X[1.0,c] X[1.0,c] X[1.0,c] X[1.0,c] X[1.0,c] X[1.0,c] X[1.0,c]}
    \hline
    \hline
    Ratio & Xception & rResNet & BayarNet & HeNet & H-VGG & YeNet & LFNet & SRNet & ILFNet\\
    \hline
    $10\%$ & 68.96 & 73.46 & 65.73 & 64.30 & 75.50 & 76.00 & 88.27 & 87.40 & 88.17\\
    20$\%$ & 81.06 & 83.97 & 80.86 & 80.40 & 85.23 & 86.29 & 94.46 & 94.60 & 94.93\\
    30$\%$ & 89.23 & 90.93 & 87.89 & 87.83 & 91.03 & 92.09 & 97.06 & 98.23 & 98.40\\
    40$\%$ & 93.19 & 92.27 & 92.83 & 91.67 & 93.59 & 94.27 & 98.07 & 99.29 & 99.43\\
    50$\%$ & 95.97 & 94.33 & 94.82 & 92.99 & 95.00 & 94.96 & 98.16 & 99.47 & 99.53\\
    \cline{1-10}
    Mixed & 85.93 & 87.06 & 84.83 & 83.96 & 88.66 & 88.83 & 95.39 & 95.69 & 96.56\\
    \hline
    \hline
    \end{tabu}
    \label{table1}
\end{table*}

\begin{table*}[t]
\caption{Performance evaluation of ILFNet and comparative convolutional neural networks for two-class classification on various retargeting ratios $(\%)$.}
    \centering
    \footnotesize
    \begin{tabu} to \linewidth{X[1.0,c] X[1.0,c] X[1.0,c] X[1.0,c] X[1.0,c] X[1.0,c] X[1.0,c] X[1.0,c]  X[1.0,c] X[1.0,c] X[1.0,c]}
    \hline
    \hline
    Class & Ratio & Xception & rResNet & BayarNet & HeNet & H-VGG & YeNet & LFNet & SRNet & ILFNet\\
    \hline
    \multirow{6}{*}{OR, SI} & $10\%$ & 84.05 & 79.84 & 76.30 & 70.11 & 82.95 & 83.25 & 91.15 & 92.99 & 93.50\\
    & 20$\%$ & 90.95 & 86.61 & 86.05 & 82.80 & 89.85 & 89.64 & 95.24 & 97.20 & 97.45\\
    & 30$\%$ & 94.50 & 89.30 & 90.59 & 88.25 & 92.69 & 92.30 & 96.80 & 98.78 & 98.94\\
    & 40$\%$ & 95.34 & 90.29 & 93.00 & 90.14 & 93.94 & 93.24 & 97.19 & 99.14 & 99.29\\
    & 50$\%$ & 95.65 & 90.40 & 93.45 & 90.19 & 94.25 & 93.25 & 97.25 & 99.20 & 99.45\\
    \cline{2-11}
    & Mixed & 91.84& 87.94 & 87.69 & 84.20 & 90.29 & 89.64 & 95.68 & 97.59 & 97.64\\
    \hline
    \multirow{6}{*}{OR, SR} & $10\%$ & 63.59 & 70.89 & 66.55 & 68.20 & 74.85 & 74.45 & 88.50 & 87.29 & 88.20\\
    & 20$\%$ & 76.20 & 79.90 & 78.00 & 79.65 & 82.56 & 83.50 & 93.69 & 93.90 & 94.39\\
    & 30$\%$ & 85.05 & 86.14 & 85.54 & 85.35 & 88.40 & 89.55 & 96.05 & 97.74 & 98.10\\
    & 40$\%$ & 90.14 & 88.65 & 90.55 & 89.20 & 91.00 & 91.84 & 97.15 & 98.99 & 99.14\\
    & 50$\%$ & 94.00 & 90.15 & 93.10 & 91.14 & 92.79 & 92.90 & 97.24 & 99.18 & 99.32\\
    \cline{2-11}
    & Mixed & 81.90 & 83.94 & 82.00 & 82.90 & 85.04 & 86.20 & 94.24 & 94.89 & 95.90\\
    \hline
    \hline
    \end{tabu}
    \begin{tablenotes}
    \small
    \item \textit{Notes:} OR, SI, and SR denote the abbreviations of original, seam insertion, and seam removal, respectively.
    \end{tablenotes}
    \label{table2}
\end{table*}

\subsection{Baselines}
To demonstrate the effectiveness of the proposed ILFNet, we designed an experiment to analyze the classification performance of our work versus the comparative approaches.
For CNN-based approaches of classifying manipulation artifacts and low-level signals, Xception \cite{xception}, rResNet \cite{resnet}, BayarNet \cite{bayar}, HeNet \cite{henet}, H-VGG \cite{h-vgg}, YeNet \cite{yenet}, LFNet \cite{icip}, and SRNet \cite{srnet} were employed as baselines in the experiments.
For the three-class classification, the CNN components were lightly modified so that the last layer provides three predicted probabilities.
Inspired by the setting in \cite{icip}, in the case of rResNet, the ResNet-34 model \cite{resnet}, in which the initial pooling layer was excluded, was employed.
The hyperparameters of the comparative CNN-based approaches were set as described in each paper \cite{icip,xception,resnet,bayar,henet,h-vgg,yenet,sr1}, and the batch size and optimizer were set using the methodology specified in Section~\ref{seam_sec_training_setting}.
For fair experiments, the weight initialization was set equally, and heuristic adjustment of the learning rate was excluded from the training process.
The rotation-based data augmentation was used only in the training phase of the horizontal seam-carving classifier.

As illustrated in Fig.~\ref{fig7}, each model was trained until it sufficiently converged in terms of training accuracy and loss for 50 epochs.
We found that trivial improvement occurred in the performance of each model after 50 epochs.
Like ILFNet, the best model was selected as the one that maximizes the validation accuracy.
In addition, we conducted a comparative experiment using a conventional handcrafted feature-based approach \cite{seam_ryu}, and the results are provided in Section~\ref{seam_sec_conventional}.
The parameters of the conventional forensic method were set as specified in \cite{seam_ryu}.
Because traditional seam-carving classifiers only allow two-class classification (i.e., seam insertion versus original and seam removal versus original), we employed multiple classifiers in the experiments.
In the case of the proposed ILFNet, only one trained model for three-class classification was used.

\subsection{Evaluation Metrics}
We employed classification accuracy as the base evaluation criterion, which is defined as follows: $\frac{n_c}{n_t} \times 100$ $(\%)$, where $n_t$ and $n_c$ indicate the total number of testing samples and the number of correctly predicted samples, respectively.
When calculating the accuracy of the testing set, the ratio of data corresponding to each class was maintained at $1:1:1$, except for the cases of tests with specially designed two-class classification.
In addition, the receiver operating characteristic (ROC) curves were computed to evaluate the performance of the proposed and comparative models.
The ROC curve is defined as a plot of the true positive rate against the false positive rate, and the area under the curve (AUC) is further used as a metric for performance evaluation.
In the case of the AUC, a model is considered to have outstanding performance if the value of AUC is close to 1.

\subsection{Performance Evaluation of Networks}
We first evaluated the performance of the proposed ILFNet and eight comparative networks \cite{xception,resnet,bayar,henet,h-vgg,yenet,icip,sr1} by measuring the accuracy of the three-class classification (i.e., original, seam insertion, and seam removal).
The results of measuring the accuracy of the retargeting ratio parameters are listed in Table~\ref{table1}.
The bottom part of the table lists the classification results for a mixed test set that contains a retargeting ratio of 10\% to 50\%.
The accuracy values of ILFNet for a 10\% to 50\% retargeting ratio are 88.17\%, 94.93\%, 98.40\%, 99.43\%, and 99.53\%, respectively.
The classification performance for the mixed set is 96.56\%, which is the state-of-the-art performance when compared to CNN-based baselines.
The proposed work achieved 0.87\% higher performance compared to SRNet \cite{sr1}, which exhibits outstanding performance for steganalysis in the spatial and JPEG domains.
Compared with the performance of LFNet \cite{icip}, 1.17\% higher accuracy was achieved, which confirms that the architecture refinement applied to ILFNet was effective.
Other networks in \cite{xception,resnet,bayar,henet,h-vgg,yenet} demonstrated acceptable performance, but each accuracy value was less than 90\%.
As listed in Table~\ref{table1}, all networks exhibited lower accuracy when the retargeting ratio was 10\% because a smaller ratio results in fewer traces of forgery remaining in the image.

Next, we applied the models for three-class classification to the tasks of two-class classification.
Table~\ref{table2} lists the results of the two-class classification (i.e., seam insertion versus original and seam removal versus original) on various retargeting ratios.
The proportion of data corresponding to each class was maintained at $1:1$.
As listed in Table~\ref{table2}, our work demonstrated outstanding performance for two types of classification tasks.
In the experiments, we found that our work and baselines reveal the tendency of better exploration of the traces due to seam insertion than due to the artifacts caused by seam removal.
Capturing artifacts of seam removal may be more difficult than a task for seam insertion because forensic feature extraction proceeds by focusing on the differences between adjacent pixels due to the loss of information.
For a mixed set, the proposed ILFNet showed 1.74\% higher accuracy for the task of seam insertion compared to the task of seam removal.

\begin{figure*}[t]%
\centering%
\subfigure[Xception]{%
  \includegraphics[width=2.2in, height=1.5in]{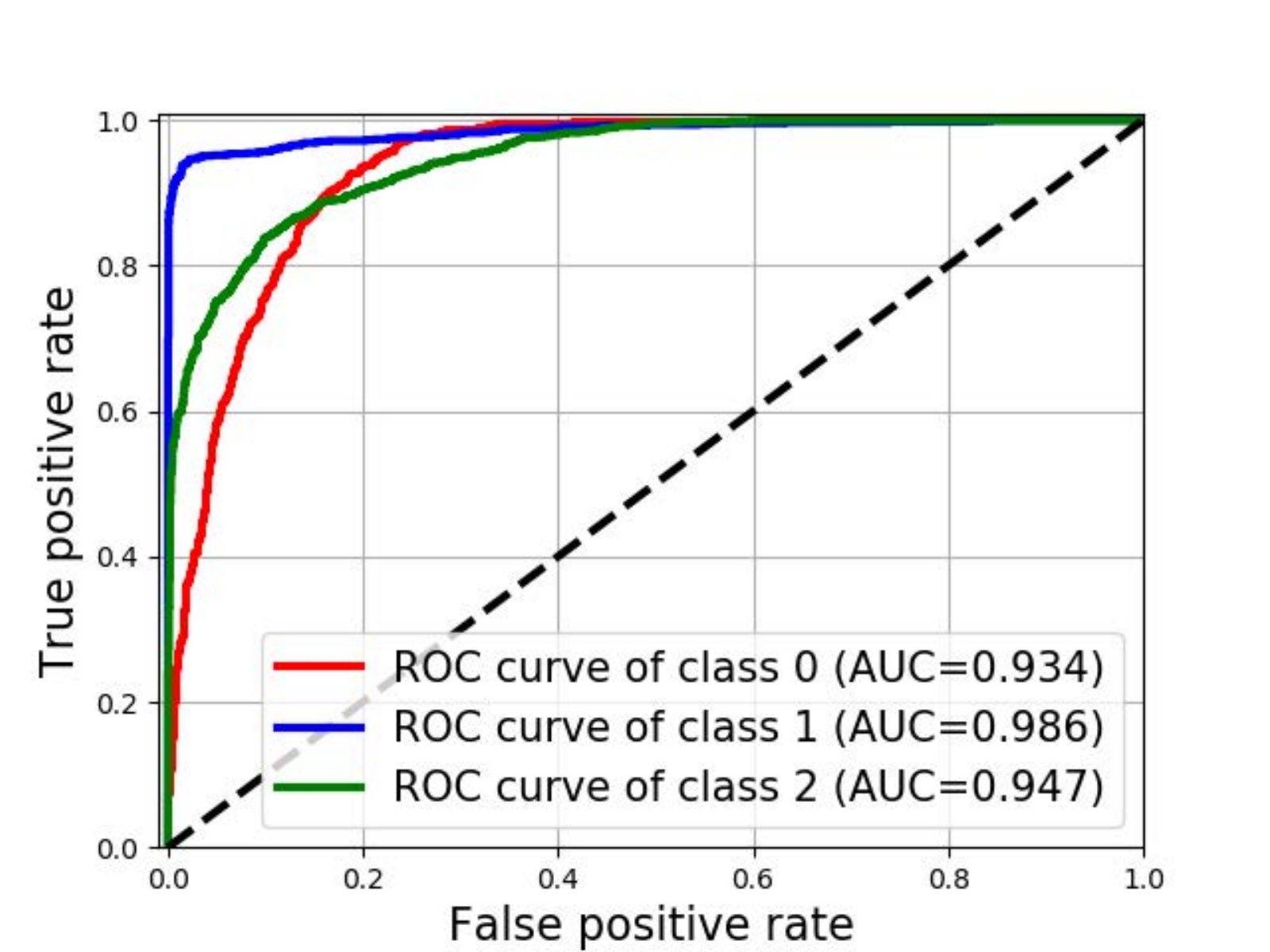}%
  \label{fig8:ex:a}
}
\subfigure[rResNet]{%
  \includegraphics[width=2.2in, height=1.5in]{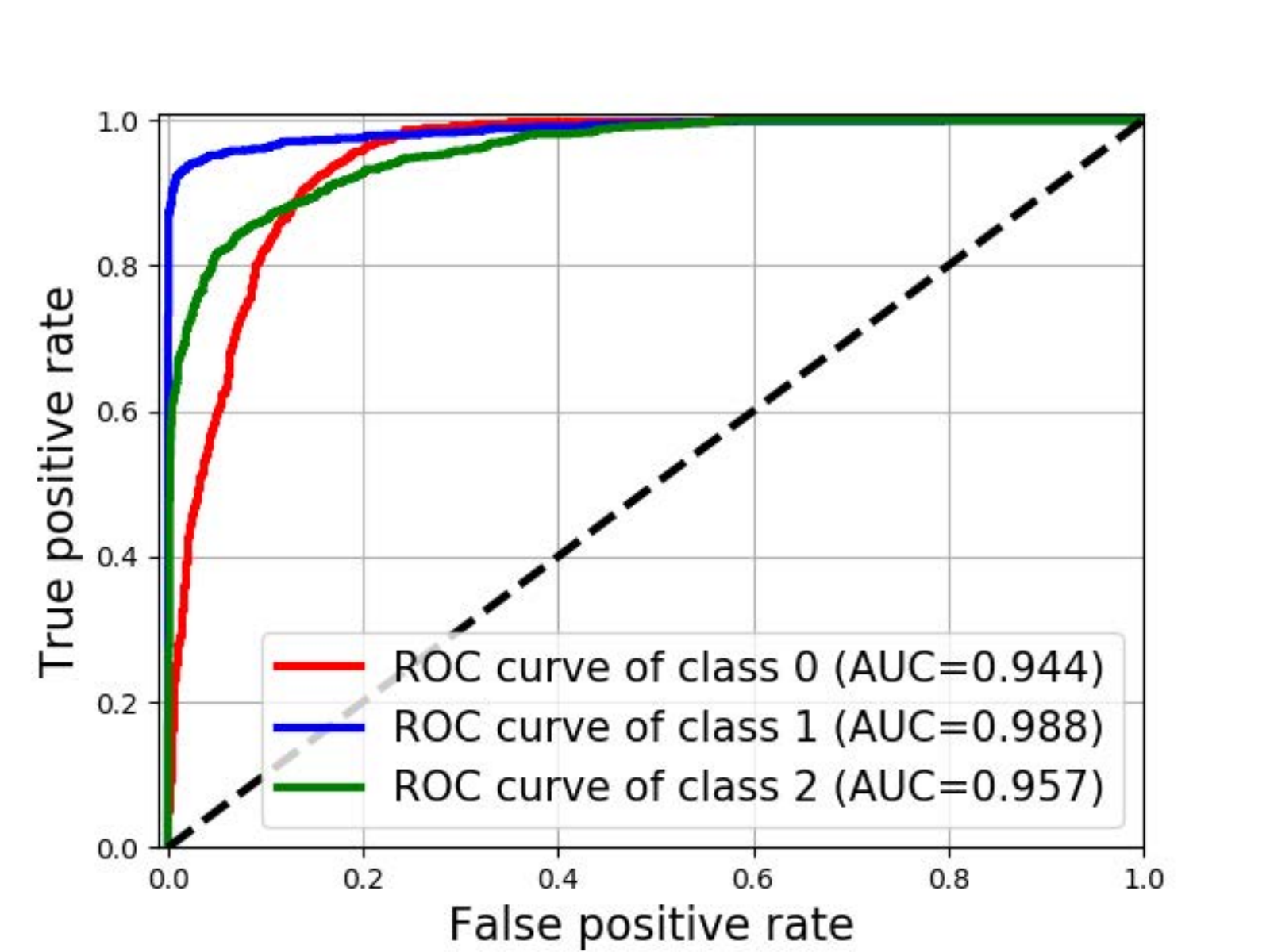}%
  \label{fig8:ex:b}
}
\subfigure[BayarNet]{%
  \includegraphics[width=2.2in, height=1.5in]{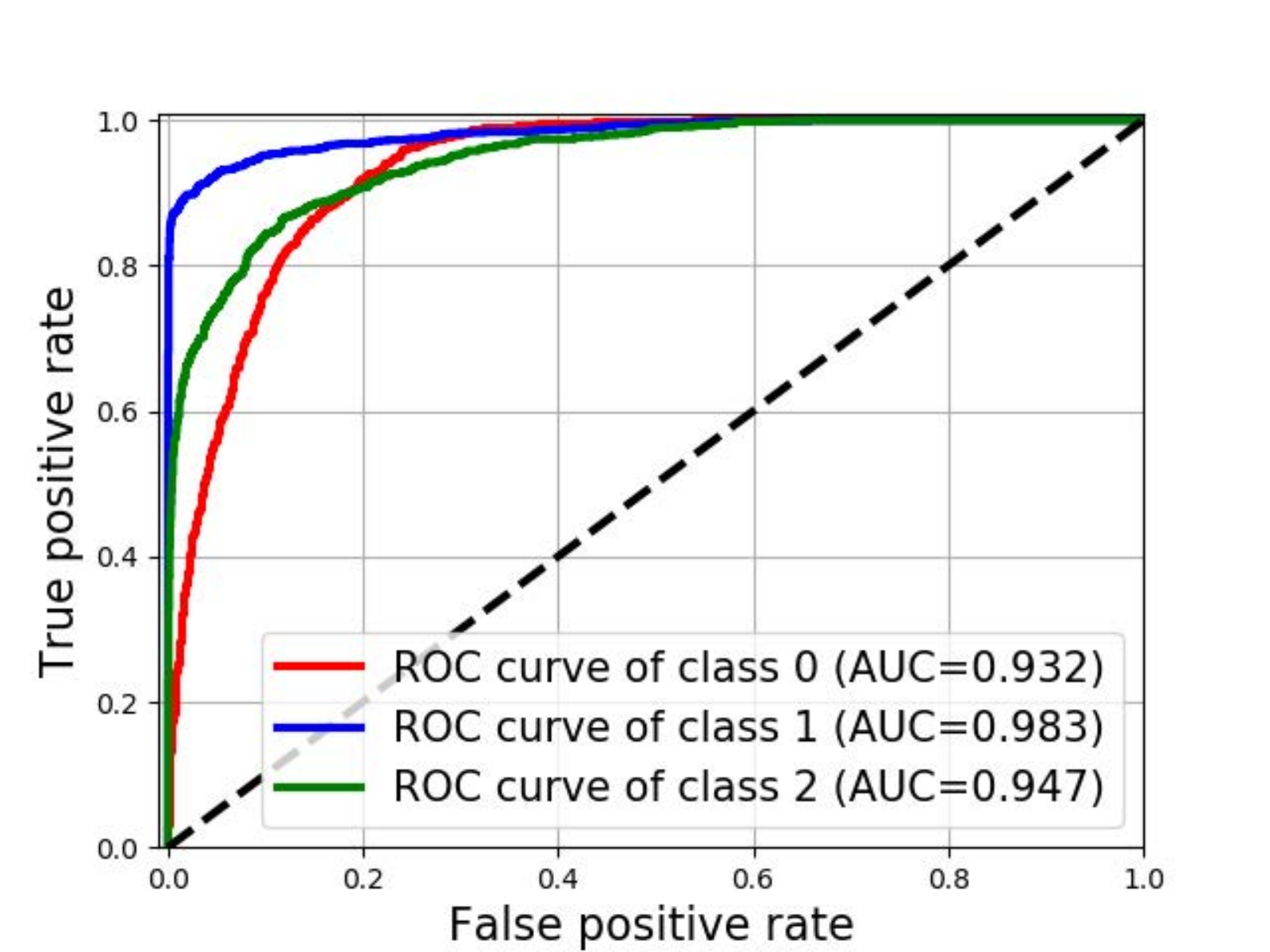}%
  \label{fig8:ex:c}
}

\subfigure[HeNet]{%
  \includegraphics[width=2.2in, height=1.5in]{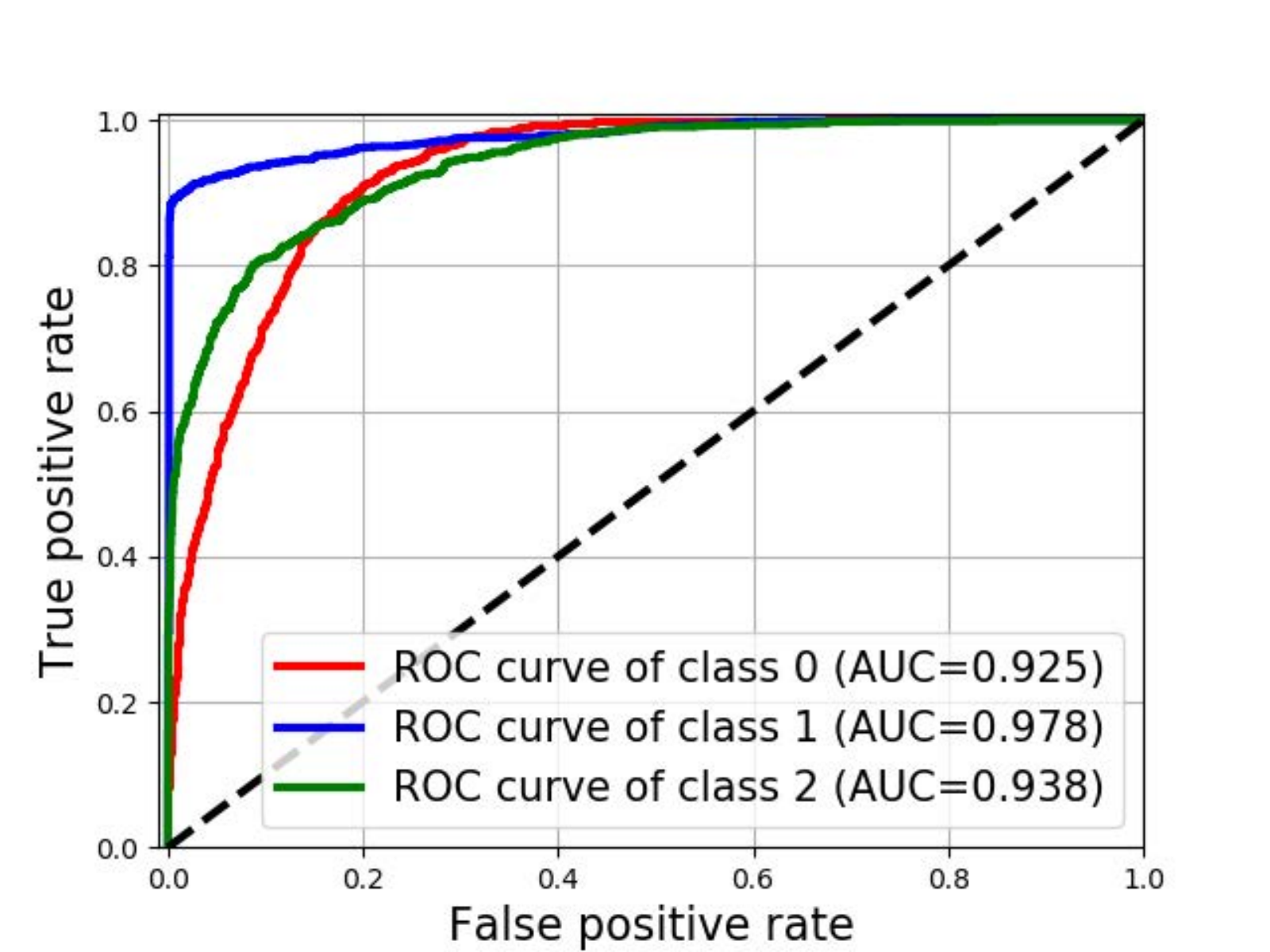}%
  \label{fig8:ex:d}
}
\subfigure[H-VGG]{%
  \includegraphics[width=2.2in, height=1.5in]{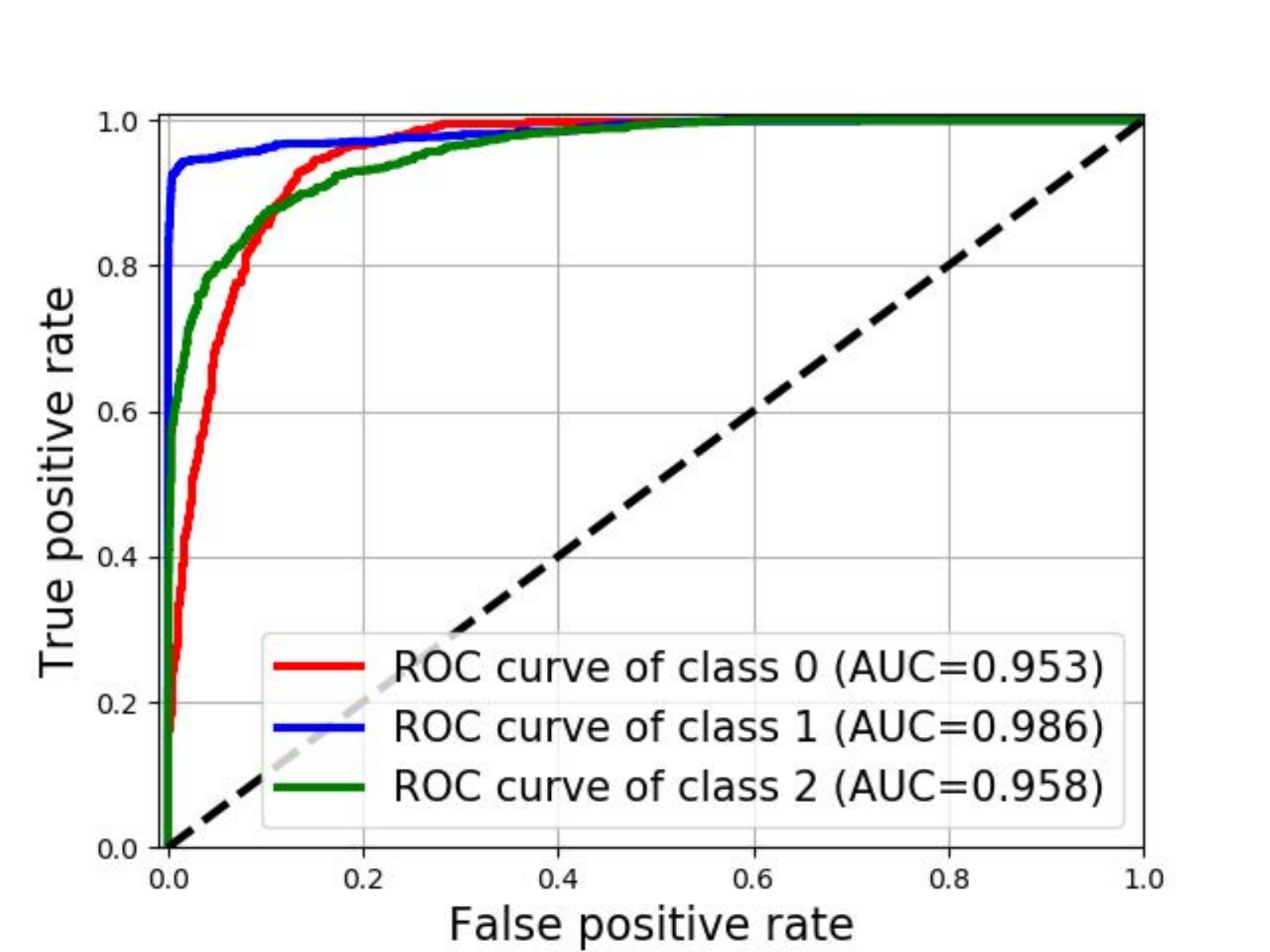}%
  \label{fig8:ex:e}
}
\subfigure[YeNet]{%
  \includegraphics[width=2.2in, height=1.5in]{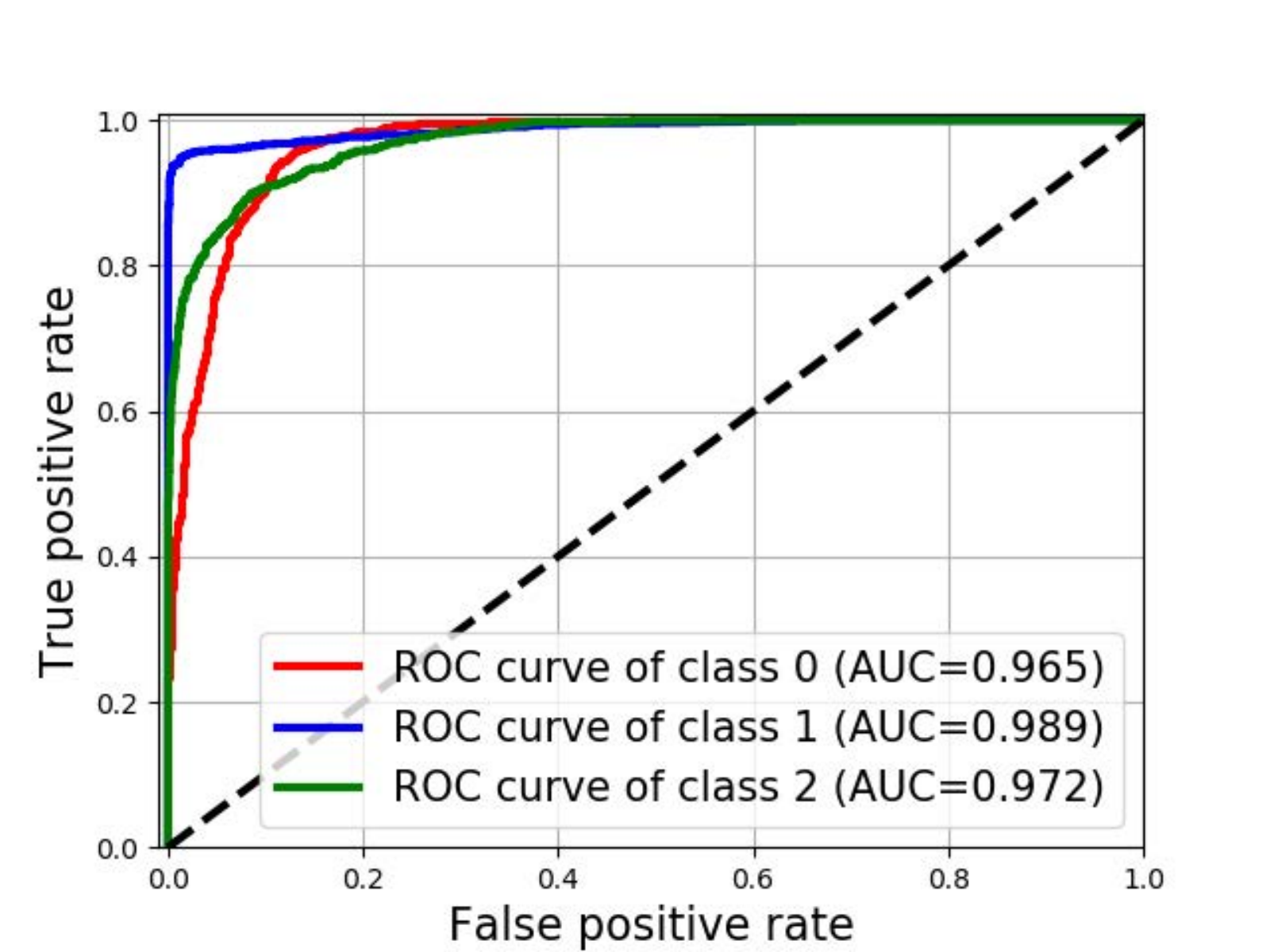}%
  \label{fig8:ex:f}
}

\subfigure[LFNet]{%
  \includegraphics[width=2.2in, height=1.5in]{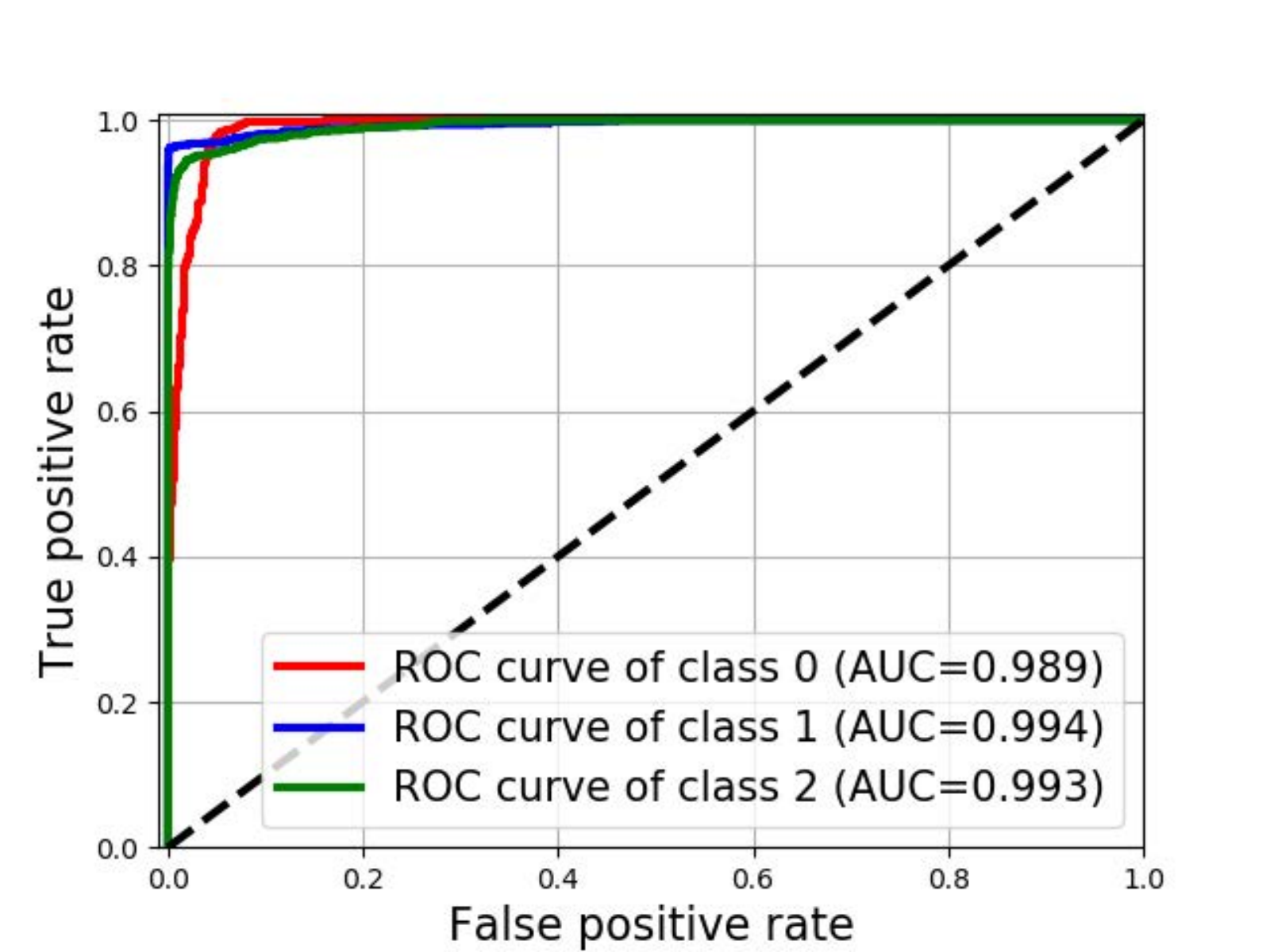}%
  \label{fig8:ex:g}
}
\subfigure[SRNet]{%
  \includegraphics[width=2.2in, height=1.5in]{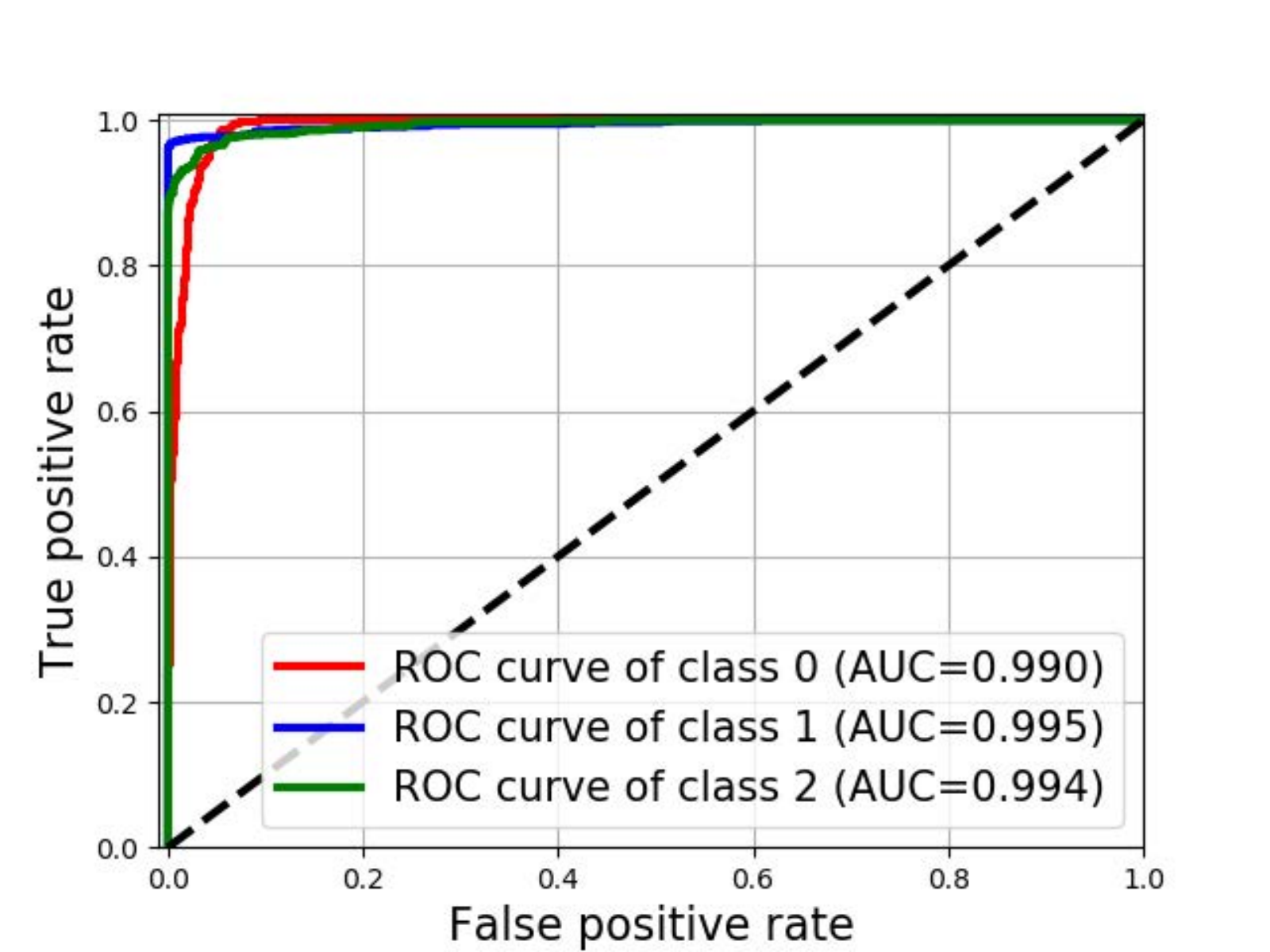}%
  \label{fig8:ex:h}
}
\subfigure[ILFNet]{%
  \includegraphics[width=2.2in, height=1.5in]{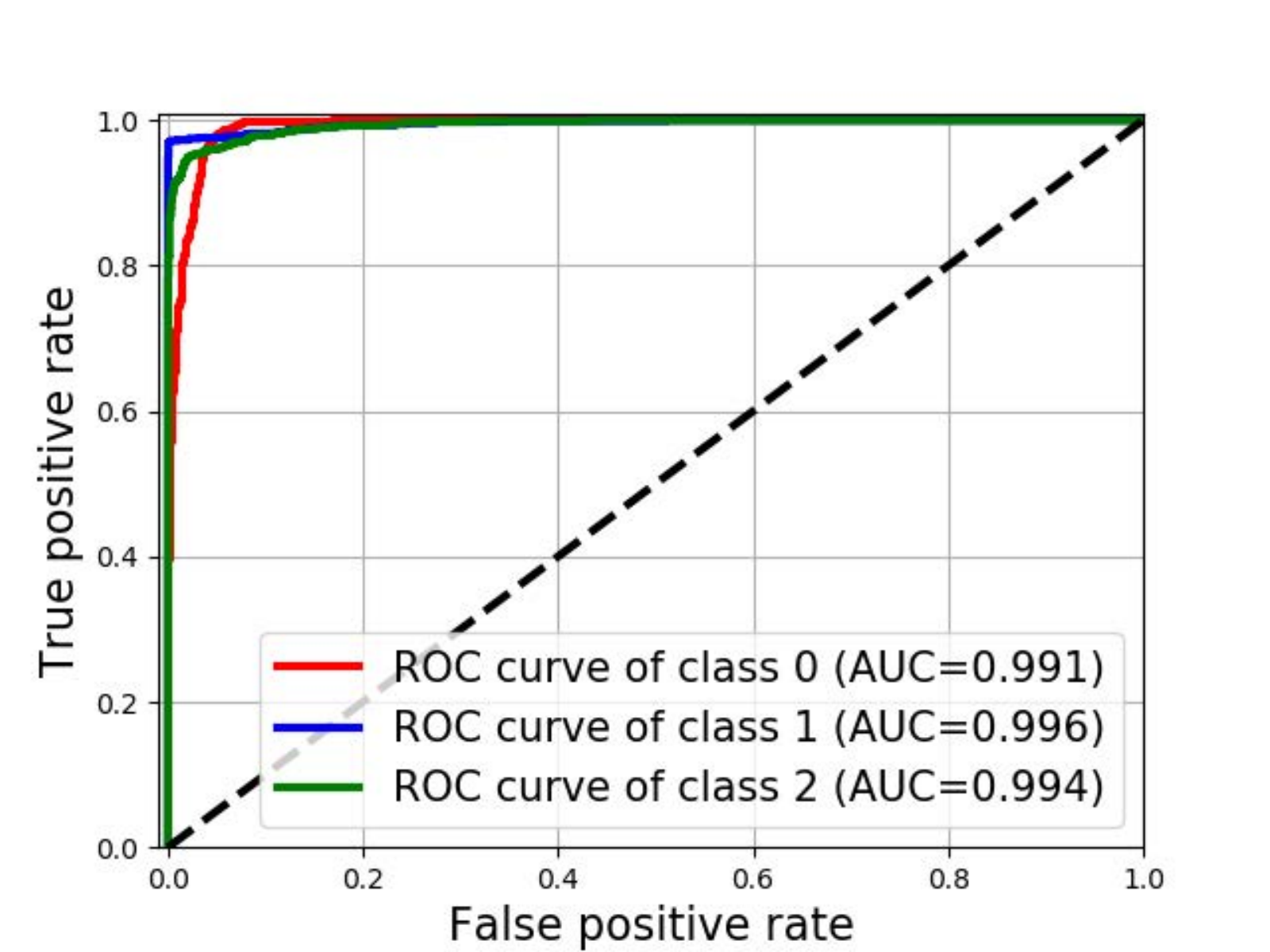}%
  \label{fig8:ex:i}
}
\caption{Receiver operating characteristic (ROC) curves of each network and the computed area under the curve (AUC) values for each class: (a)-(i) show the results for Xception, rResNet, BayarNet, HeNet, H-VGG, YeNet, LFNet, SRNet, and the proposed ILFNet, respectively.}
\label{fig8}%
\end{figure*}

To evaluate the performance of ILFNet and the comparative networks in detail, we computed the ROC curve for the three-class classification.
As presented in Fig.~\ref{fig8}, the ROC curve for each class was generated, and each AUC value for the ROC curves was calculated (see the legend of each subfigure in Fig.~\ref{fig8}).
Because the ROC curves of ILFNet, LFNet, and SRNet were closer to the top left corner than those of the other comparative networks \cite{xception,resnet,bayar,henet,h-vgg,yenet}, we conclude that ILFNet, LFNet, and SRNet perform better than the other networks.
Furthermore, our work has AUC values of 0.991, 0.996, and 0.994 for each class, which are higher than those of both LFNet and SRNet.
The comprehensively analyzed results of Table~\ref{table1} and Fig.~\ref{fig8} reveal that the proposed ILFNet performs better than the baselines in terms of three-class classification.

\begin{table}[t]
\caption{Classification accuracy of the proposed ILFNet on types of pooling layer $(\%)$.}
    \centering
    \footnotesize
    \begin{tabu} to \linewidth{X[1.0,c] X[1.5,c] X[1.5,c]}
    \hline
    \hline
    Ratio &  MaxPool layer & AvgPool layer  \\
    \hline
    10$\%$ & 88.69 & 88.17 \\
    20$\%$ & 95.10 & 94.93 \\
    30$\%$ & 97.73 & 98.40 \\
    40$\%$ & 98.63 & 99.43 \\
    50$\%$ & 98.73 & 99.53 \\
    \cline{1-3}
    Mixed &95.99 & 96.56 \\
    \hline
    \hline
    \end{tabu}
    \label{table3}
\end{table}

\begin{table*}[t]
\caption{Performance evaluation of ILFNet and comparative convolutional neural networks with an ensemble module for three-class classification of various retargeting ratios $(\%)$.}
    \footnotesize
    \centering
    \begin{tabu} to \linewidth{X[0.3,c] X[1.0,c] X[1.0,c] X[1.0,c] X[1.0,c] X[1.0,c] X[1.0,c] X[1.0,c] X[1.0,c] X[1.0,c] X[1.0,c]}
    \hline
    \hline
    $\Theta$ & Ratio & Xception & rResNet & BayarNet & HeNet & H-VGG & YeNet & LFNet & SRNet & ILFNet\\
    \hline
    \multirow{5}{*}{1}&$10\%$ & 68.96 & 73.46 & 65.73 & 64.30 & 75.50 & 76.00 & 88.27 & 87.40 & 88.17\\
    & 20$\%$ & 81.06 & 83.97 & 80.86 & 80.40 & 85.23 & 86.29 & 94.46 & 94.60 & 94.93\\
    & 30$\%$ & 89.23 & 90.93 & 87.89 & 87.83 & 91.03 & 92.09 & 97.06 & 98.23 & 98.40\\
    & 40$\%$ & 93.19 & 92.27 & 92.83 & 91.67 & 93.59 & 94.27 & 98.07 & 99.29 & 99.43\\
    & 50$\%$ & 95.97 & 94.33 & 94.82 & 92.99 & 95.00 & 94.96 & 98.16 & 99.47 & 99.53\\
    \cline{2-11}
    & Mixed & 85.93 & 87.06 & 84.83 & 83.96 & 88.66 & 88.83 & 95.39 & 95.69 & 96.56\\
    \hline    
    \multirow{5}{*}{5}& $10\%$ & 70.10 & 77.80 & 66.06 & 69.13 & 76.63 & 79.56 & 90.74 & 89.93 & 89.97\\
    & $20\%$ & 83.13 & 88.22 & 82.10 & 84.03 & 87.10 & 90.83 & 96.73 & 96.60 & 96.23\\
    & $30\%$ & 91.12 & 91.93 & 89.52 & 89.26 & 92.43 & 94.40 & 98.36 & 98.50 & 98.57\\
    & $40\%$ & 95.25 & 94.14 & 93.66 & 92.43 & 95.21 & 96.31 & 99.23 & 99.26 & 99.56\\
    & $50\%$ & 97.96 & 95.17 & 96.01 & 92.60 & 96.90 & 96.90 & 99.46 & 99.86 & 99.87\\
    \cline{2-11}
    & Mixed & 87.62 & 89.46 & 85.67 & 85.75 & 89.75 & 91.36 & 96.96 & 96.82 & 97.02\\
    \hline
    \multirow{5}{*}{10}& $10\%$ & 70.63 & 78.86 & 66.20 & 70.73 & 77.02 & 81.30 & 91.47 & 90.36 & 91.30\\
    & $20\%$ & 83.67 & 89.13 & 83.13 & 84.20 & 87.63 & 91.13 & 96.90 & 96.37 & 96.53\\
    & $30\%$ & 91.14 & 92.73 & 89.12 & 89.94 & 92.67 & 94.87 & 98.53 & 98.16 & 98.64\\
    & $40\%$ & 95.66 & 94.13 & 93.80 & 92.27 & 95.36 & 96.13 & 99.23 & 99.36 & 99.57\\
    & $50\%$ & 98.33 & 95.70 & 96.27 & 92.96 & 97.30 & 97.03 & 99.32 & 99.89 & 99.83\\
    \cline{2-11}
    & Mixed & 87.99 & 90.18 & 86.31 & 86.12 & 89.89 & 92.09 & 97.08 & 97.02 & 97.18\\
    \hline
    \hline
    \end{tabu}
    \label{table4}
\end{table*}

We conducted a performance analysis to determine the pooling layer suitable for ILFNet (Table~\ref{table3}).
As mentioned in Section~\ref{seam_sec_diff}, we applied the pooling layer for dimensional reduction to BT-3 for hierarchical feature learning.
To determine the pooling layer for ILFNet, we analyzed the three-class classification performance of two trained models employing the MaxPool layer or AvgPool layer.
Here, the kernel and stride for the pooling layer were set to $3\times 3$ and 2, respectively.
As listed in Table~\ref{table3}, the AvgPool-based model exhibits 96.56\% accuracy for the mixed ratio set, which is greater than that of the MaxPool-based model by 0.57\%.
When the retargeting ratios were 10\% and 20\%, the MaxPool-based model demonstrated an accuracy slightly above the AvgPool-based model, but overall, the model with the AvgPool layer performed better.
Based on the results in the table, we decided to place the AvgPool layer on BT-3, which comprises ILFNet.

\begin{figure*}[t]%
\centering%
\subfigure[]{%
  \includegraphics[width=2.2in, height=1.6in]{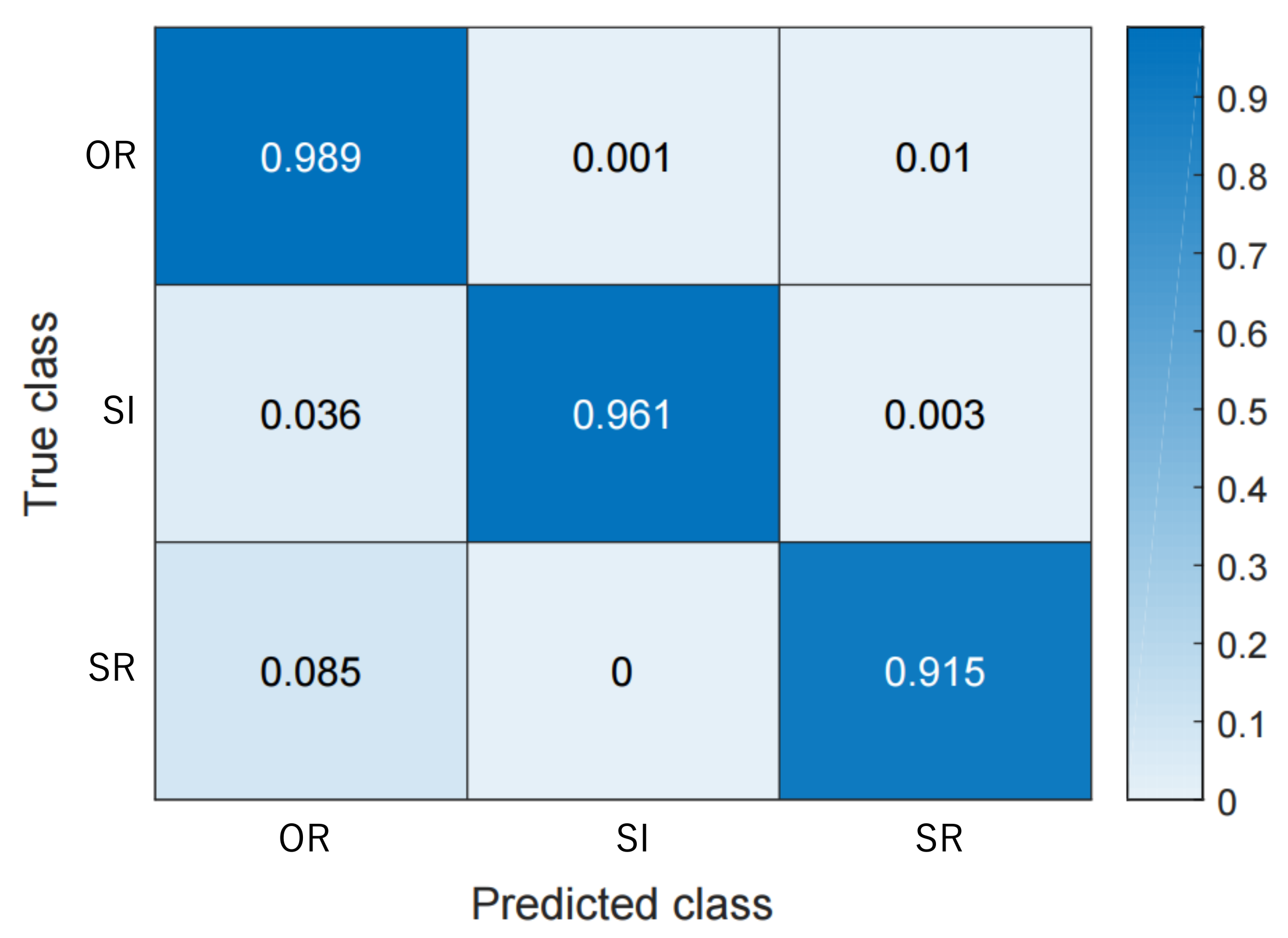}%
  \label{fig9:ex:a}
}\hfil
\subfigure[]{%
  \includegraphics[width=2.2in, height=1.6in]{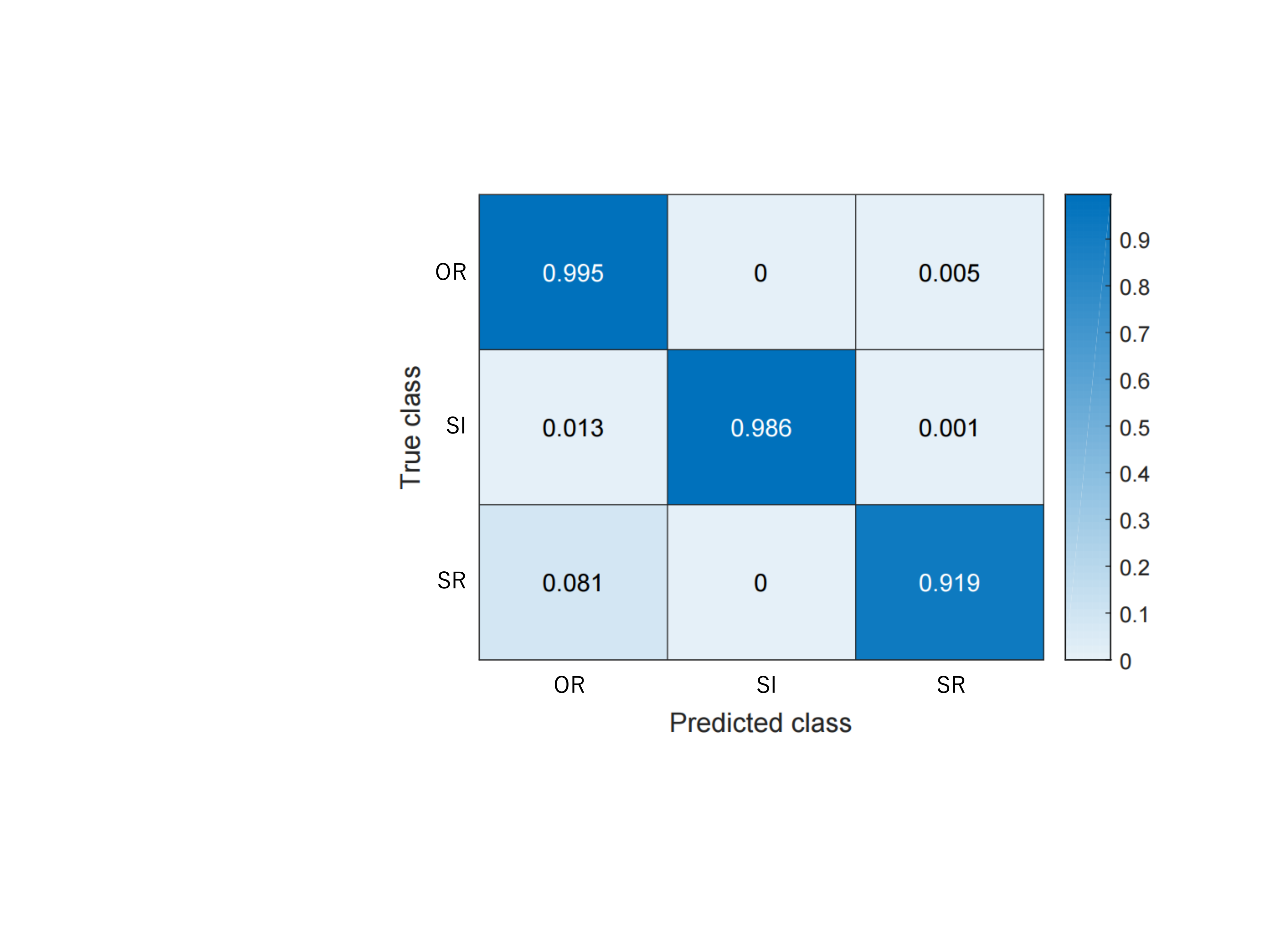}%
  \label{fig9:ex:b}
}\hfil
\subfigure[]{%
  \includegraphics[width=2.2in, height=1.6in]{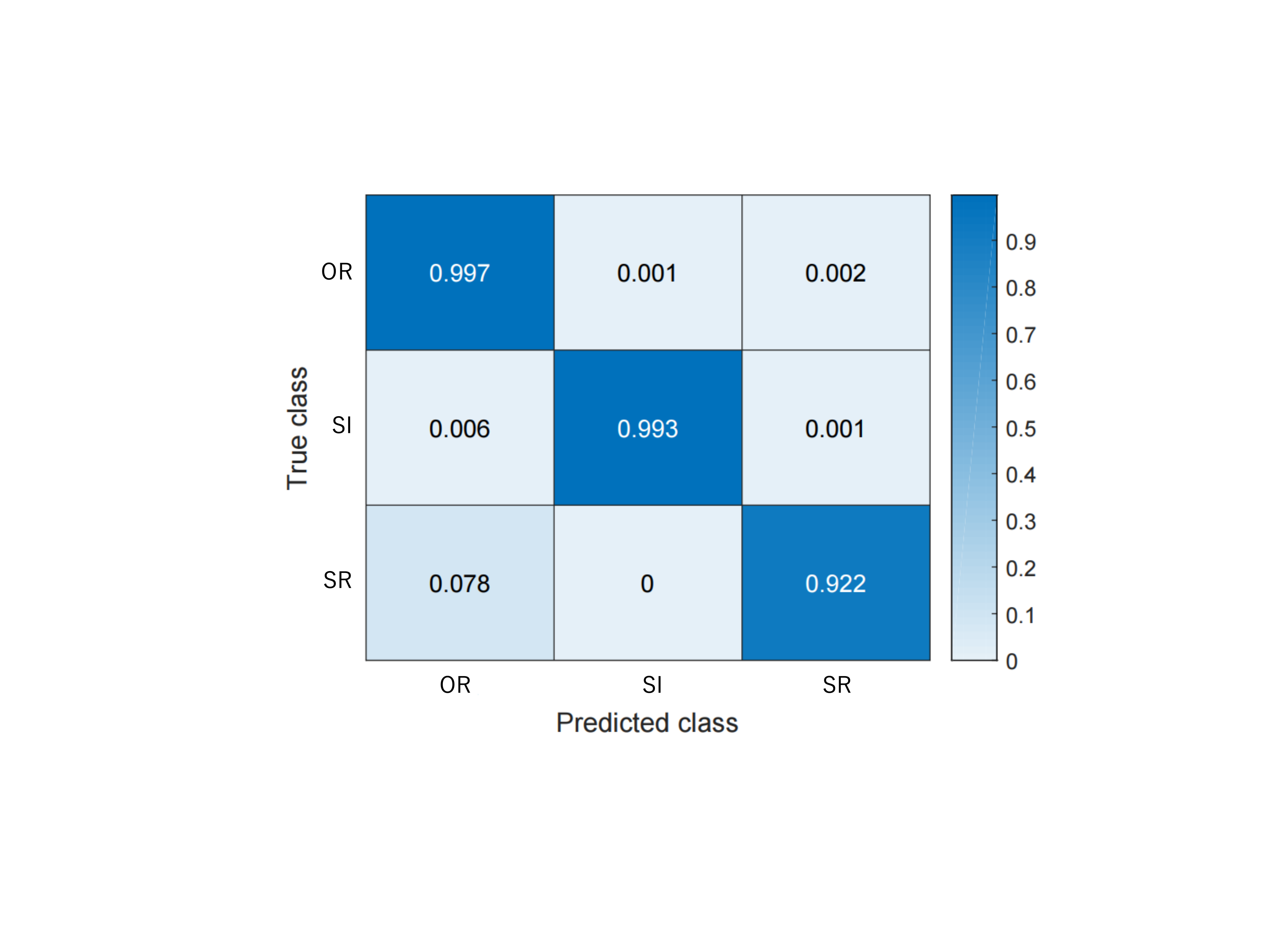}%
  \label{fig9:ex:c}
}
\caption{Confusion matrix of the proposed ILFNet for classifying seam carving with an ensemble module: (a) result of ILFNet with $\Theta=1$, (b) result of ILFNet with $\Theta=5$, and (c) result of ILFNet with $\Theta=10$. Here, OR, SI, and SR denote the abbreviations of original, seam insertion, and seam removal, respectively.}
\label{fig9}%
\end{figure*}

\subsection{Performance Evaluation of Networks with Ensemble Module}
In Section~\ref{seam_sec_ensemble}, we presented a methodology to improve the test performance of the trained model using the ensemble module, which does not require additional training.
Through the ensemble module, we expected the trained CNN-based model with the ensemble module and multiple samples to comprehensively explore local texture artifacts due to seam carving scattered throughout a given suspicious image.
As mentioned, $\Theta$ indicates the number of patches sampled by applying cropping to the suspicious image.
When the value of $\Theta$ is equal to 1, it indicates models that attempt to classify forgery from a single sample in the previous section.
In this experiment, we applied the ensemble module with $\Theta=1, 5, 10$ to our trained and comparative models and analyzed the classification performance of each model against seam-carving forgery.

Table~\ref{table4} lists the performance evaluation of ILFNet and comparative CNNs with an ensemble module for three-class classification on various retargeting ratios.
When the number of samples ($\Theta$) provided to the ensemble module was set to 1, 5, and 10, the classification accuracies of the proposed ILFNet were 96.56\%, 97.02\%, and 97.18\%, respectively.
Likewise, for the comparative models with ensemble modules, performance improved due to the ensemble module, as presented in Table~\ref{table4}.
For the case in which multiple samples were provided in the ensemble module (i.e., $\Theta=5,10$), the proposed ILFNet achieved the highest accuracy, and the LFNet exhibited second-best performance.
The YeNet achieved a large improvement in performance with the adoption of the ensemble module, and when the value of $\Theta$ increases from 1 to 10, performance improves by 3.26\%.

In the seam-carving-based retargeting process, the distribution of computed seams is affected by the content of the image; thus, the local region generally expands or reduces throughout the given image.
Therefore, by providing samples of various local areas in the seam-carved image in the model, the probability that samples containing abundant forensic traces caused by seam-carving forgery are provided to the model increases.
The results in Table~\ref{table4} support why ensemble modules should be adopted to the proposed forensic framework for classifying seam carving.
Based on the results in Table~\ref{table4}, the ensemble module can provide additional performance gain in terms of classifying seam-carving artifacts.

For a more detailed analysis, we generated a confusion matrix for three-class classification using ILFNet with the ensemble module.
As the number of samples provided to ILFNet increased (i.e., $\Theta=1 \rightarrow \Theta= 5,10$), the number of correct predictions for the corresponding true class increased (see Fig.~\ref{fig9}).
In the case of ILFNet with $\Theta=10$, the correctly predicted probability values for the true classes of original, seam insertion, and seam removal were 0.997, 0.993, and 0.922, respectively.
Furthermore, we found that seam-removed images are often misclassified as original images, as observed in Fig.~\ref{fig9}.

\begin{table*}[t]
\caption{Performance evaluation of ILFNet and comparative convolutional neural networks with an ensemble module for three-class classification on unseen seam-carving algorithms $(\%)$.}
\footnotesize
    \centering
    \begin{tabu} to \linewidth{X[2.0,c] X[0.5,c] X[1.0,c] X[1.0,c] X[1.0,c] X[1.0,c] X[1.0,c] X[1.0,c] X[1.0,c] X[1.0,c] X[1.0,c]}
    \hline
    \hline
    SC method & $\Theta$ & Xception & rResNet & BayarNet & HeNet & H-VGG & YeNet & LFNet & SRNet & ILFNet\\
    \hline
    \multirow{3}{*}{Avidan \emph{et al.}\cite{seam1}}& 1 &85.93 & 87.06 & 84.83 & 83.96 & 88.66 & 88.83 & 95.39 & 95.69 & 96.56\\
    & 5 & 87.62 & 89.46 & 85.67 & 85.75 & 89.75 & 91.36 & 96.96 & 96.82 & 97.02\\
    & 10 & 87.99 & 90.18 & 86.31 & 86.12 & 89.89 & 92.09 & 97.08 & 97.02 & 97.18\\\hline
    \multirow{3}{*}{Rubinstein \emph{et al.}\cite{seam2}}&1 & 79.33 & 83.91 & 75.62 & 75.67 & 81.66 & 83.00 & 89.53 & 89.80 & 90.10\\
    & 5 & 80.70 & 87.10 & 78.97 & 79.50 & 83.53 & 86.26 & 89.77 & 90.13 & 90.73\\
    & 10 & 81.14 & 88.53 & 79.40 & 80.43 & 83.46 & 86.71 & 90.20 & 89.92 & 91.18\\\hline
    \multirow{3}{*}{Achanta \emph{et al.}\cite{seam3}}&1 & 56.83 & 60.93 & 58.73 & 69.57 & 74.43 & 70.03 & 79.91 & 76.53 & 79.27\\
    & 5 & 60.77 & 63.82 & 61.20 & 75.10 & 79.13 & 76.73 & 81.00 & 76.32 & 81.37\\
    & 10 & 60.46 & 64.50 & 61.93 & 74.98 & 79.23 & 75.62 & 81.33 & 76.90 & 81.71\\\hline
    \multirow{3}{*}{Frankovich \emph{et al.}\cite{seam4}}&1 & 80.27 & 83.03 & 77.20 & 77.46 & 81.33 & 84.45 & 88.52 & 90.13 & 89.57\\
    & 5 & 84.43 & 86.37 & 79.23 & 80.06 & 84.15 & 87.60 & 89.55 & 90.23& 90.37\\
    & 10 & 83.70 & 87.80 & 80.53 & 80.47 & 85.50 & 87.82 & 90.73 & 90.45& 91.38\\
    \hline
    \hline
    \end{tabu}
    \label{table5}
\end{table*}

\subsection{Performance Evaluation of Networks for Unseen Cases}

In this section, the results of the robustness experiments for the unseen cases that were not considered in the training process of the models are provided.
It is important in multimedia forensics to ensure robustness against unconsidered environments, such as digital watermarking \cite{deep_wm,dibr_wm1,dibr_wm2}, which is robust from various attacks in the distribution process.
From this perspective, it is beneficial for a CNN-based forensic approach to be robust against unseen cases.
To demonstrate the effectiveness of the proposed ILFNet, we further conducted extended experiments for the testing set of unseen cases, meaning testing environments that were not considered in the training phase.
We employed the trained models for the three-class classification described in the previous sections and conducted the robustness test without additional training for the unseen cases.
In this section, the following unseen cases are covered.

\begin{figure*}[t!]
\centering{\includegraphics[width=0.95\linewidth]{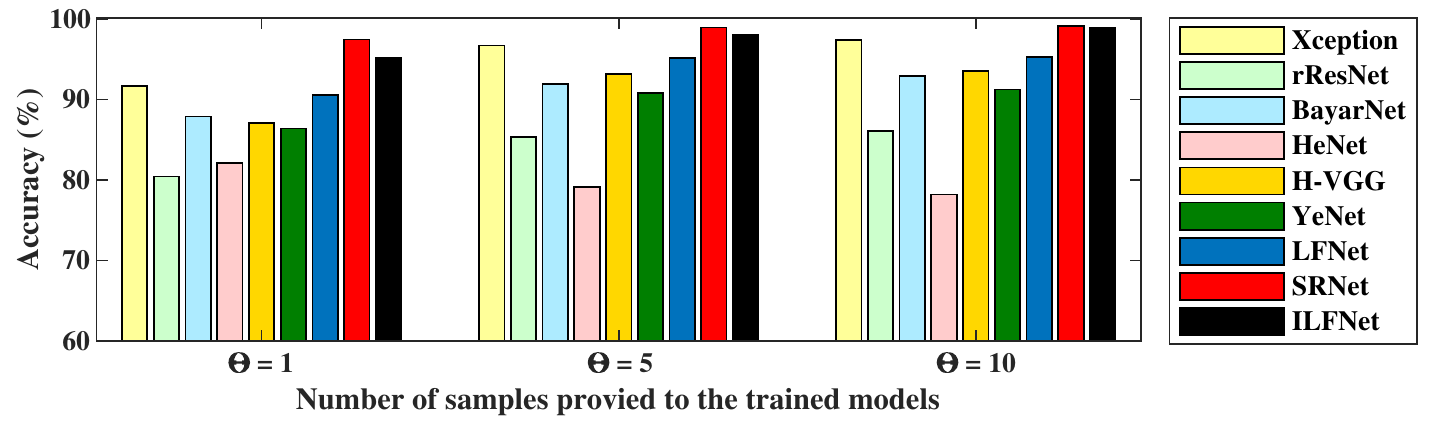}}
\caption{Performance evaluation of ILFNet and comparative convolutional neural networks with an ensemble module on the unseen retargeting ratio of 0\%. In the experiment, the models were applied to the testing set consisting of single-compressed and double-compressed original images.}
\label{fig10}
\end{figure*}

\begin{itemize}
\item Unseen seam-carving algorithms in \cite{seam2,seam3,seam4},
\item Unseen retargeting ratios of 0\%, 4\%, 6\%, and 8\%,
\item Unseen post-processing of noise addition,
\item Unseen uncompressed image format such as BMP.
\end{itemize}

First, we conducted the performance evaluation of ILFNet and comparative CNNs for unseen seam-carving algorithms \cite{seam2,seam3,seam4}.
As described in Section~\ref{seam_sec_seamcarving}, the computed seams have various characteristics due to the inherent properties of the content (e.g., the shape of the object and background) and the predefined function of each the seam-carving algorithm.
To address this issue, we designed a network architecture for learning forensic features, even in areas with few artifacts, and proposed an ensemble module-based methodology to improve performance by comprehensively analyzing multiple local samples.
In this experiment, we verified whether our attempts were effective, and models trained using only a seam-carving algorithm \cite{seam1} were used to measure the accuracy of the unseen cases \cite{seam2,seam3,seam4}.

Table~\ref{table5} indicates the classification accuracy of ILFNet and comparative CNNs on unseen seam-carving algorithms \cite{seam2,seam3,seam4}.
The table contains the results for a mixed test set that contains a retargeting ratio of 10\% to 50\% for each algorithm.
In the comprehensive analysis, the proposed ILFNet extracted artifacts due to seam-carving algorithms \cite{seam2,seam3,seam4} that were not considered in the training process better than other comparative CNNs.
With an ensemble module of $\Theta=10$, ILFNet achieved a classification accuracy of 91.18\%, 81.71\%, and 91.38\% for each unseen algorithm \cite{seam2,seam3,seam4}.
As listed in Table~\ref{table5}, models trained on the training set for \cite{seam1} tended to detect forensic traces in the test set for \cite{seam2,seam4} better than the test set for \cite{seam3}.
The seam-carving algorithms in \cite{seam2,seam4}, which were extended from \cite{seam1} using $e_{g}$, generally calculate seams similar to the results of \cite{seam1}, whereas an approach in \cite{seam3} using the newly defined $e_{s}$ calculates relatively different forms of seams compared to \cite{seam1,seam2,seam4} (Section~\ref{seam_sec_seamcarving}).
Therefore, the models appear to have higher classification accuracy for unseen algorithms \cite{seam2,seam4}, and the proposed ILFNet achieved a relatively acceptable performance for the testing set of \cite{seam3} than other comparative models.

We further conducted experiments for unlearned retargeting ratios of seam carving.
As described in Section~\ref{seam_sec_dataset}, the original images were retargeted employing a seam-carving process \cite{seam1} from 10\% to 50\% in 10\% steps for training the models.
For the experiments, we generated a testing set of unseen retargeting ratios of 0\%, 4\%, 6\%, and 8\% based on the algorithm \cite{seam1}.
When the ratio is equal to 0\%, the original image is subject to JPEG compression (quality factor = 100) without enlargement or reduction.
Thus, in this case, the models were applied to the testing set consisting of single-compressed and double-compressed original images.
Because rounding and truncation errors occur during encoding and decoding in JPEG compression with the fixed quality factor, respectively, the differences between single-compressed and double-compressed original images are not exactly zero \cite{deep_forensic1}.
Fig.~\ref{fig10} reveals the classification results of this experiment, and we aimed to prove that the proposed model classifies the single- and re-compressed images, in which the seam-carving forgery is not applied to the original content.
In this experiment, the proposed ILFNet and SRNet achieved outstanding performance, and the classification accuracies of our model were 96.12\%, 98.05\%, and 98.9\%, when $\Theta$ in ensemble module was set to 1, 5, and 10, respectively.

\begin{table*}[t]
\caption{Performance evaluation of ILFNet and comparative convolutional neural networks with an ensemble module for three-class classification on unseen retargeting ratios $(\%)$.}
\footnotesize
    \centering
    \begin{tabu} to \linewidth{X[1.0,c] X[0.5,c] X[1.0,c] X[1.0,c] X[1.0,c] X[1.0,c] X[1.0,c] X[1.0,c] X[1.0,c] X[1.0,c] X[1.0,c]}
    \hline
    \hline
    Ratio & $\Theta$ & Xception & rResNet & BayarNet & HeNet & H-VGG & YeNet & LFNet & SRNet & ILFNet\\
    \hline
    \multirow{3}{*}{4\%}&1 & 50.02 & 56.87 & 49.66 & 48.43 & 58.74 & 59.73 & 72.20 & 70.90 & 72.44\\
    & 5 & 50.83 & 58.10 & 49.02 & 49.63 & 58.50 & 59.90 & 72.47 & 71.26 & 72.53\\
    & 10 & 51.86 & 58.56 & 50.03 & 49.77 & 59.22 & 59.96 & 73.50 & 71.77 & 72.70\\
    \hline
    \multirow{3}{*}{6\%}&1 & 57.26 & 62.14 & 54.47 & 54.23 & 64.13 & 64.50 & 77.33 & 75.83 & 77.14\\
    & 5 & 57.33 & 64.95 & 54.86 & 56.20 & 65.83 & 67.03 & 78.78 & 77.06 & 77.40\\
    & 10 & 58.07 & 65.53 & 55.42 & 57.26 & 64.57 & 66.60 & 79.86 & 77.74 & 78.10\\
    \hline
    \multirow{3}{*}{8\%}&1 & 61.20 & 66.07 & 58.23 & 58.47 & 68.30 & 68.43 & 80.76 & 78.60 & 80.95\\
    & 5 & 61.86 & 69.96 & 59.70 & 61.30 & 69.03 & 70.53 & 82.00 & 79.50 & 82.13 \\
    & 10 & 62.50 & 70.43 & 60.13 & 62.16 & 69.60 & 71.70 & 82.06 & 79.96 & 82.30\\
    \hline
    \hline
    \end{tabu}
    \label{table6}
\end{table*}

\begin{table*}[!t]
\caption{Performance evaluation of ILFNet and comparative convolutional neural networks for three-class classification on post-processing of additive white Gaussian noise $(\%)$.}
    \footnotesize
    \centering
    \begin{tabu} to \linewidth{X[0.6,c] X[1.0,c] X[1.0,c] X[1.0,c] X[1.0,c] X[1.0,c] X[1.0,c] X[1.0,c] X[1.0,c] X[1.0,c]}
    \hline
    \hline
    $\sigma$ & Xception & rResNet & BayarNet & HeNet & H-VGG & YeNet & LFNet & SRNet & ILFNet\\
    \hline
    0.1 & 85.25 & 86.76 & 84.62 & 84.50 & 88.77 & 88.83 & 95.06 & 95.38 & 96.10 \\
    0.2 & 84.50 & 85.41 & 83.43 & 81.40 & 85.67 & 87.29 & 92.73 & 93.29 & 93.43\\
    0.3 & 83.34 & 84.97 & 81.02 & 73.21 & 79.03 & 85.93 & 87.23 & 91.36 & 90.37 \\
    0.4 & 82.81 & 84.80 & 78.57 & 68.51 & 72.26 & 79.63 & 82.80 & 87.56 & 86.63\\
    0.5 & 82.23 & 83.77 & 75.23 & 63.40 & 67.70 & 74.36 & 80.53 & 81.40 & 83.93 \\
    \hline
    \hline
    \end{tabu}
    \label{table7}
\end{table*}

\begin{table}[t]
\caption{Performance evaluation of ILFNet with an ensemble module for three-class classification on post-processing of additive white Gaussian noise $(\%)$.}
    \centering
    \footnotesize
    \begin{tabu} to \linewidth{X[1.0,c] X[1.5,c] X[1.5,c] X[1.5,c]}
    \hline
    \hline
    \multirow{2}{*}{$\sigma$} & \multicolumn{3}{c}{ILFNet} \\
    \cline{2-4}
    & $\Theta=1$ & $\Theta=5$ & $\Theta=10$ \\
    \hline
    0.1 & 96.10 & 96.73 & 97.01 \\
    0.2 & 93.43 & 95.60 & 96.30 \\
    0.3 & 90.37 & 93.26 & 93.83 \\
    0.4 & 86.63 & 91.87 & 92.03 \\
    0.5 & 83.93 & 87.53 & 88.50 \\
    \hline
    \hline
    \end{tabu}
    \label{table9}
\end{table}

In addition, based on the seam-carving algorithm \cite{seam1}, we conducted a performance evaluation on the testing set with retargeting corresponding to the 4\%, 6\%, and 8\% ratios of the image width.
As listed in Table~\ref{table6}, all networks exhibited lower accuracy when the unseen retargeting ratio was smaller, which may be caused by the fewer traces of forgery remaining in the given samples.
When analyzing the results in Table~\ref{table6}, ILFNet, SRNet, and LFNet demonstrated acceptable performance, whereas the accuracy values of the other networks \cite{xception,resnet,bayar,henet,h-vgg,yenet} were less than 72\%, even with the ensemble module.
In the experiments, the accuracy values of ILFNet using an ensemble module with $\Theta$ $=$ 10 were 72.70\%, 78.10\%, and 82.30\%, when the retargeting ratio was 4\%, 6\%, and 8\%, respectively.
In particular, when the unseen ratios were set at 4\% and 6\%, LFNet achieved outstanding performance, and when the ratio was 8\%, ILFNet exhibited the highest accuracy.
When analyzing the results for the unseen retargeting ratios of 0\%, 4\%, 6\%, and 8\% comprehensively, our work demonstrated stable and outstanding performance.

Next, we conducted experiments for unseen post-processing of additive white Gaussian noise (AWGN).
The AWGN can be applied in the distribution and manipulation of images \cite{dibr_wm1,dibr_wm2, deep_wm}; hence, it is important to ensure robustness against AWGN regarding practical forensics.
In this experiment, we applied AWGN with a $\sigma$ value of 0.1 to 0.5 to a mixed test set based on a seam-carving algorithm \cite{seam1} containing a retargeting ratio of 10\% to 50\%.
Table~\ref{table7} lists the classification accuracy obtained by applying the trained models ($\Theta=1$) to the test images with AWGN that were not considered in the training phase.
In the experiments, the proposed ILFNet achieved outstanding performance, and our model achieved accuracy values of 96.10\%, 93.43\%, 90.37\%, 86.63\%, and 83.93\% for $\sigma$ ranged from 0.1 to 0.5.
We also found that the accuracy values of all networks decreased as $\sigma$ increased and that low-level features caused by seam-carving forgery may have been affected by noise signal addition.
In addition, we analyzed the robustness against AWGN in the proposed ILFNet using the ensemble module with $\Theta=1, 5, 10$.
As listed in Table~\ref{table9}, our work exhibited performance improvement using the ensemble module.
In particular, when the results for the $\Theta$ values of 1 and 10 were compared, performance improved by 0.91\%, 2.87\%, 3.46\%, 5.4\% and 4.57\% for each $\sigma$ value.

Finally, we experimented with the unlearned case of the image format.
As mentioned in Section~\ref{seam_sec_dataset}, inspired by the various uses of JPEG compression reported in \cite{deep_forensic2,deep_forensic3}, we saved the training images in the JPEG format.
In the process, we minimized the additional distortion (i.e., the rounding and truncation errors of JPEG compression \cite{deep_forensic1}) on the training set by setting the quality factor of the JPEG compression to 100 to induce the proposed network to focus on the deformation by seam carving.
The BMP is a representative uncompressed image format used as the standard image format for various applications.
To analyze the robustness of the unseen image format of our model that learned the local artifacts of seam carving applied to JPEG images with a quality of 100, we created a testing set for the BMP format.
To do this, we saved data (i.e., original, seam-removed, and seam-inserted images before JPEG compression) corresponding to the testing set used in the experiment in Table~\ref{table1} in BMP format.

\begin{table*}[t]
\caption{Performance evaluation of ILFNet and comparative convolutional neural networks with an ensemble module for three-class classification on the unseen image format $(\%)$.}
\footnotesize
    \centering
    \begin{tabu} to \linewidth{X[1.0,c] X[0.5,c] X[1.0,c] X[1.0,c] X[1.0,c] X[1.0,c] X[1.0,c] X[1.0,c] X[1.0,c] X[1.0,c] X[1.0,c]}
    \hline
    \hline
    Format & $\Theta$ & Xception & rResNet & BayarNet & HeNet & H-VGG & YeNet & LFNet & SRNet & ILFNet\\
    \hline
    \multirow{3}{*}{JPEG}& 1 &85.93 & 87.06 & 84.83 & 83.96 & 88.66 & 88.83 & 95.39 & 95.69 & 96.56\\
    & 5 & 87.62 & 89.46 & 85.67 & 85.75 & 89.75 & 91.36 & 96.96 & 96.82 & 97.02\\
    & 10 & 87.99 & 90.18 & 86.31 & 86.12 & 89.89 & 92.09 & 97.08 & 97.02 & 97.18\\\hline
    \multirow{3}{*}{BMP} & 1 & 76.73 & 80.06 & 73.73 & 75.63 & 71.57 & 78.33 & 81.57 & 79.50 & 80.26 \\
    & 5 & 79.27 & 82.74 & 77.60 & 79.23 & 73.00 & 81.80 & 83.97 & 81.27 &82.93  \\
    & 10 & 79.56 & 83.77 & 78.92 & 80.73 & 73.52 & 81.93 & 84.97 & 82.60 & 84.26 \\
    \hline
    \hline
    \end{tabu}
    \label{table8}
\end{table*}

\begin{figure*}[t]
\centering{\includegraphics[width=0.98\linewidth]{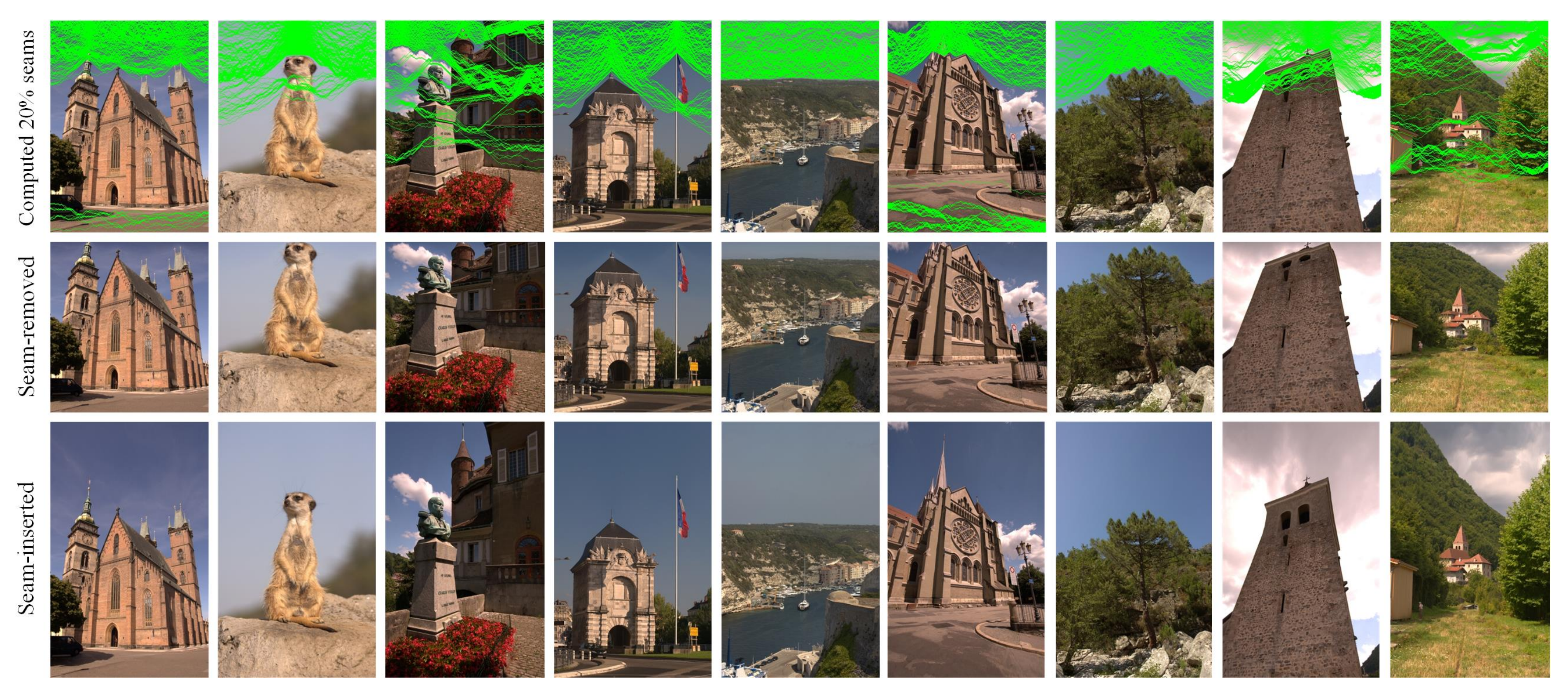}}
\caption{Examples of enlargement and reduction of image height based on computed horizontal seams. Examples are the results of seam removal and seam insertion corresponding to 20\% of the height by applying the seam-carving algorithm \cite{seam1} to the original image ($384\times512$). From top to bottom, original images with visualized computed seams, seam-removed images, and seam-inserted images are provided.}
\label{fig11}
\end{figure*}

Table~\ref{table8} presents the classification accuracies obtained by applying the models trained for the JPEG format to the testing set for the BMP format.
In the experiment, LFNet achieved the highest accuracy for each $\Theta$, and LFNet exhibited the second-best performance.
The accuracy values of our model were 80.26\%, 82.93\%, and 84.26\%, when $\Theta$ in the ensemble module was set to 1, 5, and 10, respectively.
In particular, all networks tended to improve performance as the provided number of sample data points increased.
Compared to the results for the JPEG format in Table~\ref{table8}, the classification performance for the BMP format of all networks is relatively low.
We estimated that this performance degradation was caused by rounding and truncation errors of JPEG compression remaining in the images of the training set used for training the models.
In other words, pixel distortions caused by JPEG compression with a quality factor of 100 are minor, but ILFNet, which is specialized in extracting fine-grained signals, learns these artifacts in addition to forensic features caused by seam carving.
Therefore, the performance of the model trained on JPEG images may deteriorate in the testing set consisting of an uncompressed BMP image.

Summarizing the results in this section, the proposed ILFNet achieved stable and high performance for four types of unseen cases.
Therefore, this work is suitable for practical forensics (i.e., real-world approaches) compared to other comparative networks \cite{xception,resnet,bayar,henet,h-vgg,yenet,icip,srnet}.

\subsection{Performance Evaluation of Classifying Horizontal Seam Carving}

The horizontal seam is similar to the vertical seam except for the connection being from left to right.
As stated in \cite{seam1}, a dynamic programming approach based on the energy function is performed to select the horizontal seam in the horizontal direction (i.e., left to right).
Fig.~\ref{fig11} reveals the result of image retargeting, which corresponds to 20\% of the image height, based on the horizontal seams calculated using the seam-carving algorithm \cite{seam1}.
Fig.~\ref{fig11} confirms that the image height can be adjusted while maintaining the prominent content of the image through horizontal seam carving.
Although the results in Figs.~\ref{fig1} and \ref{fig11} have a point of sameness in that image retargeting is performed using the same seam-carving algorithm, differences exist in the direction of the calculated seams; therefore, different forensic features remain in the retargeted images.
Therefore, because the CNN models employed in the previous section are trained to capture the forensic features for vertical seam carving, performance degradation occurs when the models are applied directly to the horizontal seam-carved image.

\begin{table}[t]
\caption{Performance evaluation of LFNet, SRNet, and ILFNet with an ensemble module for classifying seam-carving forgery of horizontal direction $(\%)$.}
    \centering
    \footnotesize
    \begin{tabu} to \linewidth{X[0.4,c] X[0.8,c] X[1.0,c] X[1.0,c] X[1.0,c]}
    \hline
    \hline
    \multirow{2}{*}{$\Theta$} & \multirow{2}{*}{Ratio} & \multicolumn{3}{c}{With data augmentation} \\
    \cline{3-5}
     &  & LFNet & SRNet & ILFNet \\
    \hline
    \multirow{6}{*}{1}& 10$\%$ & 77.30 & 76.06 & 80.77\\
    & 20$\%$ & 86.03 & 86.40 & 89.50 \\
    & 30$\%$ & 91.93 & 92.53 & 94.51 \\
    & 40$\%$ & 95.57 & 96.23 & 97.26 \\
    & 50$\%$ & 98.06 & 98.06 & 98.13 \\
    \cline{2-5}
    & Mixed & 89.50 & 89.90 & 92.13 \\
    \hline
    \multirow{6}{*}{5}& 10$\%$ & 83.53 & 80.80 & 86.16 \\
    & 20$\%$ & 90.52 & 90.51 & 93.70 \\
    & 30$\%$ & 95.10 & 95.20 & 97.00 \\
    & 40$\%$ & 97.83 & 98.12 & 98.75 \\
    & 50$\%$ & 99.14 & 99.27 & 99.33 \\
    \cline{2-5}
    & Mixed & 92.22 & 91.90 & 94.16\\
    \hline
    \multirow{6}{*}{10}& 10$\%$ & 84.10 & 81.66 & 87.56 \\
    & 20$\%$ & 91.03 & 90.24 & 94.23 \\
    & 30$\%$ & 95.32 & 95.23 & 97.40 \\
    & 40$\%$ & 97.88 & 98.03 & 98.91 \\
    & 50$\%$ & 99.17 & 99.34 & 99.27 \\
    \cline{2-5}
    & Mixed & 93.20 & 93.03 & 95.52\\
    \hline
    \hline
    \end{tabu}
    \label{table10}
\end{table}

In this section, a training methodology using a data augmentation approach to classify horizontally seam-carved artifacts is introduced.
To prevent the consumption of creating a new training set related on horizontal seam carving, we conducted the training process by applying rotations of 90$^{\circ}$ and 270$^{\circ}$ to the training set, which was created using vertical seam carving in Section~\ref{seam_sec_dataset}.
In the training phase for CNNs, such as LFNet, SRNet, and ILFNet, the images in the mini-batch obtained from the training set were subjected to a rotation of 90$^{\circ}$ or 270$^{\circ}$ degrees before being input into the network.
For a fair performance evaluation, a new testing set was created by applying horizontal seam carving to the original images corresponding to the testing set in Table~\ref{table1}.
Table~\ref{table10} lists the accuracy values obtained by applying the newly trained models to the testing set containing horizontal seam-carved artifacts.
Compared to LFNet and SRNet, ILFNet achieved higher performance for all cases.
In particular, the accuracy values of ILFNet were 92.13\%, 94.16\%, and 95.52\%, when $\Theta$ was set to 1, 5, and 10, respectively.
Thus, we confirmed that seam-carved artifacts can be explored in different directions from the data constituting the training set through a data augmentation-based training methodology.

\subsection{Performance Evaluation with a Conventional Approach}
\label{seam_sec_conventional}

To further demonstrate the effectiveness of our work, we evaluated the performance of ILFNet and the conventional handcrafted feature-based method \cite{seam_ryu}, referred to as the Ryu method, by measuring the accuracy of two-class classification (i.e., seam insertion versus original and seam removal versus original).
The Ryu method consists of two algorithms for detecting seam insertion and seam removal, whereas only one trained model enables classifying two types of seam-carved forgery in the case of ILFNet.
For the Ryu method, seam insertion is detected using a candidate map focused on the relationship between adjacent pixels, and seam removal detection is performed using learning feature vectors with an SVM classifier.
The weighting factor $t$ was set to 0.85, and a LIBSVM classifier with the radial basis function of $r$ = 0.125 is employed.
The testing set that was used to derive the results listed in Table~\ref{table2} was also exploited in this experiment.
Table~\ref{table11} reveals the classification accuracy of ILFNet and the Ryu method, as measured by changes in the retargeting ratio.
For the mixed testing set, the proposed ILFNet achieved higher accuracy values of 97.64\% and 95.90\%, respectively, for seam-insertion and seam-carving classifications than the Ryu method.
For two-class classification, the Ryu method performed worse than the proposed ILFNet but achieved slightly higher accuracy over some CNN-based approaches.

\begin{table}[t]
\caption{Performance evaluation of ILFNet and the conventional handcrafted feature-based method on various retargeting ratios $(\%)$.}
    \centering
    \footnotesize
    \begin{tabu} to \linewidth{X[0.8,c] X[1.2,c] X[1.0,c] X[1.2,c] X[1.0,c] }
    \hline\hline
    \multirow{2}{*}{Ratio} & \multicolumn{2}{c}{Seam insertion}& \multicolumn{2}{c}{Seam removal}\\
    \cmidrule(r){2-3} \cmidrule(r){4-5}
   & Ryu method & ILFNet & Ryu method & ILFNet \\
    \hline
    10$\%$ & 84.25 & 93.50 & 68.05 & 88.20 \\
    20$\%$ & 89.85 & 97.45 & 79.07 & 94.39 \\
    30$\%$ & 90.25 & 98.94 & 87.60 & 98.10 \\
    40$\%$ & 90.45 & 99.29 & 94.05 & 99.14 \\
    50$\%$ & 90.65 & 99.45 & 96.15 & 99.32 \\
    \hline
    Mixed & 89.21 & 97.64 & 85.02 & 95.90 \\
    \hline
    \hline
    \end{tabu}
    \label{table11}
\end{table}

\subsection{Feature Map Visualization}

\begin{figure*}[t]
\centering{\includegraphics[width=1.0\linewidth]{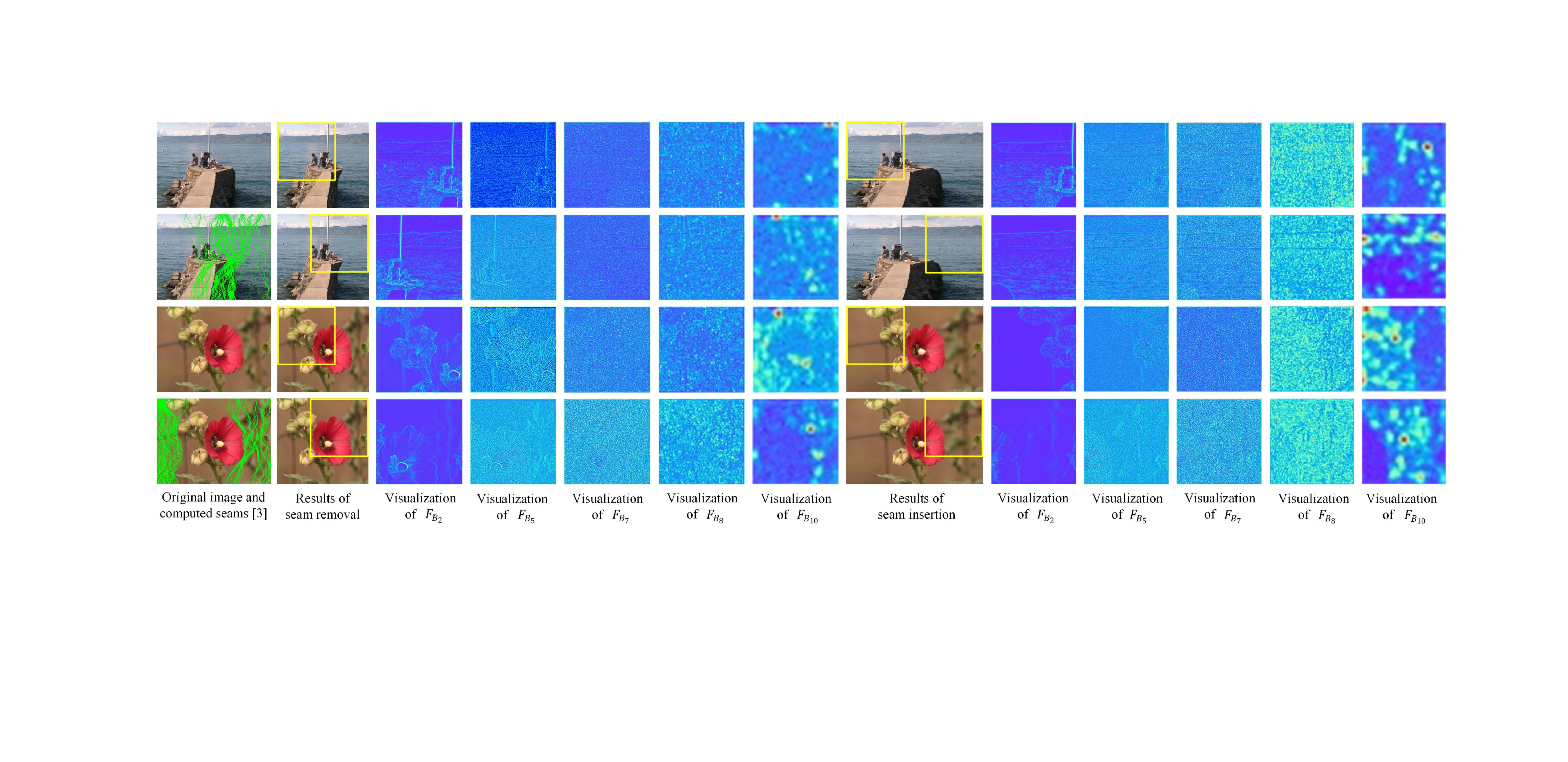}}
\caption{Visualization of feature maps obtained from the network blocks of ILFNet. Image retargeting corresponds to 20\% of the width of the original image ($512\times384$) by employing vertical seam carving \cite{seam1}, and cropped samples represented in yellow are input into ILFNet.}
\label{fig12}
\end{figure*}

In this section, feature maps obtained from the network blocks, including BT-1, BT-2, BT‑3, and BT-4, were visualized to analyze the ability of ILFNet to explore and extract seam-carving artifacts.
To do this, we input samples cropped from images with 20\% seam removal and seam insertion to the trained model.
Fig.~\ref{fig12} presents the results of visualizing the feature maps.
Moreover, $B_i$ and $F_{B_i}$ represent the $i$-th network block that constitutes ILFNet and the feature maps obtained from $B_i$, respectively, where $i=\{1,...,10\}$.
As illustrated in Fig.~\ref{fig12}, we visualized $F_{B_2}$, $F_{B_5}$, $F_{B_7}$, $F_{B_8}$, and $F_{B_{10}}$ by applying averaging to the channel of each feature map.
Lighter-colored areas refer to areas with higher energy values. Based on the visualization results of a specific $F_{B_i}$, an analysis of whether each block constituting the ILFNet learns and operates as intended is introduced.

\begin{figure*}[t]%
\centering%
\subfigure[]{%
  \includegraphics[height=1.7in]{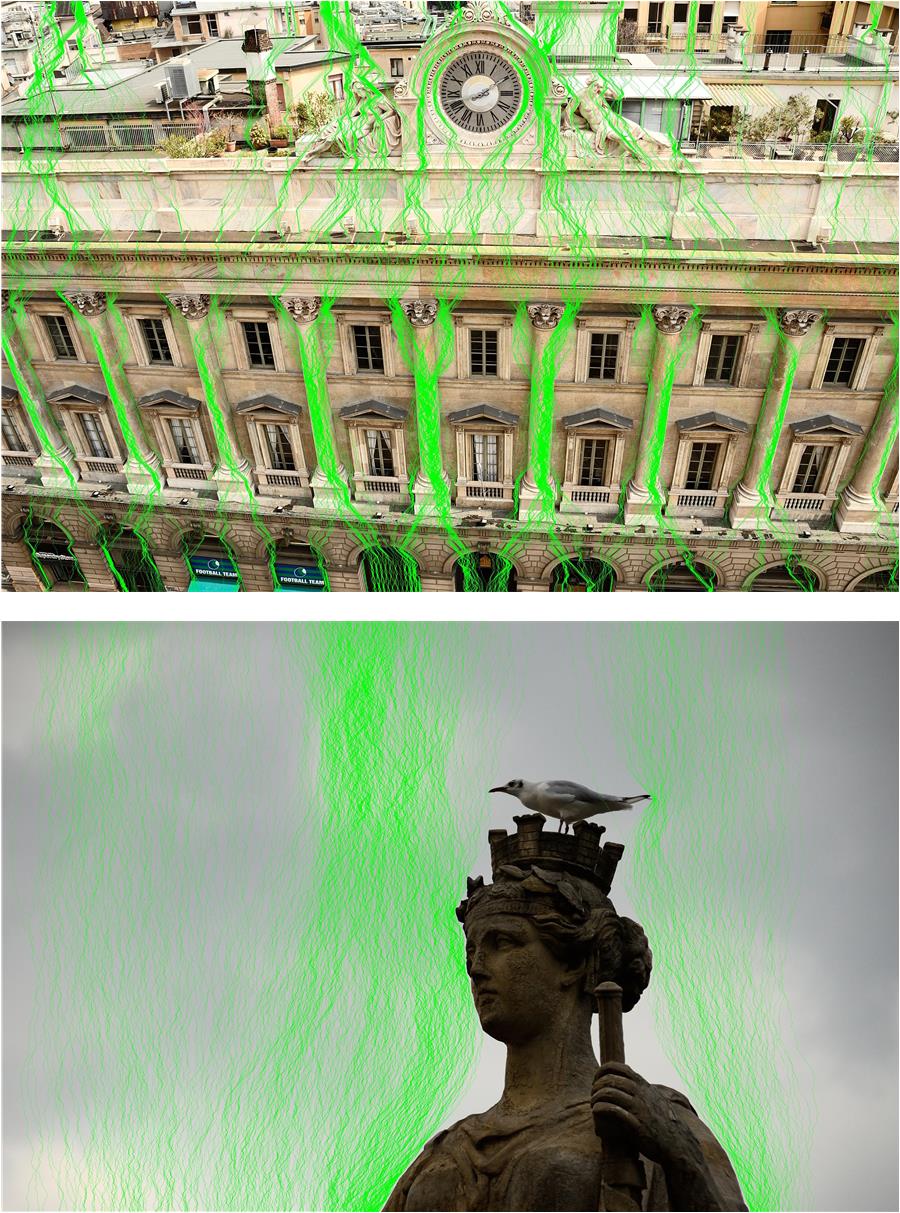}
  \label{fig13:ex:a}
}\hspace{-2mm}
\subfigure[]{%
  \includegraphics[height=1.7in]{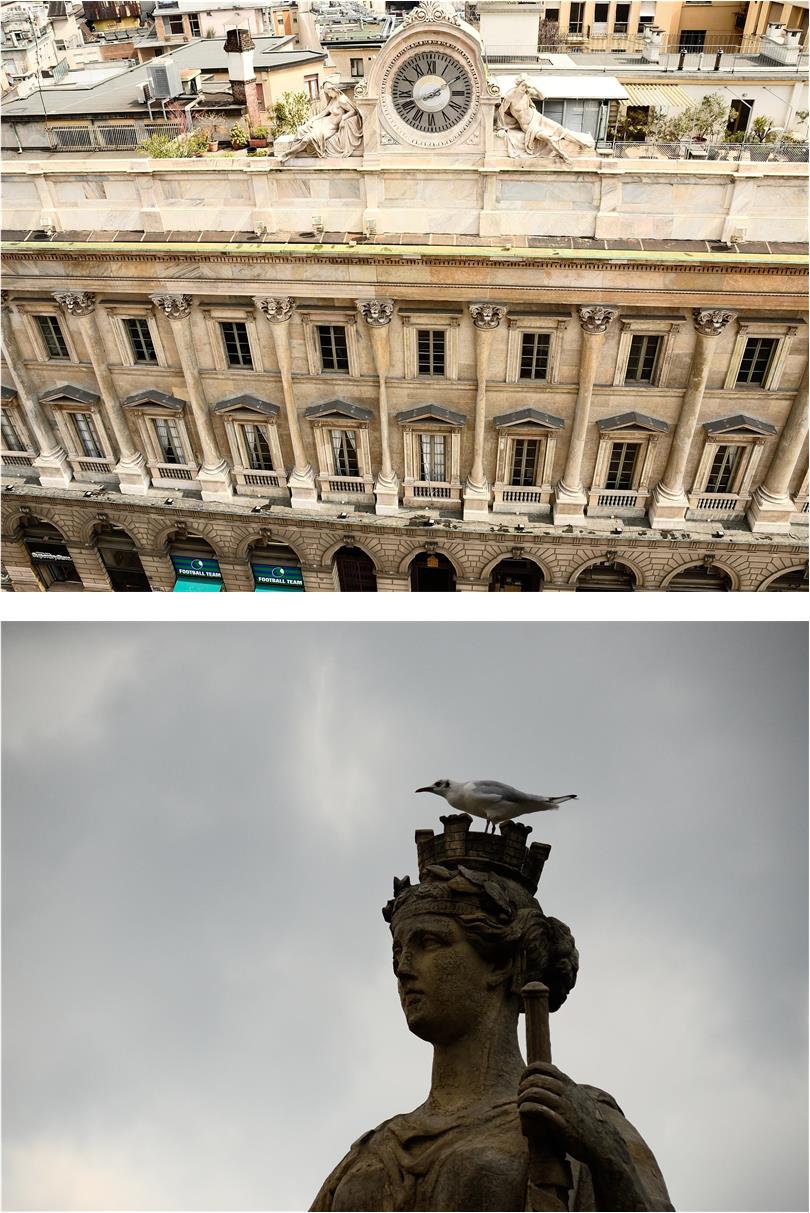}
  \label{fig13:ex:b}
}\hspace{-2mm}
\subfigure[]{%
  \includegraphics[height=1.7in]{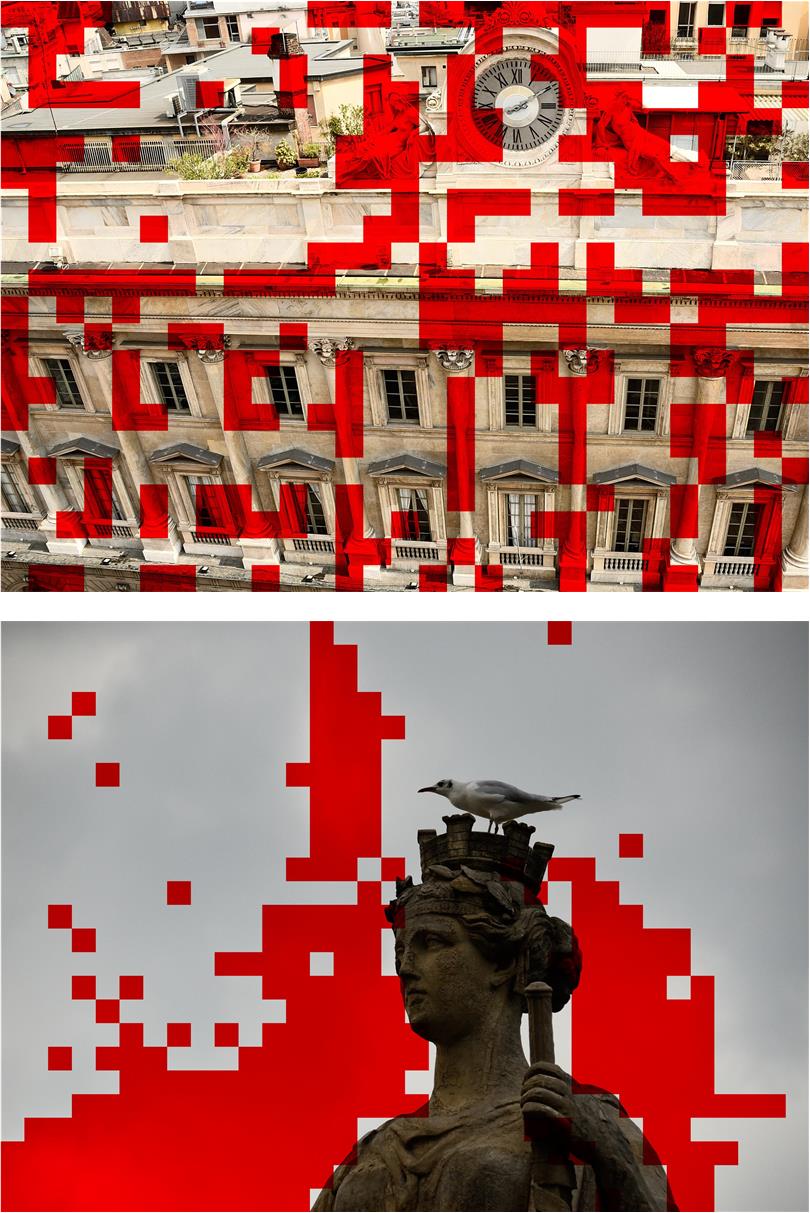}%
  \label{fig13:ex:c}
}\hspace{-2mm}
\subfigure[]{%
  \includegraphics[height=1.7in]{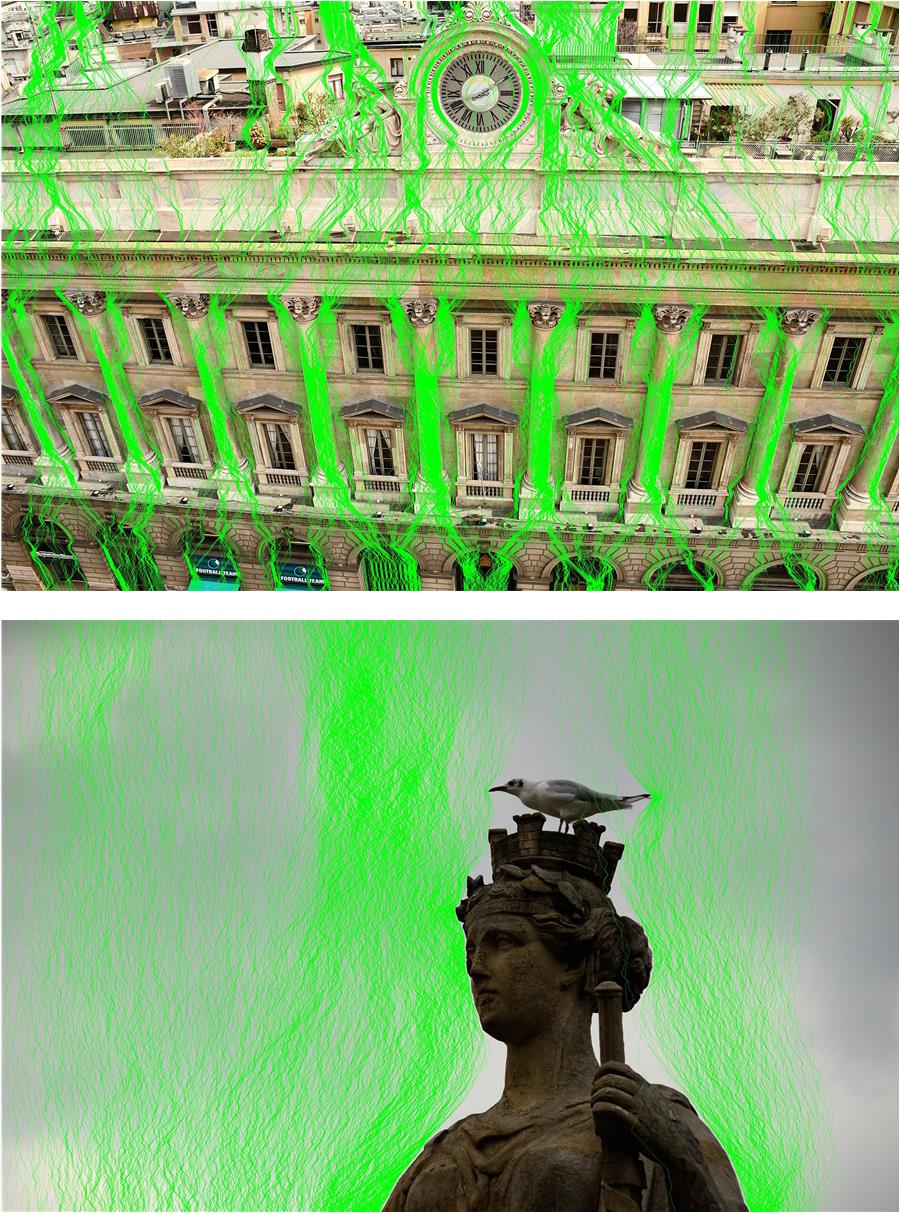}%
  \label{fig13:ex:d}
}\hspace{-2mm}
\subfigure[]{%
  \includegraphics[height=1.7in]{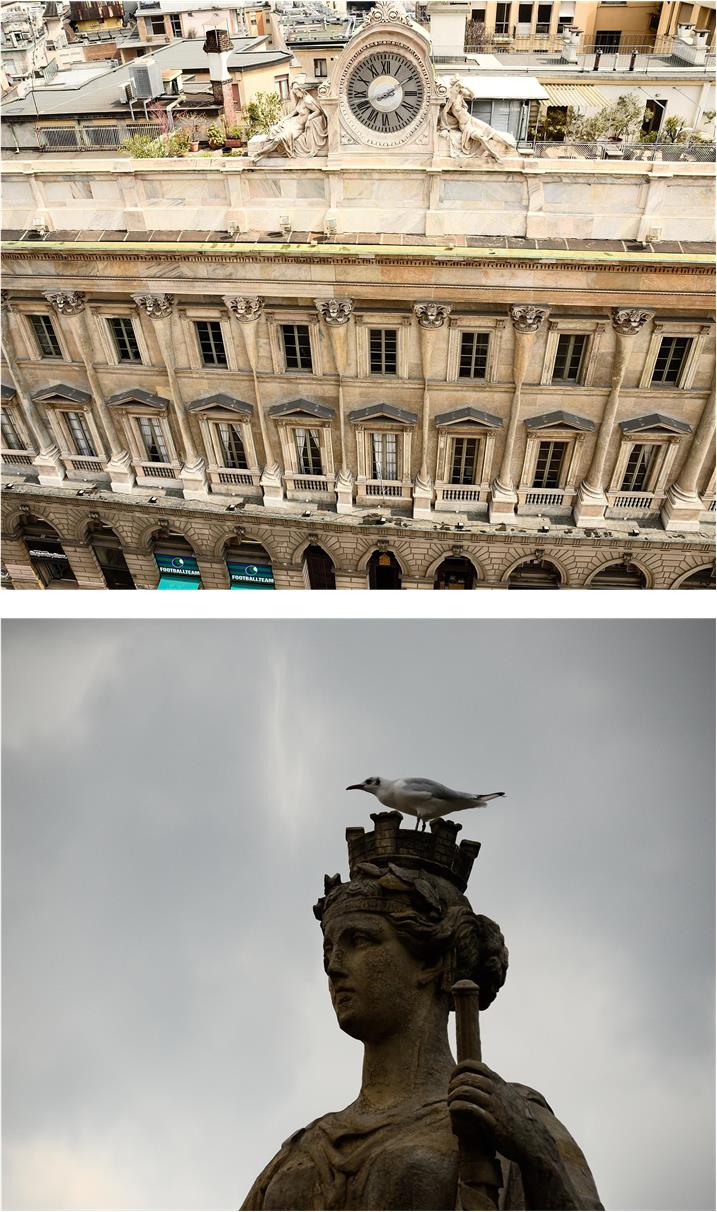}%
  \label{fig13:ex:e}
}\hspace{-2mm}
\subfigure[]{%
  \includegraphics[height=1.7in]{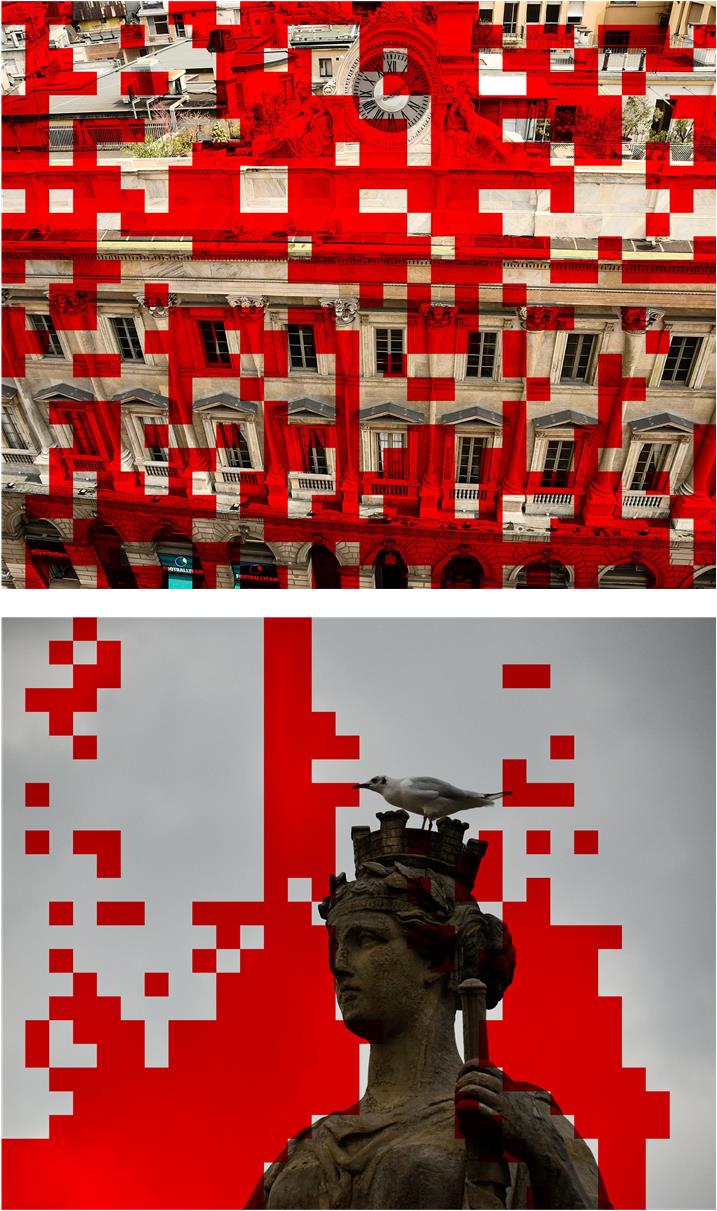}%
  \label{fig13:ex:f}
}
\caption{Localization results for seam-removed artifacts of the proposed network: (a) original images ($4224\times2816$) with seams corresponding to 5\% of the width marked in green, (b) 5\% seam-removed images, (c) results of seam removal localization, (d) original images ($4224\times2816$) with seams corresponding to 10\% of the width marked in green, (e) 10\% seam-removed images, (f) results of seam removal localization. For (c) and (f), seam-removed regions are marked in red.}
\label{fig13}%
\end{figure*}

First, BT-1 and BT-2 constituting the front segment of ILFNet were induced to extract noise-like signals by focusing on the differences between adjacent pixels of the sample.
From the results of visualization on $F_{B_2}$ and $F_{B_5}$, BT-1 and BT-2 are activated on subtle differences between adjacent pixels, as we intended.
Next, we induced higher and refined features to be extracted and learned from the feature maps (i.e., $F_{B_5}$) through a middle segment comprising consecutive BT-3s.
Unlike the visualization results for $F_{B_2}$ and $F_{B_5}$, where the energy was concentrated on the edges of the prominent object, the energy is globally distributed in the entire region of visualization results of $F_{B_7}$.
We propose that these results are due to local residual learning and the ability of refined and higher feature learning of the local feature fusion-based BT-3 from the feature maps generated from the previous block.

Finally, we induced hierarchical feature learning by placing BT-4, which contains the AvgPool layer for the dimensionality reduction of feature maps, in deeper layers of ILFNet.
The visualization results of $F_{B_{10}}$ have larger energy values in areas similar to the areas where local artifacts are generated due to seam carving (see examples for the 1$st$, 7$th$, and 13$th$ columns in Fig.~\ref{fig12}).
Therefore, ILFNet can focus on the local area where seam carving is applied.
In particular, visualization results based on $F_{B_{10}}$ for seam-inserted images exhibit a higher contrast and more meaningful prediction than those for seam removal.
Capturing artifacts of seam removal is presumed to be more difficult than seam insertion because forensic feature extraction proceeds by focusing on the differences between adjacent pixels due to the loss of information.

\subsection{Localization Results}

This section presents the results of seam-removed and seam-inserted region localization.
For this experiment, we newly trained ILFNet using sample images of size $128\times 128$ cropped from the training set described in Section~\ref{seam_sec_dataset}, where $W$ $=$ $H$ $=$ $128$.
This was considered for more sophisticated localization and to evaluate the scalability of ILFNet against image size. In particular, localization using a small patch is more effective when the size of the test image is small.
Thus, based on the training methodology in Section~\ref{seam_sec_training_setting}, ILFNet has been newly trained on $128\times 128$ images, and the performance of the trained model is specified in Table~\ref{table12}.
The accuracy of ILFNet for the mixed testing set is 91.17\%.
This performance is 5.39\% lower than that of the model based on $256\times256$ samples, which occurs because the trace of seam carving decreases as the sample size decreases.

\begin{table}[t]
\caption{Classification accuracy of the proposed ILFNet trained on $128\times128$ images $(\%)$.}
    \centering
    \footnotesize
    \begin{tabu} to \linewidth{X[1.2,c] X[1.0,c] X[1.0,c] X[1.0,c] X[1.0,c] X[1.0,c] X[1.0,c]}
    \hline
    \hline
    \multirow{2}{*}{Model} & \multicolumn{6}{c}{Retargeting ratio} \\
    \cline{2-7}
    & 10$\%$ & 20$\%$ & 30$\%$ & 40$\%$ & 50$\%$ & Mixed \\
    \hline
    ILFNet & 80.83 & 88.79 & 92.87 & 95.53 & 96.27 & 91.17\\
    \hline
    \hline
    \end{tabu}
    \label{table12}
\end{table}

\begin{figure*}[t]%
\centering%
\subfigure[]{%
  \includegraphics[height=2.2in]{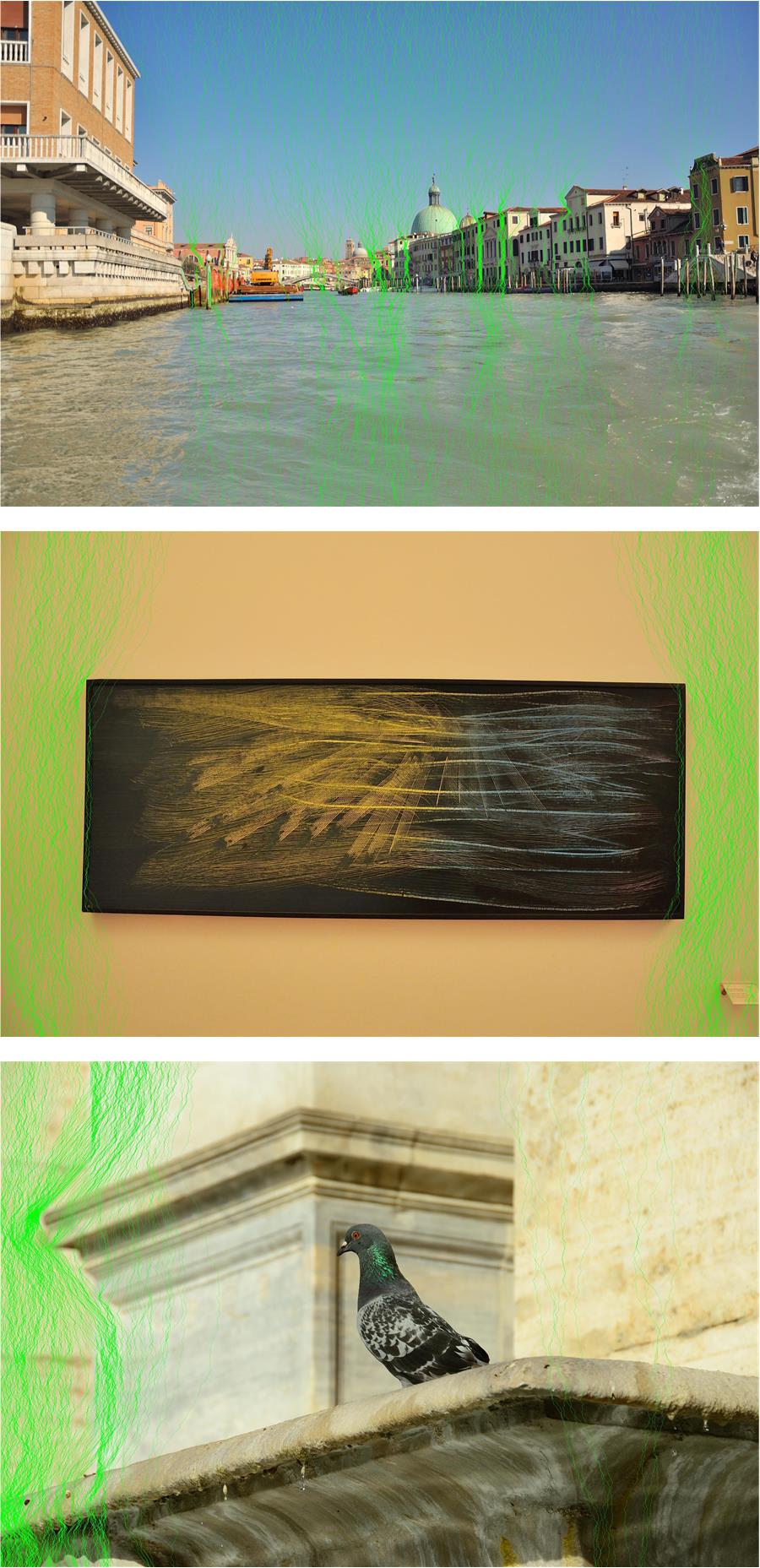}
  \label{fig14:ex:a}
}\hspace{-2mm}
\subfigure[]{%
  \includegraphics[height=2.2in]{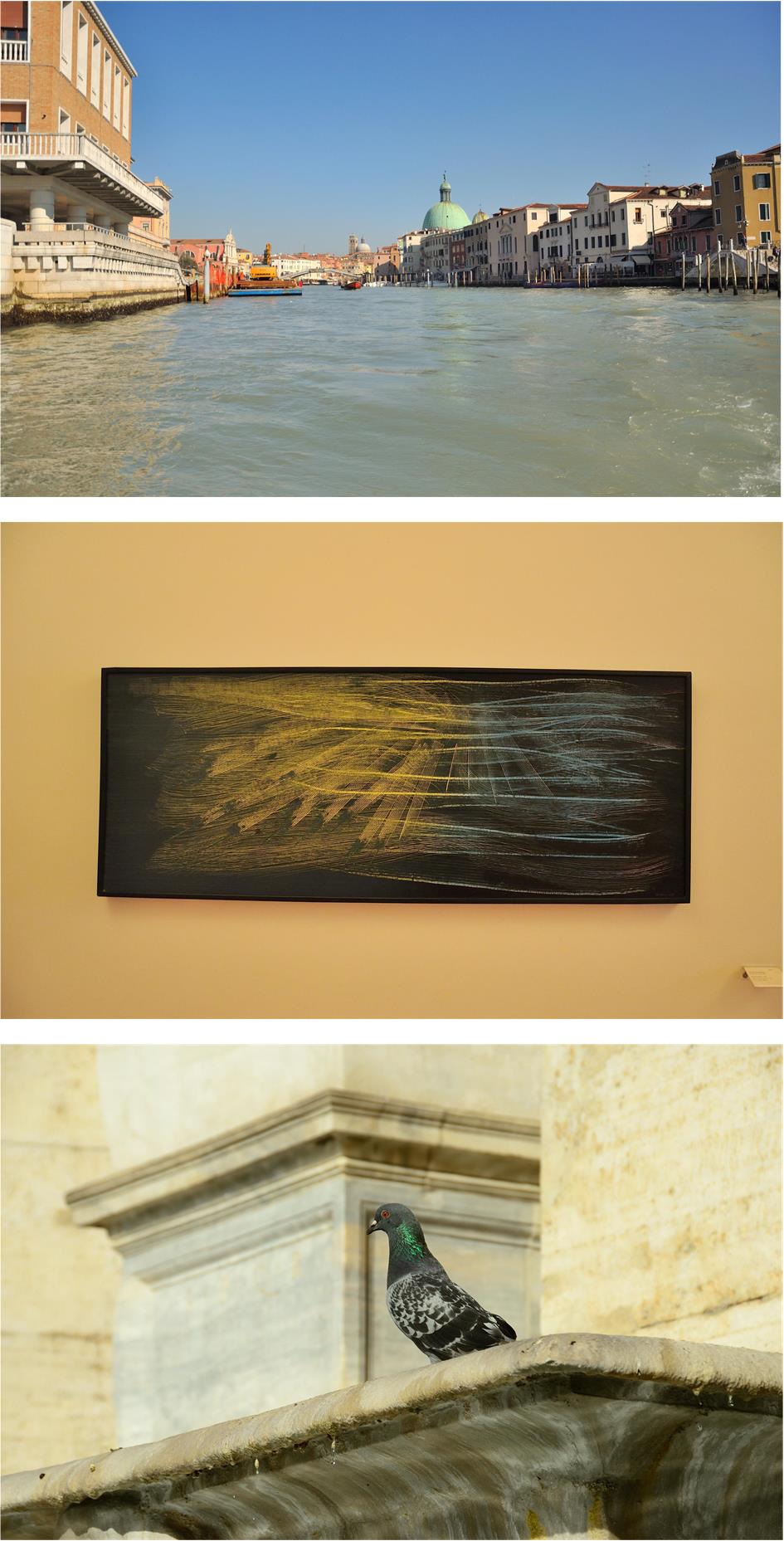}
  \label{fig14:ex:b}
}\hspace{-2mm}
\subfigure[]{%
  \includegraphics[height=2.2in]{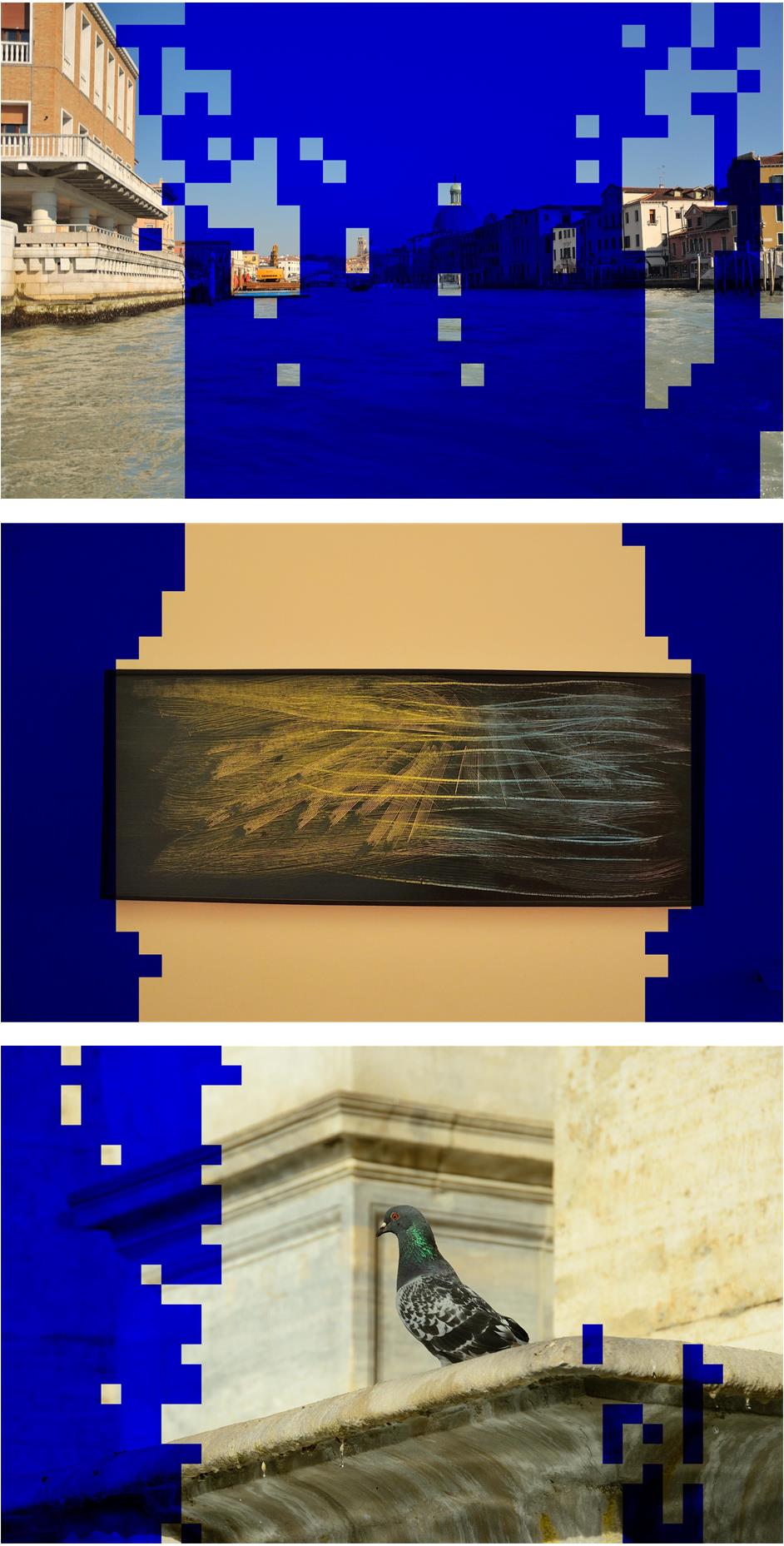}%
  \label{fig14:ex:c}
}\hspace{-2mm}
\subfigure[]{%
  \includegraphics[height=2.2in]{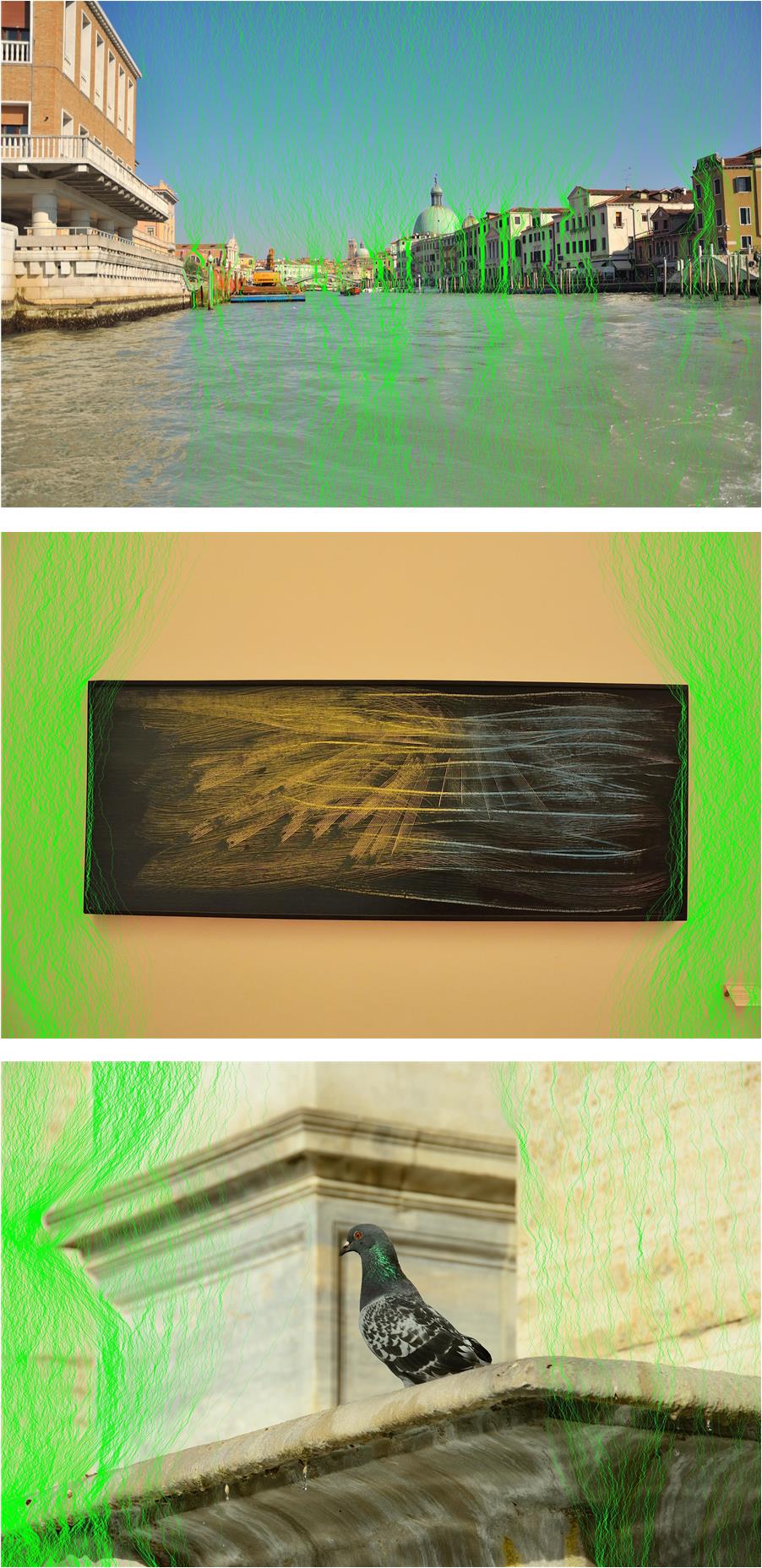}%
  \label{fig14:ex:d}
}\hspace{-2mm}
\subfigure[]{%
  \includegraphics[height=2.2in]{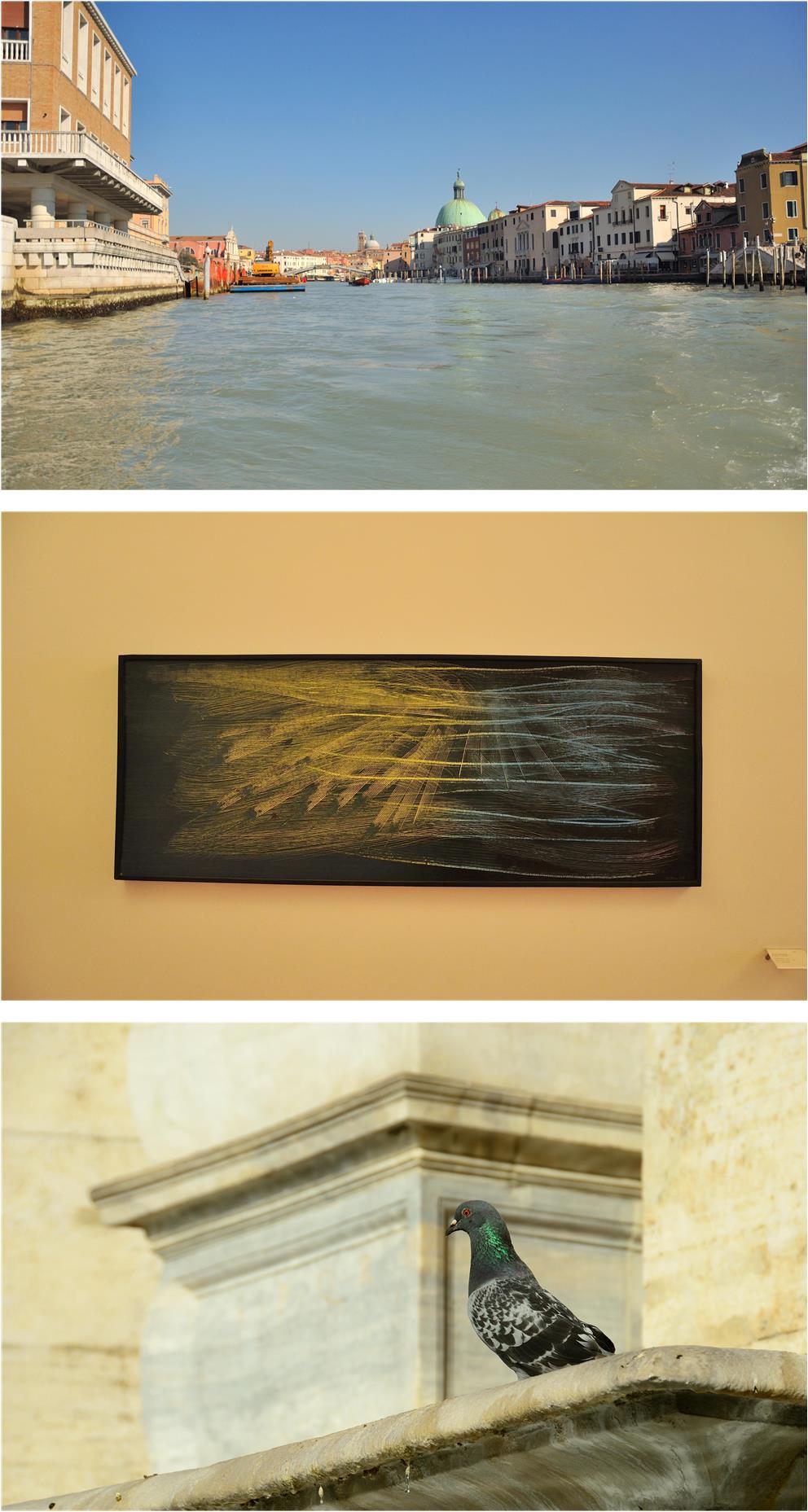}%
  \label{fig14:ex:e}
}\hspace{-2mm}
\subfigure[]{%
  \includegraphics[height=2.2in]{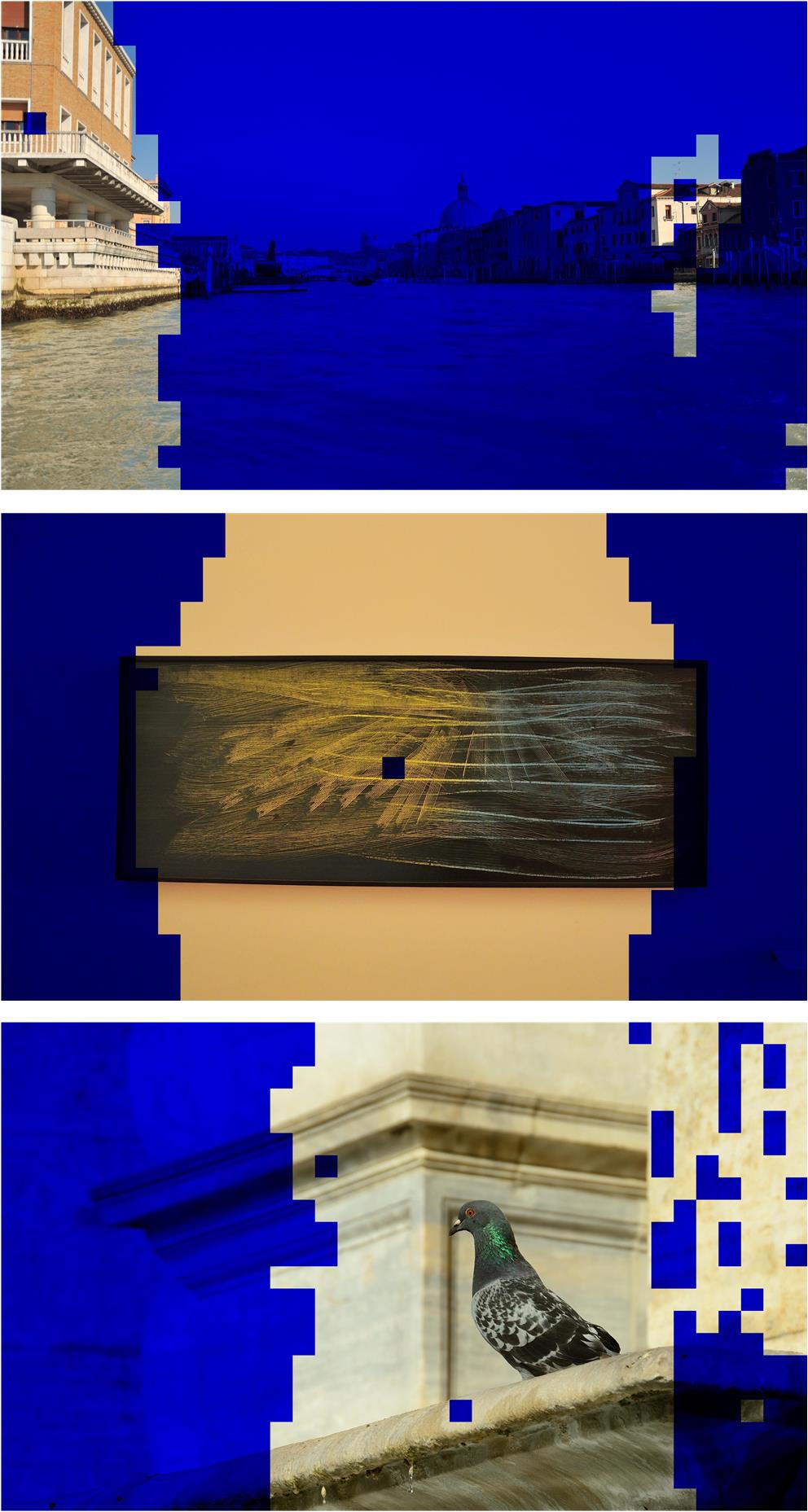}%
  \label{fig14:ex:f}
}
\caption{Localization results for seam-inserted artifacts of the proposed network: (a) original images ($4224\times2816$) with seams corresponding to 5\% of the width marked in green, (b) 5\% seam-inserted images, (c) results of seam-insertion localization, (d) original images ($4224\times2816$) with seams corresponding to 10\% of the width marked in green, (e) 10\% seam-inserted images, (f) results of seam-insertion localization. For (c) and (f), seam-inserted regions are marked in blue.}
\label{fig14}%
\end{figure*}

We conducted localization for local seam-carved areas by applying the trained model to large test images. In the experiments, image retargeting corresponding 5\% and 10\% of the width was applied to the test image $(4224\times2816)$ in the RAISE \cite{raise} dataset.
The seam-carved images, including both enlargement and reduction, were divided into patches with a stride of $128$. Then, we performed a patch-level classification to localize the manipulated regions.
Figs.~\ref{fig13} and \ref{fig14} illustrate the proposed model-based localization results for seam removal and seam insertion, respectively.
The figures reveal that ILFNet localizes the manipulated local area relatively accurately.
For seam insertion, fewer false positives were found that for seam removal.
In addition, more accurate localization is possible when the retargeting ratio increases, which may be because, as the ratio increases, the traces of seam carving in the image are more enriched.
Although some error cases exist, the proposed ILFNet effectively explores and captures the artifacts of seam-carving forgery.

\section{Conclusion}
This paper proposes a CNN-based forensic framework that learns and captures local texture artifacts caused by seam-carving forgery.
Learning low-level forensic features requires a different approach from the general CNN for learning content-dependent features.
To address this issue, we designed the proposed ILFNet comprising five types of network blocks, which are specialized for learning forensic features.
Furthermore, an ensemble module for enhancing classification performance and comprehensively analyzing the features in the local areas of the given test images was presented.
To demonstrate the effectiveness of the proposed ILFNet, extensive experiments were conducted with comparative CNNs and a non-CNN-based approach.
Compared to the comparative classifiers, our work exhibits state-of-the-art performance in terms of classifying seam forgery artifacts.
In addition, our trained model with the ensemble module also demonstrated high performance for the testing set of unseen cases.
The experimental results also demonstrate that our method can be applied to localize both seam-removed and seam-inserted areas.
In future work, we will apply ILFNet to datasets with various JPEG quality factors and improve the classification performance by refining the network architecture.

\ifCLASSOPTIONcaptionsoff
  \newpage
\fi

\end{document}